\documentclass{aa}  
\usepackage{graphicx}
\usepackage{txfonts}

\usepackage{natbib}
\bibpunct{(}{)}{;}{a}{}{,} % to follow the A&A style

\usepackage[usenames]{color} % See ~/Misc/color_latex/color-package-demo.tex

\newcommand{\vpecrest}{v_\mathrm{pec}^\mathrm{rest}}
\newcommand{\zcl}     {z_\mathrm{cl}}
\newcommand{\zest}    {z_\mathrm{est}}
\newcommand{\sigmacl} {\sigma_\mathrm{cl}}
\newcommand{\sigmaspec} {\sigma_\mathrm{spec}}
\newcommand{\sigmalens} {\sigma_\mathrm{lens}}
\newcommand{\Nmemzero}{N_\mathrm{mem,0}}
\newcommand{\Nmemzeroone}{N_\mathrm{mem,01}}
\newcommand{\Nmemzeroonetwo}{N_\mathrm{mem,012}}
\newcommand{\zspec}     {z_\mathrm{spec}}
\newcommand{\zphot}     {z_\mathrm{phot}}
\newcommand{\zphotlow}  {z_\mathrm{phot,low}}
\newcommand{\zphothigh} {z_\mathrm{phot,high}}
\newcommand{\Ngal}      {N_\mathrm{gal}}
\newcommand{\Nstar}     {N_\ast}
\newcommand{\zclP}      {z_\mathrm{cl}^\mathrm{P}}
\newcommand{\zclfinal}  {z_\mathrm{cl}^\mathrm{final}}
\newcommand{\FWHMmin}   {\mathrm{FWHM}_\mathrm{min}}
\newcommand{\Itot}      {I_\mathrm{tot}}

\newcommand{\Ione}      {I_{1}}
\newcommand{\Ionebright}{I_{1,\mathrm{bright}}}
\newcommand{\Ionefaint} {I_{1,\mathrm{faint}}}
\newcommand{\sigmaCCD}{\sigma_\mathrm{CCD}}
\newcommand{\sigmaCCDunrebinned}{\sigma_\mathrm{CCD}^\mathrm{unrebinned}}
\newcommand{\sigmaCCDrebinned}{\sigma_\mathrm{CCD}^\mathrm{rebinned}}
\newcommand{\sigmaCCDrebinnedtilde}{\tilde{\sigma}_\mathrm{CCD}^\mathrm{rebinned}}
\newcommand{\nave}{n_\mathrm{ave}}
\newcommand{\Iskysubtr}{I_\mathrm{skysubtr}}
\newcommand{\Inonskysubtrunrebinned}{I_\mathrm{non\,skysubtr}^\mathrm{unrebinned}}
\newcommand{\Inonskysubtrrebinned}{I_\mathrm{non\,skysubtr}^\mathrm{rebinned}}

\newcommand{\xr}        {x_\mathrm{r}}
\newcommand{\yt}        {y_\mathrm{t}}
\newcommand{\VRsynth}{(V-R)_\mathrm{synth}}
\newcommand{\VRphot} {(V-R)_\mathrm{phot}}

\newcommand{\procedurename}[1]{\texttt{#1}}

\newlength{\thirdwidth}                    % \thirdwidth is now 0in
\newlength{\fourthwidth}                    % \fourthwidth is now 0in
\begin{document}
   \title{Spectroscopy of clusters in the ESO distant cluster survey
          (EDisCS).II.\thanks{%
   Based on observations collected at the European Southern Observatory, Chile,
   as part of large programme 166.A--0162 (the ESO Distant
   Cluster Survey).}\thanks{%
   Full Table 4 is only available in electronic form at the CDS via
   anonymous ftp to \texttt{cdsarc.u-strasbg.fr (130.79.125.5)}
   or via \texttt{http://cdsweb.u-strasbg.fr/Abstract.html}}
   }

   \subtitle{Redshifts, velocity dispersions, and substructure for clusters in the last 15 fields}

\author{
Bo Milvang-Jensen\inst{1,2,3}
\and
Stefan Noll\inst{1,4}
\and
Claire Halliday\inst{5}
\and
Bianca M. Poggianti\inst{6}
\and
Pascale Jablonka\inst{7}
\and
Alfonso Arag{\'o}n-Salamanca\inst{8}
\and
Roberto P. Saglia\inst{1}
\and
Nina Nowak\inst{1}
\and
Anja von der Linden\inst{9}
\and
Gabriella De Lucia\inst{9}
\and
Roser Pell{\'o}\inst{10}
\and
John Moustakas\inst{11}
\and
S{\'e}bastien Poirier\inst{12}
\and
Steven P. Bamford\inst{13}
\and
Douglas I. Clowe\inst{14}
\and
Julianne J. Dalcanton\inst{15}
\and
Gregory H. Rudnick\inst{16}
\and
Luc Simard\inst{17}
\and
Simon D. M. White\inst{9}
\and
Dennis Zaritsky\inst{18}
}

\authorrunning{B. Milvang-Jensen et al.}
\titlerunning{Spectroscopy of clusters in EDisCS II}

   \offprints{B. Milvang-Jensen}

\institute{
Max-Planck-Institut f{\"u}r extraterrestrische Physik,
Giessenbachstrasse, D--85748 Garching bei M{\"u}nchen, Germany
\and
Dark Cosmology Centre, Niels Bohr Institute, University of Copenhagen,
Juliane Maries Vej 30, DK--2100 Copenhagen {\O}, Denmark;
\texttt{milvang@astro.ku.dk} % \texttt{milvang@dark-cosmology.dk}
\and
The Royal Library / Copenhagen University Library, Research Dept.,
Box 2149, DK--1016 Copenhagen K, Denmark
\and
Observatoire Astronomique Marseille Provence,
Laboratoire d'Astrophysique de Marseille, Traverse du Siphon,
13376 Marseille Cedex 12, France
\and
Osservatorio Astrofisico di Arcetri, Largo E. Fermi 5, 50125 Firenze, Italy
\and
Osservatorio Astronomico di Padova, Vicolo dell'Osservatorio 5, 35122 Padova,
Italy
\and
Universit{\'e} de Gen{\`e}ve, Laboratoire d'Astrophysique
de l'Ecole Polytechnique F{\'e}d{\'e}rale de Lausanne (EPFL),
Observatoire, CH--1290 Sauverny, Switzerland
\and
School of Physics and Astronomy, University of Nottingham,
University Park, Nottingham NG7 2RD, UK
\and
Max-Planck-Institut f\"{u}r Astrophysik,
Karl-Schwarzschild-Strasse 1, % Postfach 1317,
D--85748 Garching bei M{\"u}nchen, Germany
\and
Laboratoire d'Astrophysique de Toulouse-Tarbes, CNRS, Universit\'e de
Toulouse, 14 Avenue Edouard Belin, F--31400 Toulouse, France
\and
New York University, 4 Washington Place, New York, NY 10003, USA
\and
GEPI, CNRS-UMR8111, Observatoire de Paris, section de Meudon,
5 Place Jules Janssen, F--92195 Meudon Cedex, France
\and
ICG, University of Portsmouth, Mercantile House, Hampshire Terrace,
Portsmouth, PO1 2EG, UK
\and
Ohio University, Department of Physics and Astronomy, Clippinger Labs 251B,
Athens, OH 45701, USA
\and
Department of Astronomy, University of Washington, Box 351580,
Seattle WA 98195--1580, USA
\and
NOAO, 950 North Cherry Avenue, Tucson AZ 85719, USA
\and
Herzberg Institute of Astrophysics, National Research
Council of Canada, 5071 West Saanich Road, Victoria, BC V9E 2E7,
Canada
\and
Steward Observatory, University of Arizona, 933 North Cherry Avenue,
Tucson, AZ 85721, USA
}

\date{Received 27 November 2007 / Accepted 22 January 2008}

\abstract
{}
{We present spectroscopic observations of galaxies in 15 survey fields
as part of the ESO Distant Cluster Survey (EDisCS)\@.
We determine the redshifts and velocity dispersions of
the galaxy clusters located
in these fields, and we test for possible substructure in the clusters.}
{We obtained multi-object mask spectroscopy using the FORS2 instrument
at the VLT\@.
We reduced the data with particular attention to the sky subtraction.
We implemented the method of Kelson for
performing sky subtraction prior to any rebinning/interpolation of the data.
 From the measured galaxy redshifts, we determine cluster velocity dispersions
using the biweight estimator and test for possible substructure in
the clusters using the Dressler--Shectman test.}
{The method of subtracting the sky prior to any rebinning/interpolation of
the data delivers photon-noise-limited results, whereas the traditional
method of subtracting the sky after the data have been rebinned/interpolated 
results in substantially larger noise for spectra from tilted slits.
Redshifts for individual galaxies are presented and
redshifts and velocity dispersions are presented for
21 galaxy clusters. % at redshifts between 0.40 and 0.96.
For the 9 clusters with at least 20 spectroscopically confirmed members,
we present the statistical significance of the presence of substructure
obtained from the Dressler--Shectman test, and substructure is detected
in two of the clusters.}
{Together with data from our previous paper,
spectroscopy and spectroscopic velocity dispersions are now available
for 26 EDisCS clusters with redshifts in the range 0.40--0.96 and
velocity dispersions in the range
$166\,\mathrm{km}\,\mathrm{s}^{-1}$--$1080\,\mathrm{km}\,\mathrm{s}^{-1}$.}


   \maketitle

\section{Introduction}

Galaxy clusters provide important environments for the study of galaxy
evolution out to high redshift
(e.g.\ \citealt{Aragon-Salamanca_etal:1993}). They are detectable at
optical wavelengths (e.g.\
\citealt{Abell:1958}; \citealt{Zwicky_etal:1968};
\citealt{Shectman:1985}; \citealt{Gunn_etal:1986};
\citealt{Couch_etal:1991}; \citealt{Postman_etal:1996};
\citealt{Gladders_Yee:2000}; \citealt{Gonzalez_etal:2001};
\citealt{Wilson_etal:2006}; \citealt{Scoville_etal:2007}), by the
X-ray (e.g.\ \citealt{Rosati_etal:1995};
\citealt{Finoguenov_etal:2007} and references therein) and radio
\citep{Feretti_Giovannini:2007} emission of their intracluster medium
(ICM) and point sources, and by the scattering of cosmic background
radiation by ICM free electrons via the Sunyaev-Zel'dovich effect
(e.g.\ \citealt{Sunyaev_Zeldolvich:1970};
\citealt{Carlstrom_etal:2002}; \citealt{Birkinshaw_Lancaster:2007}).

Spectroscopic surveys of cluster galaxies began with work on Coma
and Perseus by \citet{Kent_Gunn:1982} and
\citet{Kent_Sargent:1983}, and progressed to the systematic studies of
tens of local clusters by \citet{Dressler_Shectman:1988} and
\citet{Zabludoff_etal:1990}.  Currently, the most comprehensive
spectroscopic catalogues of local galaxy clusters are provided by
applying a variety of selection techniques to large-area surveys,
primarily the Sloan Digital Sky Survey (SDSS) from which the C4
\citet{Miller_etal:2005} and MaxBCG \citet{Koester_etal:2007}
catalogues are produced.

Intermediate redshift ($z \sim 0.2$--$0.7$)
spectroscopic cluster surveys arrived with the work of the CNOC
\citep{Yee_etal:1996,Balogh_etal:1997} and MORPHS
\citep{Dressler_etal:1999,Poggianti_etal:1999} collaborations.
Further kinematic studies of individual and small samples of
intermediate redshift clusters include those by
\citet{Kelson_etal:1997,Kelson_etal:2006}, \citet{Tran_etal:2003},
\citet{Bamford_etal:2005}, \citet{Seroteroos_etal:2005} and
\citet{Moran_etal:2005}.

At higher redshift ($z \sim 0.7$--$1.3$) surveys have been completed
by \citet{Postman_etal:1998,Postman_etal:2001}, and a number of
individual clusters have been studied
\citep[e.g.][]{vanDokkum_etal:2000,Jorgensen_etal:2005,Jorgensen_etal:2006,Tanaka_etal:2006,Demarco_etal:2007,Tran_etal:2007}.
However, to date our knowledge of clusters beyond $z \sim 0.5$ has
been generally limited to the highest-mass systems.

Cluster velocity dispersions provide a measure of cluster mass
(\citealt{Fisher_etal:1998}; \citealt{Tran_etal:1999};
\citealt{Borgani_etal:1999}; \citealt{Lubin_etal:2002}).
The measurement of cluster velocity dispersions should be made using
statistics insensitive to galaxy redshift outliers and the shape of
the velocity distribution, e.g.\ the biweight scale and gapper
estimators proposed by \citet{Beers_etal:1990}
(e.g.\ \citealt{Halliday_etal:2004}). Galaxy clusters may however have
cluster substructure (\citealt{Dressler_Shectman:1988};
\citealt{Geller_Beers:1982}).
Substructure can take many forms (e.g., bimodality, small clumps,
filaments) and its detection constitutes a non--trivial technical
problem. Many statistical tests have been developed and applied to
reasonably large samples of clusters over the past decades. They all
agree with the conclusion that substructure is an important
phenomenon, but often diverge quite significantly on the fraction of
clusters exhibiting significant substructure. As an example,
\citet{Dressler_Shectman:1988} adopted a method to quantify the
significance of cluster substructure using galaxy spectroscopic
redshifts and projected sky positions. 
The presence of substructure suggests that the galaxy cluster is still
relaxing.
This may imply that the cluster velocity dispersion is a less
reliable measure of cluster mass, although our limited data for 3
clusters with detected substructure and with a measured lensing mass
do not indicate this (Fig.~\ref{fig:lensing}).

In this paper
we present cluster velocity dispersion measurements and assessment of
cluster substructure for 21 galaxy clusters from the ESO Distant
Cluster Survey (EDisCS) located in 15 survey fields.
Measurements for the 5 remaining clusters located in 5 survey fields
were presented in \citet{Halliday_etal:2004}.

EDisCS \citep{White_etal:2005}
is a project to study high-redshift cluster galaxies,
as well as coeval field galaxies, in terms of their
sizes, luminosities, morphologies, internal kinematics,
star formation properties and stellar populations.
We achieve this by obtaining deep multi-band imaging of 20 survey fields
containing clusters at $z$ = 0.4--1 and deep spectroscopy of
$\sim$100 galaxies per field.
The EDisCS fields were chosen to target galaxy cluster candidates from the
Las Campanas Distant Cluster Survey (LCDCS, \citealt{Gonzalez_etal:2001}).
The LCDCS is a survey of 
  an           area of 130 square degrees
imaged in a single wide optical
filter using a 1~m telescope in drift-scan mode
with an effective exposure time of 3.2 minutes.
All detected objects are removed. For a high-redshift cluster this only
affects a few of the brightest galaxies in the cluster; the rest of the
galaxies are not detected individually.
High-redshift clusters can then be detected as diffuse light peaks
with a typical scale of 10$''$, resulting in a catalogue of
1073 cluster candidates with estimated redshifts $\zest$ = 0.3--1.0.
 From this catalogue we selected 30 of the highest surface-brightness
candidate clusters. Using moderately deep 2--band VLT/FORS2 imaging
(going 3 magnitudes deeper than the original LCDCS imaging),
28 of the candidates were found to show
a significant overdensity of red galaxies
close to the LCDCS position \citep{Gonzalez_etal:2002}.
 From these clusters, we selected
10 clusters at $\zest \approx 0.5$ (hereafter mid--$z$) and
10 clusters at $\zest \approx 0.8$ (hereafter high--$z$)
to constitute the EDisCS sample.
These fields were imaged optically in $BVI$ (mid--$z$) and $VRI$ (high--$z$)
with VLT/FORS2 using 14 nights \citep{White_etal:2005},
and in the near-infrared (NIR) in $K$ (mid--$z$) and $JK$ (high--$z$)
with NTT/SOFI  using 20 nights (Arag{\'o}n-Salamanca et al., in prep.).
Spectroscopy was obtained using VLT/FORS2 using 22 nights
(\citealt{Halliday_etal:2004} and this paper).
Follow-up observations include
MPG/ESO 2.2m/WFI wide field imaging in $VRI$ of all fields,
HST/ACS imaging in F814W of 10 fields \citep{Desai_etal:2007},
H$\alpha$ narrow-band imaging \citep{Finn_etal:2005},
XMM--Newton/EPIC X--ray observations \citep{Johnson_etal:2006}, and
Spitzer IRAC (3--8$\mu$m) and MIPS (24$\mu$m) imaging.
The legacy value of the EDisCS fields has been further increased
by another ESO Large Programme that studies galaxies at redshift 5--6
in 10 of the EDisCS fields, using the EDisCS imaging as well as
new deep VLT/FORS2 $z$--band imaging and spectroscopy
(cf.\ \citealt{Douglas_etal:2007}).

The unprecedented novelty of the EDisCS dataset stems from the range
of cluster velocity dispersions and masses covered by the sample. On
the one hand, the survey provides high-redshift counterparts to the
lower velocity dispersion clusters abundant in the local universe.  On
the other hand, it probes a range in cluster masses large enough to
allow the study of the dependency on cluster mass of the processes
affecting cluster galaxy evolution.  The scientific exploitation of
the rich EDisCS dataset is ongoing, but it has already produced
important results in this respect.  Studies have so far been completed
on the red-sequence galaxies
\citep{DeLucia_etal:2004,DeLucia_etal:2007}, the star-forming galaxies
as seen in H$\alpha$ \citep{Finn_etal:2005} and [OII]
\citep{Poggianti_etal:2006}, the cluster velocity dispersions
(\citealt{Halliday_etal:2004} and this paper),
the weak-lensing mass reconstruction of the clusters
\citep{Clowe_etal:2006}, the X--ray properties of the clusters
\citep{Johnson_etal:2006}, the HST--based visual galaxy morphologies
\citep{Desai_etal:2007}, the evolution of the early-type galaxy
fraction \citep{Simard_etal:EDisCS_bulge_disk_VLT}, and the evolution
of the brightest cluster galaxies \citep{Whiley_etal:EDisCS_BCG}.
Further studies of the properties of the cluster galaxies and the clusters
themselves will follow.

This paper is organised as follows.
Section~\ref{sec:targetsel_obs} reports the target selection and
the observations.
Section~\ref{sec:reduc} describes the data reduction using two
different methods for the sky subtraction:
sky subtraction performed after (respectively before)
any rebinning/interpolation of the data has been done
(cf.\ \citealt{Kelson:2003}).
Section~\ref{sec:galaxy_redshifts} presents the redshift measurements
and the redshift histograms.
Section~\ref{sec:completeness} examines the success rate, the failure rate
and potential selection biases.
Section~\ref{sec:zcl_sigmacl} describes the determination of
cluster redshifts and velocity dispersions.
Section~\ref{sec:substructure} discusses possible cluster substructure.
Section~\ref{sec:discussion} discusses the velocity dispersions
for the full sample of EDisCS clusters in comparison with the weak-lensing
measurements and with other samples.
Section~\ref{sec:summary} provides a summary, and
Appendix~\ref{appendix:skysub_compare} compares results from the
two sky subtraction methods.

The discussed photometry is based on Vega zero-points
unless stated otherwise.
We assume a cosmology with $\Omega_\mathrm{m} = 0.3$, $\Omega_\Lambda = 0.7$
and $H_0 = 70\,\mathrm{km}\,\mathrm{s}^{-1}\,\mathrm{Mpc}^{-1}$.

\section{Target selection and observations}
\label{sec:targetsel_obs}

The target selection strategy, mask design procedure and observations
for the EDisCS spectroscopy are described in detail in
\citet{Halliday_etal:2004}.
Here we give the main points.
We also provide a table with the target selection parameters
(Table~\ref{tab:target_selection}), and we discuss the performance
of the uphotometric redshifts used.

\subsection{Target selection strategy}
\label{sec:target_selection}

\begin{table*}
\caption{Target selection parameters for the masks with long exposures in the 20 EDisCS fields}
\label{tab:target_selection}
\begin{center}
\renewcommand{\arraystretch}{1.00} % Defalut is 1.0
\setlength{\tabcolsep}{4pt} % Default is 6pt
\begin{tabular}{llllllcllll}
\hline
\hline
Field                                    & $\zclfinal$                                     & Run & Mask numbers & $\zphotlow$ & $\zphothigh$ & Explanation & $\zclP$ & $\Ionebright$ & $\Ionefaint$ & Filters \\
\hline                                                          
\multicolumn{11}{l}{Mid--$z$ fields:} \\                            
                           1018.8$-$1211 &                            0.4734                 & 3   & 05,06,07    & 0.27   & 0.67   & $0.47\pm0.2       $ & 0.472 & 19.5 & 22 & BVIK     \\[0.7ex]
                           1059.2$-$1253 &                            0.4564                 & 3   & 05,06,07    & 0.26   & 0.66   & $0.46\pm0.2       $ & 0.457 & 19.6 & 22 & BVIK     \\[0.7ex]
                           1119.3$-$1129 &                            0.5500                 & 4   & 09,10,11    & 0.35   & 0.75   & $0.55\pm0.2       $ & 0.549 & 19.9 & 22 & BVI      \\[0.7ex]
                           1202.7$-$1224 &                            0.4240                 & 4   & 09,10,11    & 0.22   & 0.62   & $0.42\pm0.2       $ & 0.424 & 19.5 & 22 & BVIK     \\[0.7ex]
                                         &                                                   & 2   & 02,03,04    & 0.28   & 0.68   & $0.48\pm0.2       $ & 0.54  & 18.6 & 22 & BVI[J]K  \\
\raisebox{1.5ex}[0cm][0cm]{1232.5$-$1250}& \raisebox{1.5ex}[0cm][0cm]{0.5414}                & 3   & 05          & 0.34   & 0.74   & $0.54\pm0.2       $ & 0.541 & 19.7 & 22 & BVI[J]K  \\[0.7ex]
                           1238.5$-$1144 &                            0.4602                 & 4   & 09          & 0.26   & 0.66   & $0.46\pm0.2       $ & 0.465 & 19.2 & 22 & BVI      \\[0.7ex]
                           1301.7$-$1139 &                            0.4828, 0.3969         & 4   & 09,10,11    & 0.20   & 0.68   & $0.40-0.2,0.48+0.2$ & 0.485 & 19.2 & 22 & BVIK     \\[0.7ex]
                           1353.0$-$1137 &                            0.5882                 & 4   & 09,10,11    & 0.39   & 0.79   & $0.59\pm0.2       $ & 0.577 & 19.6 & 22 & BVIK     \\[0.7ex]
                           1411.1$-$1148 &                            0.5195                 & 3   & 05,06,07    & 0.32   & 0.72   & $0.52\pm0.2       $ & 0.520 & 19.4 & 22 & BVIK     \\[0.7ex]
                           1420.3$-$1236 &                            0.4962                 & 4   & 09,10,11    & 0.30   & 0.70   & $0.50\pm0.2       $ & 0.497 & 19.5 & 22 & BVIK     \\[0.7ex]
\multicolumn{11}{l}{High--$z$ fields:} \\                           
                           1037.9$-$1243 &                            0.5783, 0.4252         & 3   & 05,06,07,08 & 0.38   & 0.78   & $0.58\pm0.2       $ & 0.58  & 20.0 & 23 & VRIJK    \\[0.7ex]
                                         &                                                   & 2   & 02,03,04    & 0.5031 & 0.9031 & $0.70\pm0.2       $ & 0.55  & 19.5 & 23 & VRIJK    \\
\raisebox{1.5ex}[0cm][0cm]{1040.7$-$1155}& \raisebox{1.5ex}[0cm][0cm]{0.7043}                & 3   & 05,06       & 0.504  & 0.904  & $0.70\pm0.2       $ & 0.704 & 20.6 & 23 & VRIJK    \\[0.7ex]
                                         &                                                   & 2   & 02,03,04    & 0.494  & 0.894  & $0.70\pm0.2       $ & 0.69  & 19.5 & 23 & VRIJK    \\
\raisebox{1.5ex}[0cm][0cm]{1054.4$-$1146}& \raisebox{1.5ex}[0cm][0cm]{0.6972}                & 3   & 05          & 0.50   & 0.90   & $0.70\pm0.2       $ & 0.697 & 20.2 & 23 & VRIJK    \\[0.7ex]
                                         &                                                   & 2   & 02,03,04    & 0.546  & 0.946  & $0.75\pm0.2       $ & 0.75  & 19.5 & 23 & VRI[J]K  \\
\raisebox{1.5ex}[0cm][0cm]{1054.7$-$1245}& \raisebox{1.5ex}[0cm][0cm]{0.7498}                & 3   & 05          & 0.55   & 0.95   & $0.75\pm0.2       $ & 0.748 & 20.4 & 23 & VRI[J]K  \\[0.7ex]
                           1103.7$-$1245 &                            0.9586, 0.6261, 0.7031 & 3;4 & 05,06;09,10 & 0.50   & 0.90   & $0.70\pm0.2       $ & 0.704 & 20.2 & 23 & VRIJK    \\[0.7ex]
                           1122.9$-$1136 &                            \ldots                 &\ldots& \ldots     & \ldots & \ldots &  \ldots             & \ldots&\ldots&\ldots&\ldots  \\[0.7ex]
                           1138.2$-$1133 &                            0.4796, 0.4548         & 3   & 05,06,07,08 & 0.28   & 0.68   & $0.48\pm0.2       $ & 0.480 & 19.7 & 23 & VRIJK    \\[0.7ex]
                                         &                                                   & 2   & 02,03,04    & 0.597  & 0.997  & $0.80\pm0.2       $ & 0.79  & 19.5 & 23 & VRIJK    \\
\raisebox{1.5ex}[0cm][0cm]{1216.8$-$1201}& \raisebox{1.5ex}[0cm][0cm]{0.7943}                & 3   & 05          & 0.60   & 1.00   & $0.80\pm0.2       $ & 0.794 & 20.4 & 23 & VRIJK    \\[0.7ex]
                           1227.9$-$1138 &                            0.6357, 0.5826         & 3   & 05,06,07,08 & 0.44   & 0.84   & $0.64\pm0.2       $ & 0.64  & 20.4 & 23 & VRIJK    \\[0.7ex]
                           1354.2$-$1230 &                            0.7620, 0.5952         & 3   & 05,06,07,08 & 0.40   & 0.96   & $0.60-0.2,0.76+0.2$ & 0.76  & 20.3 & 23 & VRIJK    \\
\hline
\end{tabular}
\end{center}

\vspace*{0.1ex}

Notes --
The 66 masks listed in this table are the masks with long exposures
(with the exception of the listed 1238.5$-$1144 mask,
cf.\ Sect.~\ref{sec:obs}).
Each field has between 1 and 5 such masks. The masks are numbered
from 01 to 11 as described in \citet{Halliday_etal:2004}.
For reference the column $\zclfinal$ lists the
spectroscopic cluster redshifts
(\citealt{Halliday_etal:2004} and this paper);
these cluster redshifts were determined from all the obtained spectroscopy
and were not as such part of the spectroscopic target selection.
Where more than one redshift is listed the order is as follows:
main cluster, secondary cluster ``a'', secondary cluster ``b''
(cf.\ Sect.~\ref{sec:zhist}).
The ``Explanation'' column gives the same information as the
$\zphotlow$ and $\zphothigh$ columns, except for a possible rounding to two decimals.
The ``Filters'' column indicates what photometric data were
employed to calculate
the photometric redshifts used in the spectroscopic target selection.
Filters in brackets indicate data \emph{not} employed;
these data were obtained
later and used in subsequent studies, e.g.\ to calculate the final EDisCS
photometric redshifts (Pell{\'o} et al., in prep.).
No data are listed for the 1122.9$-$1136 field, since it only has
an initial short mask, no long masks, cf.\ Sect.~\ref{sec:obs}.
\end{table*}

\begin{figure*} % Two column figure
\sidecaption
\includegraphics[width=12cm,bb = 6 430 569 751]
  {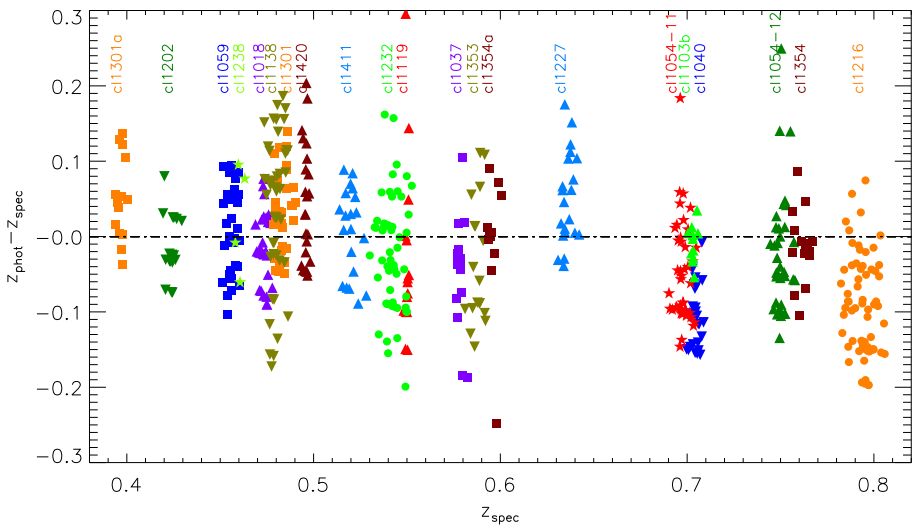}
\caption[]{%
Performance of the photometric redshifts used for the spectroscopic
target selection
(not the final ones from Pell{\'o} et al., in prep.).
Plotted are the members of the cluster(s) at the targeted redshift(s)
for the given field (21 clusters in 19 fields).
For example, if the target selection was
$\zphot$ = 0.58$\pm$0.2 for the given field
(cf.\ Table~\ref{tab:target_selection}), then the $z = 0.58$ cluster in
that field is plotted, but not the $z = 0.43$ cluster.
Four galaxies out of 568 are outside the plotted $y$--range.
Abbreviated cluster names are given on the plot.
The 5 clusters not plotted are
cl1037a ($z = 0.43$),
cl1103a ($z = 0.63$),
cl1103  ($z = 0.96$),
cl1138a ($z = 0.45$) and
cl1227a ($z = 0.58$).
}
\label{fig:zphot_zspec}
\end{figure*}

The target selection was based on the available
VLT/FORS2 optical photometry \citep{White_etal:2005} and the
NTT/SOFI NIR photometry (Arag{\'o}n-Salamanca et al., in prep.).
The optical data cover $6.5' \times 6.5'$ and are well-matched to
the FORS2 spectrograph field-of-view.
The NIR data cover a somewhat smaller region of
$4.2' \times 6.0'$ (mid--$z$ fields) and
$4.2' \times 5.4'$ (high--$z$ fields).
The photometry was used as input to a modified version of the photometric
redshifts code \procedurename{hyperz}\footnote{%
\texttt{http://webast.ast.obs-mip.fr/hyperz}}
\citep*{Bolzonella_etal:2000},
see also \citet{Halliday_etal:2004}
and the EDisCS photo--$z$ paper (Pell{\'o} et al., in prep.).
A combined photometry and photo--$z$ catalogue (hereafter photometric
catalogue) used for the target selection,
was created for each field prior to each spectroscopic observing run.
The aim of the target selection strategy was to keep all galaxies at
the cluster redshift (brighter than a certain $I$--band magnitude),
while removing objects that were almost certainly not
galaxies at the cluster redshift.
We will see in Sect.~\ref{sec:completeness} that this aim
was successfully achieved.
The selection criteria are explained in \citet{Halliday_etal:2004}.
They can be summarised as follows.
\begin{enumerate}
\item The I--band magnitude (not corrected for Galactic extinction)
within a circular aperture of radius 1$''$,
$\Ione$, had to be in the range $[\Ionebright,\Ionefaint]$, see
Table~\ref{tab:target_selection}.
For the initial short masks (not listed in Table~\ref{tab:target_selection})
the bright limit was conservatively set to 18.6 for the mid--$z$ fields
and 19.5 for the high--$z$ fields, which is about 1 magnitude brighter
than the expected magnitude of the brightest cluster galaxy (BCG)
\citep{Aragon-Salamanca_etal:1998}.
For subsequent masks, the bright limit was either kept or set to
0.2 magnitudes brighter than the identified BCG\@.
\item (a)~The best-fit photometric redshift $\zphot$ had to be in the range
$[\zphotlow,\zphothigh]$ \emph{or}
(b)~the $\chi^2$-based probability of the best-fit template placed at
the estimated or known cluster redshift $\zclP$ had to be greater than 50\%
(see Table~\ref{tab:target_selection}).
The $\zphot$ interval was set to $\pm0.2$ from the
estimated or known cluster redshift.
For the long exposures (i.e.\ those in Table~\ref{tab:target_selection}),
the cluster redshift was usually known from a preceding short exposure
(cf.\ Sect.~\ref{sec:obs}).
For two fields (1301.7$-$1139 and 1354.2$-$1230), clusters at two redshifts
had been identified in the initial short exposure, and here the union of the
$\pm0.2$ intervals was used, giving a wider interval, e.g.\
$0.60-0.2,0.76+0.2$.
The $\pm0.2$ limit was designed to be fairly conservative:
the expectation was that the photo--$z$ dispersion would be 0.1,
which would make the selection be $\pm2\sigma$.
The performance of the used photometric redshifts is discussed below.
\item The \procedurename{hyperz} star--galaxy separation parameters
$\Ngal$ and $\Nstar$, based on SED fitting minimization,
had to have values as follows:
\newline
$\Ngal  = 1$ (``the object could be a galaxy'') \emph{or}
\newline
$\Ngal  = 2$ (``the object is almost certainly a galaxy'') \emph{or}
\newline
$\Nstar = 0$ (``the object is almost certainly not a star'').
\newline
In other words, in the $3\times3$ grid of $(\Ngal,\Nstar)$, the only two
grid points \emph{not} selected are those at (0,1) and (0,2).
\item The FWHM had to be greater than a limit $\FWHMmin$
\emph{or} the ellipticity $\epsilon$ had to be greater than 0.1,
with FWHM and $\epsilon$ being measured in the $I$--band image.
This requirement was applied to runs~3 and 4 only.
The value of $\FWHMmin$ was determined using
20--30 manually-identified stars in the given field:
based on their measured FWHM values, the limit was calculated as
$\FWHMmin$ = $\langle$FWHM$\rangle$ + 2$\sigma$(FWHM),
with $\langle$FWHM$\rangle$ being the seeing of the image and
2$\sigma$(FWHM) amounting to about 0.1$''$.
The value of $\FWHMmin$ was in the range 0.58--0.85$''$ with a typical
value of 0.69$''$.
\end{enumerate}
Applying these 4 rules to the photometric catalogue,
we derived a target catalogue for the given field and run.
Additional constraints of geometrical nature were imposed by the
mask design, cf.\ Sect.~\ref{sec:mask_creation}.

Table~\ref{tab:target_selection} lists the main target selection parameters
for each field and observing run for the 66 long exposure masks.
The table does not list the parameters for the short initial masks,
since these masks were used only to determine a good guess of the
cluster redshift; this observing strategy is described in Sect.~\ref{sec:obs}.
For reference the table also lists the spectroscopic cluster redshift(s)
for the given field derived after all the spectroscopy had been completed.
Most fields contain a single main cluster, but a few fields contain
one or two secondary clusters in addition to the main cluster;
this is discussed in Sect.~\ref{sec:zhist}.

Having obtained all the spectroscopy, we can check how the
photometric redshifts used in the spectroscopic target selection
have performed.
In Fig.~\ref{fig:zphot_zspec} we plot $\Delta z \equiv \zphot-\zspec$
vs $\zspec$ for
members of the clusters at the targeted redshifts.
The vast majority of the plotted galaxies were selected using rules 1--4
(the exception being serendipitously-observed galaxies; here we have
imposed the same magnitude cuts as in rule~1).
This implies that Fig.~\ref{fig:zphot_zspec}
provides a good indication of the photo--$z$
performance for $\Delta z$ in the range $\pm0.2$. It does not give an
unbiased indication of the fraction of `catastrophic' photo--$z$ failures
($|\Delta z| > 0.2$);
however, that is derived in Sect.~\ref{sec:failure_rate}, where the
target selection failure rate is found to be about 3\%.
It is seen from Fig.~\ref{fig:zphot_zspec} that the dispersion of $\Delta z$
for the given cluster is quite small,
namely in the range 0.03--0.10 (with a typical value 0.06)
with the largest value being for cl1138
which at $z = 0.48$ would need $B$--band data to achieve a better accuracy.
The quoted dispersions are computed as a robust (biweight) estimate.
The median value of $\Delta z$ differs somewhat from zero
(range $-0.11$ to $+0.07$).
This is probably due to minor problems in the photometry used at the time:
imperfect photometric zero-points
and the lack of seeing-matched photometry.
(These problems have been dealt with in the final EDisCS photometric redshifts,
Pell{\'o} et al., in prep.)
When all the clusters are considered together, the dispersion of
$\Delta z$ is 0.08 and the median value is $-0.02$.

We can also check the $\chi^2$--based probabilities of the best-fit template
placed at the cluster redshift used in branch~(b) of rule~2.
For the photo--$z$ catalogues used in the spectroscopic target selection
these probabilities are often too low due to imperfect photometric zero points
and the lack of seeing matched photometry.
However, since rule~2 has a logical \emph{or} between branch (a) and (b),
and since branch~(a) in itself is very good at selecting cluster members,
little harm is done by imperfections in branch~(b).
Furthermore,
the inclusion of branch~(b) only increased the number of objects in the
target catalogues by about 10\%,
and it made the target selection failure rate be
3\% (Sect.~\ref{sec:failure_rate}) instead of 5\%.

We present now some representative statistics about the number of targets.
If we were to apply only the magnitude cuts (i.e.\ rule~1),
the average number of objects per field would be 470 (range 160--860). % median 410
If we apply the full target selection (i.e.\ rules 1--4),
the average is 260 (range 100--440). % median 240
The full target selection thus on average rejects almost 50\% of
the objects that meet the magnitude cuts.
The price for this substantial efficiency increase turns out to
be missing about 3\% of the cluster members
(i.e.\ the failure rate, Sect.~\ref{sec:failure_rate}).

To the target catalogues, we added 3 galaxies that did not meet rule 2,
since these galaxies had been observed in the initial short mask, and had
been found to have redshifts that were close to the estimated cluster redshift.
Two of these galaxies were found to be cluster members, and these
are counted when computing the failure rate for the target selection for
the 66 long masks (Sect.~\ref{sec:failure_rate}).

\subsection{Automatic creation of the masks}
\label{sec:mask_creation}

We developed a programme \citep{Poirier:2004} to design the spectroscopic
slit masks (called ``MXU masks'' after the Mask eXchange Unit
in the FORS2 spectrograph).
A fuller description of how the programme works
is found in \citet{Halliday_etal:2004}; the main points are as follows.
The programme starts by placing a slit on the BCG
unless it has already been observed in a previous long mask.
Slits (10$''$ long) are then placed on objects above and below the BCG\@.
At a given location along the $y$--axis (north--south axis), the brightest
object from the target catalogue is chosen (avoiding targets that have already
been observed in a previous long mask).
Once a slit has been placed at a given location along the $y$--axis,
no other slits can be placed at that location
(i.e.\ to the left and right of that slit), since that would cause
overlapping spectra.
Objects taken from the target catalogue are noted as having
targeting flag 1 (cf.\ Table~\ref{tab:targeting_flag_values}).
If no objects from the target catalogue are available at a given location,
an object not from the target catalogue is chosen (still imposing a
faint magnitude cut in $\Ione$ of 22 and 23 for mid--$z$ and high--$z$ fields),
possibly using a somewhat shorter slit (6--8$''$).
These objects acquire a targeting flag 2.
Additionally, 2--3 short slits (5$''$) are placed on manually-identified
stars. These are used to aid the acquisition and to enable the seeing
in the spectral data of a given mask to be accurately measured.
These objects are given a targeting flag 4.
(All the 108 targeting flag 4 objects observed in the 66 long masks
were indeed found to be stars spectroscopically.)
Some slits happen to go directly or partially through an object that is
not the target. These serendipitously-observed objects are noted as
having targeting flag 3.

The achieved wavelength range of the spectra depends on the $x$--location of
the slit in the mask. For example, for the runs~3 and 4 setup,
a slit at the left, centre and right of the mask would cover an
observed-frame wavelength range of approximately
6200--9600$\,${\AA}, 5200--8500$\,${\AA} and 4150--7400$\,${\AA},
respectively.
If possible, slits were only placed on objects that were in the $x$--interval
that would produce a spectrum that contained the rest-frame wavelength range
3670--4150$\,${\AA} (covering [OII] and H$\delta$)
for the assumed cluster redshift.
The width of that $x$--interval depends on the cluster redshift.
Slits on stars (targeting flag 4 objects) were not subjected to this
restriction.

The masks were inspected, and occasionally objects that had been assigned
a slit were removed from the target catalogue and the mask redesigned.
This happened in two cases. One was when an object from the 
photometric catalogue clearly appeared to consist of two distinct physical
objects seen partially in projection, but where 
SExtractor \citep{Bertin_Arnouts:1996} had not been able to deblend them.
This seems like a wise choice: in such a situation the photometric redshift
(calculated from the combined light of the two physical objects) is
meaningless,
and it is also not clear on which of the two physical objects to
place the slit.
The second case was when the object appeared so bright or big that
it was perceived to be a foreground field galaxy,
which was the case for 10 objects.
In retrospect this seems less advisable:
5 of these objects were observed anyway to fill the mask
(i.e.\ as targeting flag 2 objects), and 2 of them did in fact
turn out to be cluster members (specifically of cl1059.2$-$1253 at $z=0.46$).
The remaining 3 objects were foreground field galaxies, at
$z$ = 0.07, 0.35 and 0.46.

The slits were aligned with the major axis of the targeted object if that
involved tilting the slit by no more than $\pm45^\circ$.
(In run~2, this was only performed for objects that the
photometric redshift code identified as late-type galaxies.)
Occasionally, when the programme assigned an untilted slit to a target,
we manually tilted the slit either to be able to observe a second object
`serendipitously' in that slit,
or to avoid/reduce signal from very bright, nearby objects.

All slits were 1$''$ wide in the dispersion direction,
which means that the spectral resolution for all slits, tilted or not,
is practically the same, cf.\ Sect.~\ref{sec:wlcal_verification}.

\begin{table}
\caption{Targeting flag values}
\label{tab:targeting_flag_values}
\begin{center}
\begin{tabular}{cl}
\hline
\hline
Value & Explanation \\
\hline
1     & Targeted as a candidate cluster member  \\
2     & Targeted as a field galaxy (filling object)  \\
3     & Serendipitous (not targeted)  \\
4     & Targeted as a star to aid acquisition and to measure seeing  \\
\hline
\end{tabular}
\end{center}
\end{table}

\subsection{Observations}
\label{sec:obs}

\begin{figure*} % Two column figure
\includegraphics[width=1.00\textwidth,bb = 61 396 516 585]
  {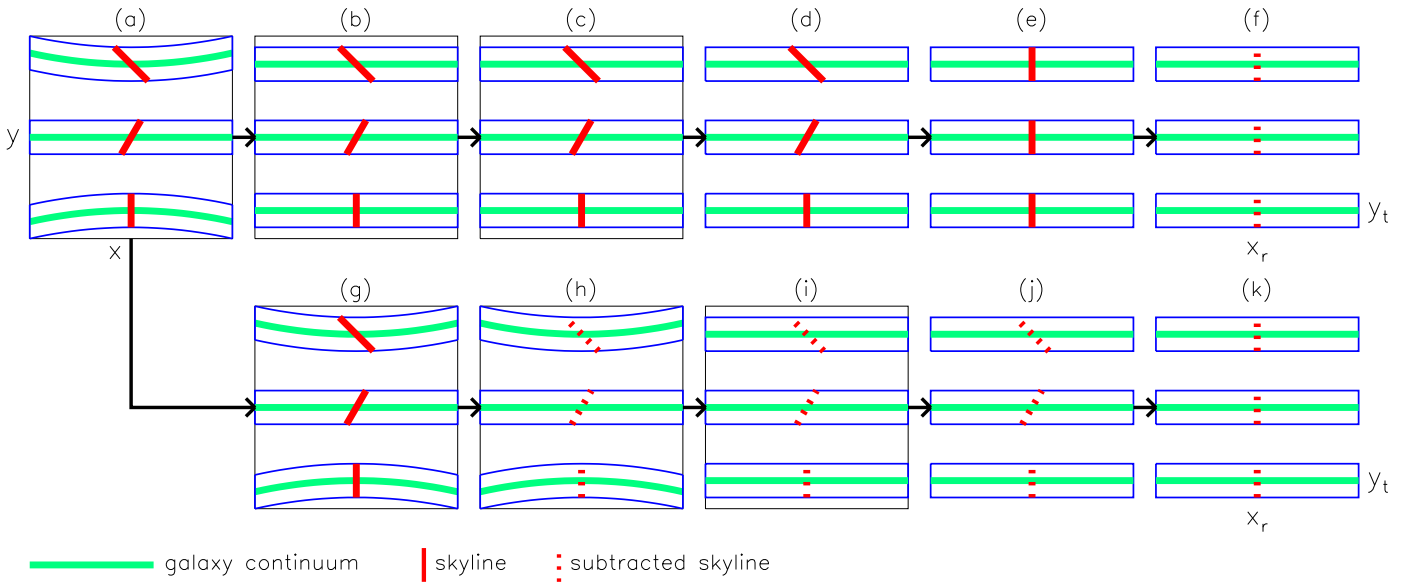}
\caption[]{%
Flowchart for the science frames.
The upper branch is for traditional sky subtraction:
(a)~`raw' frame;
(b)~after removal of spatial curvature;
(c)~after flat fielding;
(d)~after cutting-out the individual spectra;
(e)~after application of the 2D wavelength calibration;
(f)~after sky subtraction.
The lower branch is for the improved sky subtraction:
(a)~`raw' frame;
(g)~after flat fielding;
(h)~after sky subtraction;
(i)~after removal of spatial curvature;
(j)~after cutting out the individual spectra;
(k)~after application of the 2D wavelength calibration.
See text for details.
We note that the figure is schematic:
it is not to scale, and only
3 spectra per mask (instead of $\sim$30) and
1 skyline per spectrum (instead of $>$100) are shown.
}
\label{fig:flowchart}
\end{figure*}

Spectroscopic observations were completed using % VLT/FORS2
the FORS2 spectrograph\footnote{%
\texttt{http://www.eso.org/instruments/fors}}
(cf.\ \citealt{Appenzeller_etal:1998}) on the VLT,
during 4 observing runs from 2002 to 2004,
spanning 25 nights (22 nights were usable, while 3 nights were lost due to
bad weather and technical problems); see Table~1 in \citet{Halliday_etal:2004}.
The same high-efficiency grism was used in all runs
(grism 600RI+19, $\lambda_\mathrm{central} = 6780\,${\AA},
resolution FWHM $\approx$ 6$\,${\AA}), but the detector system changed
between runs~2 and 3, see Table~2 in \citet{Halliday_etal:2004}.
A total of 86 masks were observed, with 1--6 masks per field;
see Table~3 in \citet{Halliday_etal:2004},
which also lists the exposure times.
The typical observing strategy was for a given field to have an initial
``short mask'' observed in runs~1 or 2, with a total exposure time
of typically 0.5--1 hour. Based on the measured redshifts, the confirmed field
galaxies and stars were removed from the target catalogue. The objects for
which no (secure) redshift could be determined were usually
kept in the target catalogue.
The confirmed cluster galaxies were given priority so
that they almost certainly were included (repeated) in the first
``long mask'' of that field (typical total exposure time 1--4 hours).
For 18 of the 20 EDisCS fields, there are 3--4 (or even 5) long masks per field.
For the 1122.9$-$1136 field, no long masks were observed because the
initial short mask did not show any convincing cluster (cf.\ the redshift
histogram in Fig.~\ref{fig:zhist_highz}).
For the 1238.5$-$1144 field, an initial short mask plus a subsequent
20~min mask from run~4 is all that was observed (this field had low priority
due to the lack of NIR imaging).
The 66 masks listed in the target selection parameter table
(Table~\ref{tab:target_selection}) are the 65 truly long masks plus the
20~min 1238.5$-$1144 mask.
The published redshifts (\citealt{Halliday_etal:2004} and this paper)
are practically all from these 66 masks ---
the data from the 20 initial short masks would only have added a few
field galaxies and stars.
Data for 5 fields (observed run~2 [mainly] and run~3)
were published in \citet{Halliday_etal:2004},
while data for 14 fields (observed in runs~3 and 4) are published in this paper.
No redshifts are published for the 1122.9$-$1136 field,
but the few redshifts from the initial short mask
are shown in the redshift histogram in Fig.~\ref{fig:zhist_highz}.

The 86 masks were exposed for a total of 183~hours
(14~hours for the 20 initial short masks and % actually:  14.5
169~hours for the 66 subsequent long masks). % actually: 168.52
Over the 22 usable nights, this amounts to 8.3~hours of net exposure per night,
showing the high efficiency of visitor mode for this type of observations.

In addition to the science observations, various night and day time
calibration frames were obtained, see \citet{Halliday_etal:2004}.

\section{Data reduction}
\label{sec:reduc}

This paper describes the data reduction of the runs~3 and 4 data,
which amounts to 51 of the 66 long masks listed in
Table~\ref{tab:target_selection}.
The reduction was performed using both
traditional sky subtraction (Sect.~\ref{sec:reduc_traditional}), and
improved sky subtraction (Sect.~\ref{sec:reduc_improved}).
We have adopted the names `traditional' and `improved' sky subtraction
for simplicity; more descriptive names would be
sky subtraction performed after (respectively before)
any rebinning/interpolation of the data has been done.

We note that redshifts based on the traditional reduction for 5
of these 51 masks were included in \citet{Halliday_etal:2004}.
The improved reduction that we have now completed, has no effect on the
measurement of redshifts.

\subsection{Reduction using traditional sky subtraction}
\label{sec:reduc_traditional}

The procedure for the reduction using traditional sky subtraction
is described in \citet{Halliday_etal:2004}, and was
developed for previous FORS2 MXU work
\citep{Milvang-Jensen:2003,Milvang-Jensen_etal:2003}.
A summary of the procedure is provided below, and
a flowchart for the science frames % in this reduction procedure
is shown in the upper branch of Fig.~\ref{fig:flowchart}.

For a given mask 3--8 individual science frames were usually available.
These frames were bias-subtracted and then combined (averaged).
At the same time, signal from cosmic-ray hits was removed
(cf.\ \citealt{Milvang-Jensen:2003,Halliday_etal:2004}).
Due to the good stability
of FORS2, the frame-to-frame shifts in the position of skylines
and object continua were so small that the frames could
be combined without applying any shifts in $x$ or $y$.
This stage is schematically shown in Fig.~\ref{fig:flowchart}(a).
Two features should be noted:
(1)~The spectra in the upper part of the frame curve like a U and the spectra
in the lower part of the frame curve like an upside-down U;
this is the spatial curvature or S--distortion.
(2)~The skylines and the spectral features, in general, are often tilted,
because $\sim$50\% of the slits in these MXU masks are tilted to be aligned
with the major axes of the galaxies 
(done if the required slit angle was within $\pm45^\circ$).

\begin{figure*} % Two column figure
\includegraphics[width=1.00\textwidth,bb = 5 470 570 733]
  {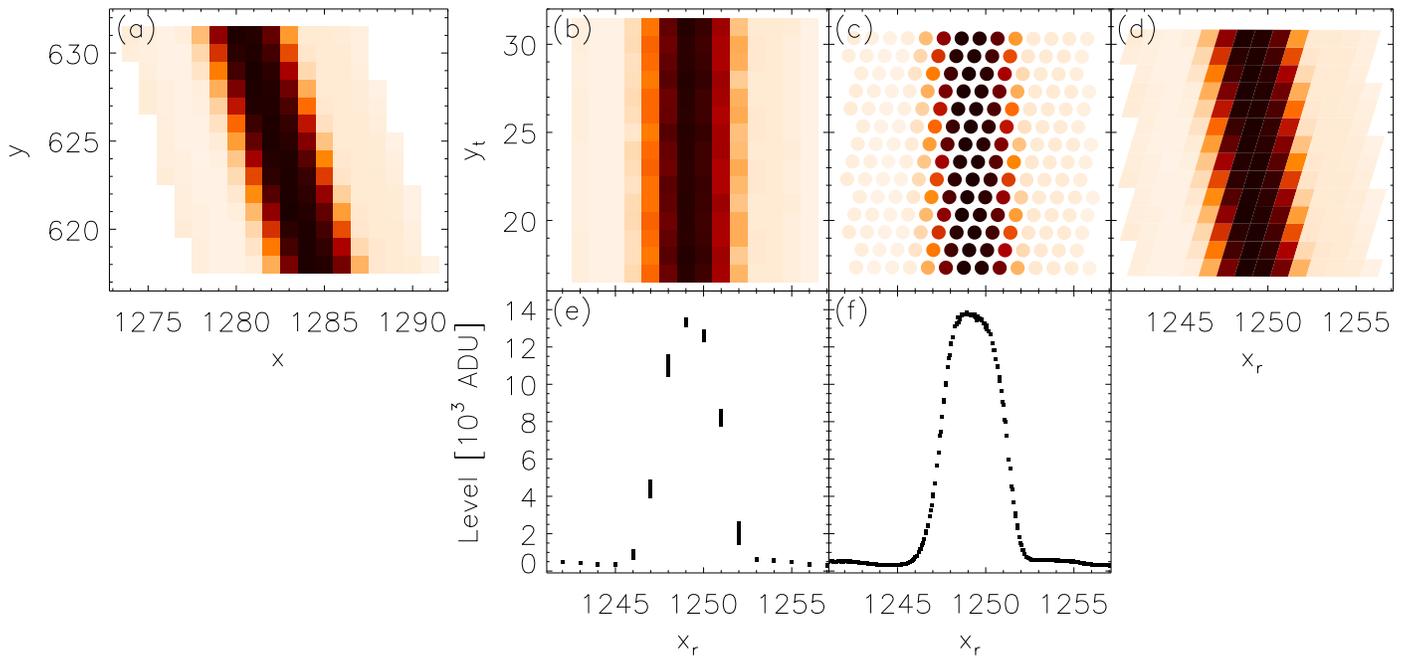}
\caption[]{%
Illustration of the irregular grid.
Panels~(a) and (b) show regular grids (the type of images usually
used in astronomy), i.e.\ with the data points located at integer-valued
coordinate positions.
Panel~(c) shows an irregular grid,
i.e.\ with the data points located at real-valued coordinate positions.
The different panels show:
(a)~Part of a `raw' frame centered on a bright skyline (no galaxy continuum is
present in the shown section), i.e.\ data that have not been interpolated
(rebinned), cf.\ Fig.~\ref{fig:flowchart}a. The image is pixelised in $(x,y)$.
(b)~The same data after interpolation in 
$y$ to remove spatial curvature and in
$x$ to apply the wavelength calibration, cf.\ Fig.~\ref{fig:flowchart}e.
The image is pixelised in $(\xr,\yt)$.
(c)~The exact data values from panel~(a), i.e.\ uninterpolated values,
but shown as an irregular grid in $(\xr,\yt)$ space.
(d)~Just a different visualisation of the irregular grid in panel~(c),
corresponding to painting the spectrum in panel~(a) on a piece of rubber
and stretching it to have axes corresponding to $(\xr,\yt)$ space.
(e)~The values from panel~(b) plotted versus $\xr$. We note the scatter caused
by the interpolation of the sharp edges of the skyline.
(f)~The values from panel~(c) plotted versus $\xr$.
We note the very low scatter.
}
\label{fig:irr_grid}
\end{figure*}

The spatial curvature (S--distortion)
was traced from the edges of the $\sim$30 individual spectra in the flat field
frames, and removed from the science frames by an interpolation in the
$y$--direction (Fig.~\ref{fig:flowchart}b).
The science frames were then flat-fielded (Fig.~\ref{fig:flowchart}c)
and cut-up into the individual spectra (Fig.~\ref{fig:flowchart}d).
The 2D wavelength calibration for each spectrum was established using
an arc frame and
applied to the science frames using an interpolation in the $x$--direction
(Fig.~\ref{fig:flowchart}e);
we note that the skylines are no longer tilted.
Each 2D spectrum now has $\xr$ on the abscissa and $\yt$ on the ordinate.
The $\xr$ coordinate is linearly related to the wavelength $\lambda$ in {\AA}
and
the $\yt$ coordinate is linearly related to the spatial position along the slit
in arcsec
(in the notation of \citealt{Kelson:2003}).
We note that our wavelengths are on the air wavelength system
(as opposed to the vacuum wavelength system), as is customary for optical work.
The sky background could now be fitted and subtracted.
The fit was made using pixels in manually-determined sky windows
(intervals in $\yt$) which are free from galaxy signal.
Finally 1D spectra were extracted from the sky-subtracted 2D spectra.
The reduction was performed mainly using IRAF\footnote{%
IRAF is distributed by the National Optical Astronomy Observatories,
which are operated by the Association of Universities for Research in
Astronomy, Inc., under cooperative agreement with the National Science
Foundation.}.

\subsection{Reduction using improved sky subtraction}
\label{sec:reduc_improved}

The traditional sky subtraction does not work well for
spectra produced by tilted slits:
strong, systematic residuals are evident where skylines have been subtracted.
To remedy this situation, we have implemented a much-improved method for the
sky subtraction.
The method is described in detail in \citet{Kelson:2003}.
We will follow the notation of Kelson.
We have implemented the method from scratch using a combination of
IRAF (with TABLES\footnote{%
TABLES is a product of the Space Telescope Science Institute,
which is operated by AURA for NASA.})
and IDL. % (with the idlutils\footnote{%

Figure~\ref{fig:irr_grid} illustrates
a concept central to the method.
Panel~(a) shows part of a `raw' frame,
i.e.\ data that have not been interpolated
(or rebinned; we will use these terms interchangeably),
and where the skylines have not been subtracted.
The image is pixelised in $(x,y)$,
i.e.\ the image is a regular grid in $(x,y)$.
Panel~(b) shows what happens in the reduction with traditional
sky subtraction: the data, with the skylines still present, have been 
interpolated in 
$y$ to remove spatial curvature, and in
$x$ to apply the wavelength calibration.
The image is pixelised in $(\xr,\yt)$.
These two interpolations are based on analytical mappings
(polynomials or spline functions).
It is seen that the interpolation of the sharp edges of the skylines has
imprinted an aliasing pattern, which prevents a good fit and subsequent
subtraction of the skylines. % to be achieved.
(This aliasing pattern is much more evident once the sky has been
subtracted, as will be shown in Fig.~\ref{fig:backsub_examples}.)
We note that it is the insufficient sampling of the edges of the skylines
that is the problem, when performing the sky subtraction in
the traditional way \citep{Kelson:2003}, not necessarily
an undersampling of the skylines. % as such.
Our data are reasonably well sampled, % as such,
with the spectral FWHM being sampled by 4~pixels.
Panel~(c) shows the irregular grid that is central to the
improved sky subtraction method. 
The panel shows the exact uninterpolated data values from panel~(a)
but as an irregular grid in $(\xr,\yt)$ space.
To turn the regular grid in $(x,y)$ (i.e.\ panel a) into
the irregular grid in $(\xr,\yt)$ (i.e.\ panel c), one simply has to use
the above-mentioned analytical mappings to compute the
correspondence between integer-valued coordinates $(x,y)$ and
real-valued coordinates $(\xr,\yt)$.
Panel~(d) is just a different visualisation of the irregular grid in panel~(c),
corresponding to painting the spectrum in panel~(a) on a piece of rubber
and stretching it to have axes corresponding to $(\xr,\yt)$ space.
To construct this visualisation, one simply has to compute the
real-valued $(\xr,\yt)$ coordinates of the corners of each pixel
in the original image (panel~a).
Panel~(e) shows the values from panel~(b) plotted versus $\xr$.
The before-mentioned aliasing pattern can here be seen as a scatter
at a given $\xr$, at the location of the edges of the skyline.\
Panel~(f) shows the values from panel~(c) plotted versus $\xr$,
and here there is very little scatter.

The essence of the improved sky subtraction method
is that the \emph{uninterpolated} values 
(for pixels that only contain sky and not object signal)
are fitted in the irregular grid in $(\xr,\yt)$. % space
The fit is subsequently evaluated for the $(\xr,\yt)$ positions
corresponding to \emph{all} pixel positions in $(x,y)$,
i.e.\ also those containing object signal,
and subtracted from the uninterpolated data.

Our implementation of the improved sky subtraction
consists of two parts.
First, the irregular grid is constructed.
In our IRAF-based reduction with traditional sky subtraction,
the analytical mappings representing the spatial curvature and the
wavelength calibration were established
using the tasks
 \procedurename{identify},
 \procedurename{reidentify} and
 \procedurename{fitcoords}
and applied using the task \procedurename{transform},
creating an interpolated image.
To calculate the coordinates of the irregular grid,
these mappings need instead to be evaluated at each integer-valued
$(x,y)$ coordinate to determine the corresponding real-valued $(\xr,\yt)$
coordinate.
This task is performed using the task \procedurename{fceval}\footnote{%
The task \procedurename{fceval} was kindly written for us
by Frank Valdes from the IRAF project.
The task is now included in the IRAF distribution.}.
Second, the irregular grid in $(\xr,\yt)$ of uninterpolated values are fitted
(only for data points located in the manually-determined sky windows,
which are intervals in $\yt$).
As fitting-functions, we use cubic-splines in $\xr$
and linear polynomials in $\yt$
(similar to the approach taken by \citealt{Kelson:2003}).
The node (or breakpoint) spacing of the cubic splines
in $\xr$ is denoted $\Delta\xr$.
\citet{Kelson:2003} was able to obtain a good fit using $\Delta\xr = 1$,
but we find that this value usually leaves systematic residuals
and that $\Delta\xr$ in the range 0.5--0.6 typically
is required to achieve a good fit\footnote{%
When we discuss the numerical values of $\Delta\xr$,
we are referring to the situation where the size in {\AA} of the pixels in
$x$ and in $\xr$ is approximately the same.
This is a natural choice, but not the only one possible.
The $\xr$ coordinate in some sense is arbitrary:
it is only defined as soon as the 2D wavelength calibration
(the slightly nonlinear mapping $(x,\yt) \mapsto \lambda$) is used to
construct an interpolated image pixelised in $(\xr,\yt)$,
where $\xr$ is linearly-related to $\lambda$, as
$\lambda = a + b\xr$.
}. % \tmp{figure??}
A node spacing $\Delta\xr$ below one pixel (say $\Delta\xr = 0.5$),
is not a problem:
in many cases the distance between the data points in $\xr$ is much smaller
than this (cf.\ the example in Fig.~\ref{fig:irr_grid}f), and where there are
cases of too few data points being located between two adjacent nodes,
the software will delete one of the nodes.
In most cases a constant function in $\yt$ could have been used
instead of a linear one;
cf.\ the example in Fig.~\ref{fig:irr_grid}(f) where any variation in
sky level with $\yt$ would have caused the points to scatter,
but occasionally a linear function is required to obtain a good fit.
The data points are weighted by the inverse of the expected variance.
The fitting is done iteratively, and sigma-clipping is applied.
We performed the fitting in IDL using modified versions of
B--spline procedures % from the idlutils library
written by Scott Burles and David Schlegel,
procedures which are part of the \procedurename{idlutils} library\footnote{%
The \procedurename{idlutils} library is developed by
Doug Finkbeiner, Scott Burles, David Schlegel, Michael Blanton, David Hogg
and others.
See the Princeton/MIT SDSS Spectroscopy Home Page at
\texttt{http://spectro.princeton.edu/}}.

In terms of the flowchart for the science frames
(the lower branch in Fig.~\ref{fig:flowchart}),
the reduction proceeds as follows.
The starting point is the combined but uninterpolated frame
(Fig.~\ref{fig:flowchart}a).
The data are flat-fielded (Fig.~\ref{fig:flowchart}g), cf.\ below.
The sky is fitted and subtracted as described above
(Fig.~\ref{fig:flowchart}h).
The spatial curvature is removed by means of an interpolation in $y$
(Fig.~\ref{fig:flowchart}i).
The individual spectra are cut-out (Fig.~\ref{fig:flowchart}j).
The 2D wavelength calibration is applied by means of an interpolation in $x$
(Fig.~\ref{fig:flowchart}k), resulting in rectified 2D spectra
(i.e.\ pixelised in $(\xr,\yt)$) that are sky-subtracted, and with
almost no systematic residuals where the skylines have been subtracted.
Finally 1D spectra are extracted.
We note that all the frames shown in the flowchart % (Fig.~\ref{fig:flowchart})
are regular grids (normal images).
The irregular grid used in fitting the sky, % in the improved sky subtraction,
i.e.\ when going
from Fig.~\ref{fig:flowchart}(g) to Fig.~\ref{fig:flowchart}(h), is not shown.

As just described, the data are flat-fielded before the sky is fitted,
which allows a better fit to be achieved. This in turn implies that
an uninterpolated flat field needs to be constructed.
The flat field corrects for pixel-to-pixel variations in sensitivity
and for the slit profile, i.e.\ possible variations in
light transmission along the slits, e.g.\ due to the slitwidth not being
exactly constant along the slit.
The flat field has a level of approximately unity,
i.e.\ there is no variation with
wavelength, since the flat field is intended to preserve the counts.
This flat field is constructed from the bias-subtracted screen flats
(similar to ``dome flats''). The wavelength dependence in the screen flats
(due to the spectral energy distribution of the used lamp and to the
spectral response of the grism, CCD, etc.) is fitted using the same software
that is used to fit the sky background, with a large node spacing
of $\Delta\xr = 50$ in $\xr$
to fit the smooth spectral features in the screen flats,
and with constants rather than linear functions in $\yt$
so that the slit profile is not altered. The screen flats are then divided
by the fit, creating the desired flat field.

\subsection{Comparison of the performance of the two sky subtraction methods}
\label{sec:skysub_compare}

\begin{figure*} % Two column figure
\makebox[1.00\textwidth]{
\includegraphics[width=0.88\textwidth,bb = 60 302 570 751]
  {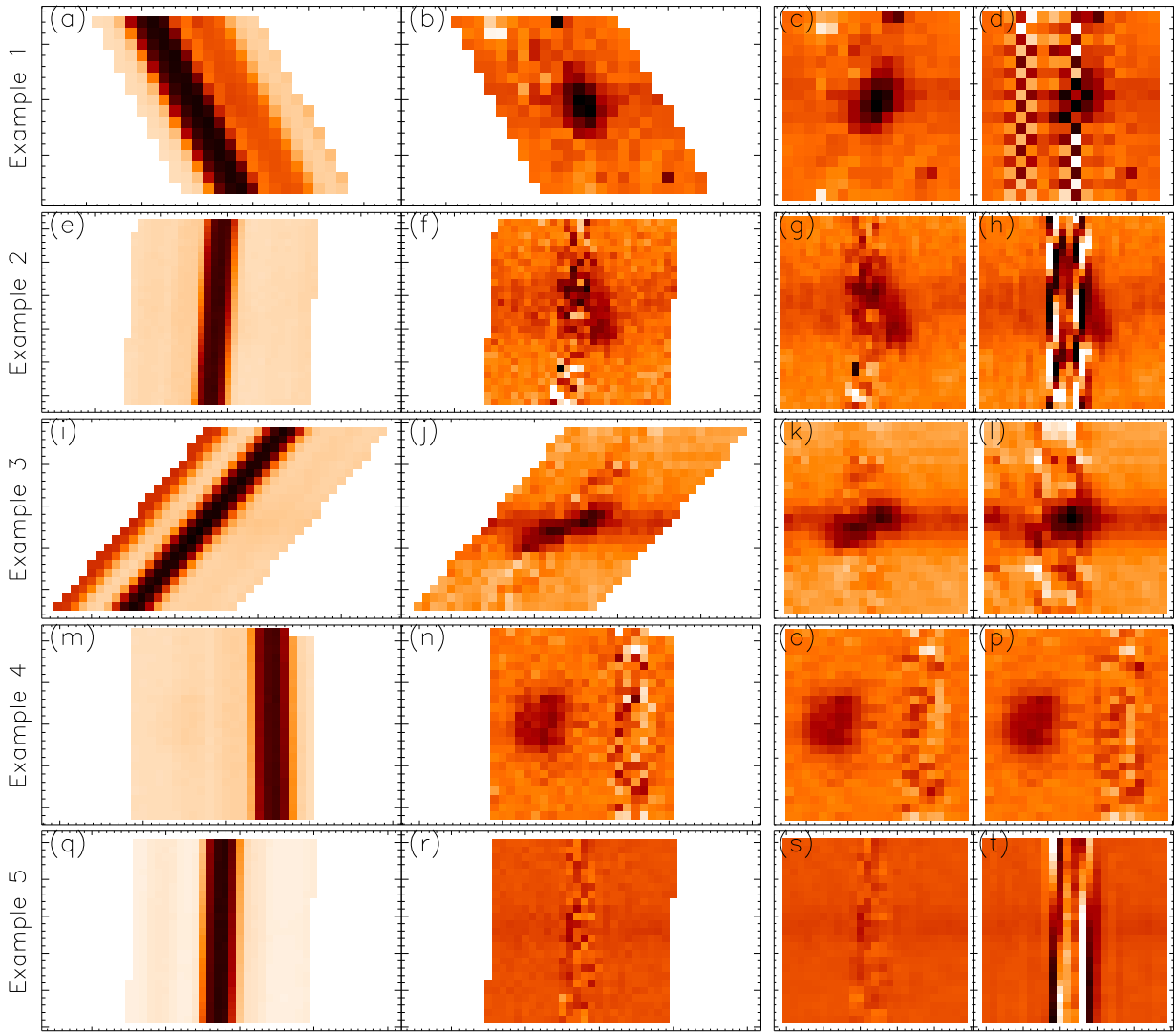}
}
\caption[]{%
Examples of the results from the improved and the traditional sky subtraction.
For each example, we show 4 panels, e.g.\ (a)--(d).
The panels show:
(a)~`Raw' data (i.e.\ combined but uninterpolated frame).
(b)~Sky-subtracted data (improved method), still uninterpolated.
(c)~Sky-subtracted data (improved method), interpolated (rectified).
(d)~Sky-subtracted data (traditional method), interpolated (rectified).
An identical greyscale is used in panels~(b)--(d).
We note that all the panels show regular grids.
The first two panels (e.g.\ a and b) are pixelised in $(x,y)$, and
the last  two panels (e.g.\ c and d) are pixelised in $(\xr,\yt)$
and are thus rectified: wavelength is on the abscissa and spatial position
is on the ordinate.
The greyscale varies from example to example.
Examples~1--3 show spectra from tilted slits
(slit angles $26.6^\circ$, $-3.4^\circ$ and $-40.0^\circ$, resp.).
Examples~4--5 show spectra from untilted slits, but in example~5 the skyline is
nevertheless slightly tilted.
For spectra with tilted skylines the improved sky subtraction method
gives better results than the traditional one. % sky subtraction method.
We note that the tilt of the emission lines, seen in the rectified frames
in examples~1--3, is due to the rotation of the galaxies.
Example~2 shows the [OII] emission line of a $z=0.7$ galaxy,
and the corresponding 1D spectrum is shown in
Fig.~\ref{fig:backsub_examples_1D}(a,b). % as object~1.
}
\label{fig:backsub_examples}
\end{figure*}

Qualitatively, for spectra produced by tilted slits,
the improved sky subtraction method
is almost always superior to the traditional one.
For spectra originating from untilted slits,
the two methods provide similar results in most cases.
Figure~\ref{fig:backsub_examples} shows 5 examples of the results using the
two sky subtraction methods.
Panel~(a) shows the `raw' spectrum, which has not been rebinned and
in which the sky background is still present.
Panel~(b) shows the immediate result from the improved sky subtraction:
a frame that has still not been rebinned, but in which the sky background
has been subtracted.
Panel~(c) shows the rebinned version of panel~(b): the spectrum is now
rectified and pixelised in $(\xr,\yt)$.
Panel~(d) shows the sky-subtracted spectrum from the traditional method
(this spectrum is also rectified).
The two methods can be directly compared in the two rightmost panels in
the figure, e.g.\ (c) and (d). % (i.e. c,d; g,h; etc.).
Examples~1--3 show spectra from tilted slits, and here the traditional method
leaves an aliasing pattern (i.e.\ a systematic error), which the improved
method does not.
Examples~4--5 show spectra from untilted slits. In example~4, the results from
the two methods are similar.
In example~5, the skyline is slightly tilted
(despite the slit not being tilted),
and here the traditional method again leaves an aliasing pattern in the
sky-subtracted spectrum.

\begin{figure} % One column figure
\centering
\makebox[0.49\textwidth]{
\includegraphics[width=0.48\textwidth,bb = 28 34 544 778,clip=]
 %{ed_show1dspec_35.ps} % Obj.2 shows a dramatic improvement
  {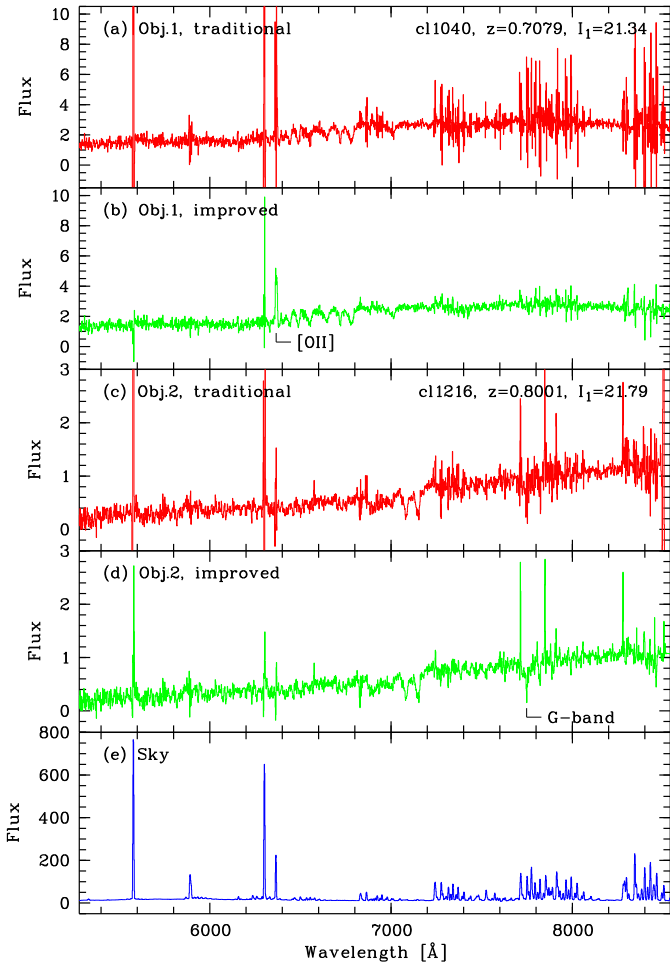} % Obj.2 shows a moderate improvement
}
\caption[]{
Comparison of the traditional and the improved sky subtraction method for two
typical, one-dimensional cluster spectra extracted from tilted slits
(slit angles of $-3.4^\circ$ and $-40.0^\circ$, resp.).
The spectra are flux-calibrated and telluric-absorption corrected (see
Sect.~\ref{sec:flux_calib}). Fluxes are given in units of
$10^{-18}\,\mathrm{erg}\,\mathrm{cm}^{-2}\,\mathrm{s}^{-1}\,
\mathrm{\AA}^{-1}$.
For objects~1 and 2 the significant improvement of [OII] and the G-band, resp.,
is indicated. The [OII] doublet of object~1 is also shown in
Fig.~\ref{fig:backsub_examples} as example~2.
The spectra are shown at their native pixel size of 1.6$\,${\AA}
(spectral resolution is 6$\,${\AA} FWHM) without any smoothing.
In order to mark the positions
of strong skylines, panel~(e) shows a representative sky spectrum.
}
\label{fig:backsub_examples_1D}
\end{figure}

Figure~\ref{fig:backsub_examples_1D} illustrates the quality improvement of
one-dimensional spectra by the improved sky subtraction method in comparison
to the traditional one for tilted slits. The residuals of
the skylines % (panel~e)
are significantly reduced for the two typical cluster galaxies shown.
For spectral lines of the object falling into the range of strong skylines,
this can mean a striking improvement as indicated for [OII]
in the case of object~1 (panels a,b) and
the G-band in the case of object~2 (panels c,d).
The [OII] doublet of object~1 is also shown in the 2D~figure
(Fig.~\ref{fig:backsub_examples}, see example~2).

A quantitative comparison of the results from the two sky subtraction methods
is carried out in Appendix~\ref{appendix:skysub_compare}.
The main conclusion is that the noise in the improved sky subtraction
is very close to the noise floor set by photon noise and
read-out noise, whereas the noise in the traditional sky subtraction
overall is larger than this (e.g.\ Fig.~\ref{fig:backsub_hist}).
This is particularly the case for tilted slits.
The difference between the two methods is found 
where the gradient in the sky background is large,
i.e.\ at the edges of the skylines
(Fig.~\ref{fig:backsub_lambda_zoom_selmask}, cf.\ \citealt{Kelson:2003}).
 % (cf.\ \citealt{Kelson:2003}).
For our data, the difference in noise can reach a factor of 10.
The difference increases with the total number of
collected sky counts, indicating that the longer the total exposure time is,
the more of a problem the excess noise in the traditional sky subtraction
becomes (Fig.~\ref{fig:backsub_collected_sky_6300}).

It should be noted that we have used linear interpolations to perform the
rebinning in $y$ and $x$. % , since they do not produce ringing.
We have also tested using higher-order interpolations.
This makes the aliasing pattern in the traditional sky subtraction
somewhat less strong, but the problem is not removed.
This indicates that not even a higher-order interpolation can recover the
detailed intrinsic shape of the skylines, even though the skylines are not
undersampled as such (FWHM $\approx$ 4~pixels).
The improved sky subtraction, on the other hand, removes the problem by
fitting and subtracting the skylines before any interpolation of the data
is done.

\subsection{Flux calibration and telluric absorption correction}
\label{sec:flux_calib}

Spectrophotometric standard stars chosen from the
ESO list\footnote{%
\texttt{http://www.eso.org/observing/standards/spectra/}}
were observed to be able to flux-calibrate the data, 
and hot stars
(specifically stars with spectral types from O9 to B3,
and with magnitude $V$ = 9--10)
chosen from the Hipparcos catalogue
\citep{ESA:1997}\footnote{%
\texttt{http://cadcwww.dao.nrc.ca/astrocat/hipparcos/}}
were observed to be able to correct the data for telluric absorption.
The star spectra % from runs 3 and 4
were reduced using standard methods implemented in the
long-slit data reduction package \procedurename{ispec2d}, which is described
in \citet{Moustakas_Kennicutt:2006a}.

\begin{figure} % One column figure
\includegraphics[width=0.48\textwidth, bb = 33 175 572 704]
  {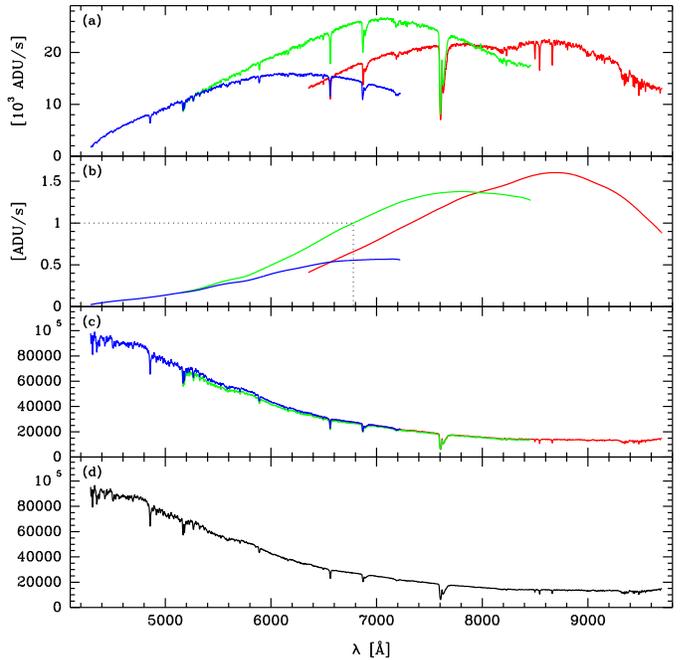}
\caption[]{%
Illustration of the method used to create a flux calibration which is valid
for all slit positions.
Panel~(a) shows 3 spectra of the same star (LTT7379), obtained through
slits placed at the extreme right (blue), the centre (green) and the
extreme left (red).
Panel~(b) shows screen flat spectra taken at the same 3 slit
positions. The spectral shape has been left untouched, and only the level
has been normalised by dividing all 3 spectra by the same constant
(cf.\ the dotted lines; see also the text).
Panel~(c) shows the star spectra (from panel~a) divided by the screen flats
(from panel~b), i.e.\ the panel shows the star SED divided by the
screen flat lamp SED, a ratio that is used as a tool ---
the screen flat lamp SED cancels out at the end of the flux calibration
procedure.
Panel~(d) shows the 3 ``spectra'' from panel~(c) merged.
}
\label{fig:flux_stars_LCR}
\end{figure}

The wavelength range of the individual MXU spectra depends on the
$x$--position of the given slit in the MXU mask
(cf.\ Sect.~\ref{sec:mask_creation}).
To be able to flux-calibrate the full spectral range of all the
MXU spectra, we observed spectrophotometric standard stars
through slits located at 3 positions:
at the far left, at the centre and at the far right. The two extreme positions
were chosen to bracket the positions that can be accommodated by the MXU masks.
The left, centre and right slits (of width 5$''$) were created using
the movable MOS arms of FORS2.
When the 3 wavelength-calibrated spectra (i.e.\ left, centre and right),
in units of ADU per second per spectral pixel, of a given star
are plotted together, see Fig.~\ref{fig:flux_stars_LCR}(a),
one problem is immediately clear:
the 3 spectra do not agree in the regions in wavelength where the spectra
overlap. The disagreement is not just in overall level, but in the shape of
the spectra.
This means that there is no universal (i.e.\ valid for all slit positions)
function that translates from ADU per second per spectral pixel
to physical flux units,
e.g.\ $\mathrm{erg}\,\mathrm{cm}^{-2}\,\mathrm{s}^{-1}\,\mathrm{\AA}^{-1}$.
We attribute this to the grism having a spectral response that depends on
the position (angle) within the field of view.
We use the same solution to this problem as in \citet{Halliday_etal:2004}.
The spectral response of the grism is recorded in all spectra,
including in the screen flat spectra
(here we are referring to screen flat spectra in which
the variation with wavelength has \emph{not} been taken out,
see Fig.~\ref{fig:flux_stars_LCR}b and below).
We divide both the standard star spectra and the MXU science spectra
by their respective screen flats.
After that, the left, centre and
right standard star spectra agree rather well (Fig.~\ref{fig:flux_stars_LCR}c).
The 3 spectra can be merged (Fig.~\ref{fig:flux_stars_LCR}d),
and a sensitivity function can be
constructed and then applied to the MXU science spectra that have also been
divided by their screen flats, creating flux-calibrated spectra.
This method assumes that all screen flats are made using the same lamp,
since otherwise the spectral energy distribution (SED)
of the screen flat lamp will not cancel out.
The screen flats used in the flux calibration only had their overall level
normalised:
the 3 standard star screen flats (left, centre, right) were normalised
by a single number, namely the level in the centre spectrum at 
6780$\,${\AA} (the central wavelength of the grism;
cf.\ Fig.~\ref{fig:flux_stars_LCR}b), and
the $\sim$30 MXU screen flats from a given mask were normalised
by a single number, namely the level in a spectrum from a slit near the
centre of the field-of-view at 6780$\,${\AA}.
It should be noted that Fig.~\ref{fig:flux_stars_LCR} is from run~3,
where the agreement between the 3 spectra is very good
(cf.\ Fig.~\ref{fig:flux_stars_LCR}c).
For run~4 the agreement is less good:
the spectra are offset in level by $\pm15$\% min--max.
We speculate that this is due to a different lamp setup 
which does not illuminate the 3 slit positions evenly when creating
the screen flats.
Since we scale the 3 spectra before we merge them (i.e.\ when going from
Fig.~\ref{fig:flux_stars_LCR}c to
Fig.~\ref{fig:flux_stars_LCR}d), no jumps are introduced
into the merged spectrum, so the relative flux calibration
within a given MXU spectrum is unaffected.

\begin{figure} % One column figure
\includegraphics[width=0.48\textwidth, bb = 57 469 276 727]
  {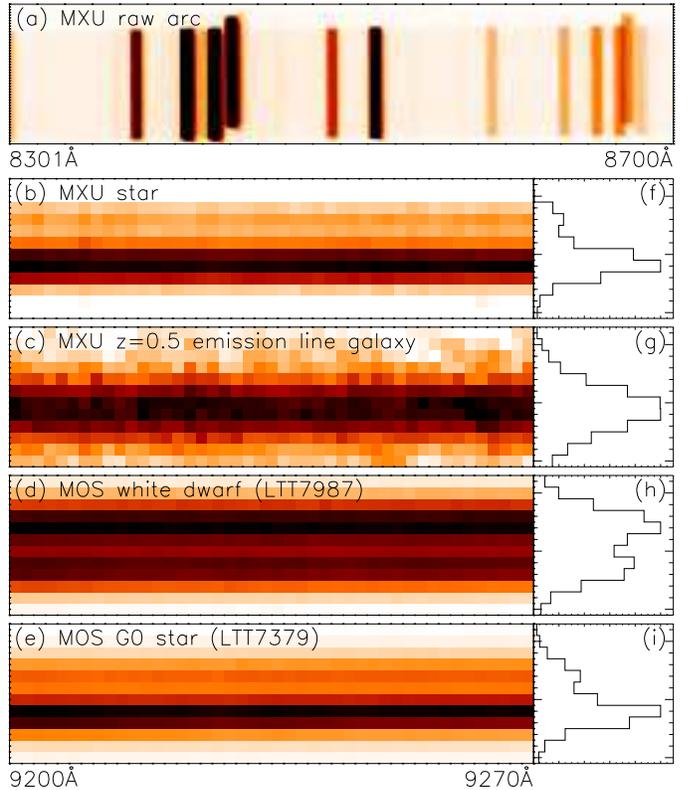}
\caption[]{%
Illustration of 2nd order contamination or lack thereof.
(a)~Part of a raw MXU arc frame. Most of the lines are from the 1st order,
but the 2 lines displaced upwards by 4$\,$px are from the 2nd order.
(b)--(e)~2D spectra of four objects.
(f)--(i)~The corresponding spatial profiles. % of the spectra.
The figure illustrates that while a modest 2nd order contamination \emph{is}
present (redwards of 8000$\,${\AA}) in the spectra of the standard star used to
establish the flux calibration (LTT7379, panel~e), no 2nd order contamination
is seen in the galaxy spectra (cf.\ panel~c) due to their redder
observed-frame SEDs.
}
\label{fig:2nd_order}
\end{figure}

It turns out that 2nd order contamination redwards of 8000$\,${\AA}
is an issue for the flux calibration,
but not for the galaxy spectra themselves.
 From the theory of optics it is known that
the different spectral orders of a grism overlap.
This implies that the grism transmits light at
wavelength $\lambda_1$ through the 1st order in the same direction as light at
wavelength $\lambda_2$ through the 2nd order. For example, light 
at
$\lambda_1 =$ 7000, 8000 and 9000$\,${\AA} could be contaminated by light at
$\lambda_2 \approx$ 3700, 4150 and 4600$\,${\AA}, respectively.
This contamination can be prevented by using a filter that blocks light below
a certain wavelength $\lambda_\mathrm{block}$.
In the above example, $\lambda_\mathrm{block}$ = 4600$\,${\AA} would prevent
2nd order contamination until $\lambda_1$ = 9000$\,${\AA}, but it would
also prevent 1st order observations below $\lambda_1$ = 4600$\,${\AA}.
When the spectral coverage is as large as in our case
(4150--9600$\,${\AA} when all MXU slit positions are considered,
Sect.~\ref{sec:mask_creation}), the choice of $\lambda_\mathrm{block}$
is necessarily a compromise between preventing 2nd order contamination
in the red and allowing observations in the blue.
We used the order-blocking filter GG435, which is the standard filter
to use with grism 600RI\@.
This is an edge filter with a transition around 4350$\,${\AA}\@.
The listed transmissions are
0.07\% at 4100$\,${\AA},
3\% at 4200$\,${\AA} and
95\% at 4500$\,${\AA}.
Our particular grism also acts as a cross disperser
(T. Szeifert, priv.\ comm.), making the 2nd order spectrum be located
3--4$\,$px (corresponding to 0.75--1$''$) above the 1st order spectrum.
This makes it easy to identify the 2nd order spectrum, where present,
in the 2D spectra.
Figure~\ref{fig:2nd_order}(a) shows part of a raw MXU arc frame.
A large number of arc lines are seen in the 1st order spectrum, and two lines
are seen in the 2nd order spectrum, displaced upwards by 4$\,$px.
Based on three such 2nd order arc lines, a linear fit provides the relation
\begin{equation}
\lambda_1 = 2.0936 \lambda_2 - 691.1\,\mathrm{\AA} \enspace ,
\label{eq:2nd_order}
\end{equation}
not unlike relations derived for other grisms
(e.g.\ \citealt{Szokoly_etal:2004,Stanishev:2007}).
We only use Eq.~(\ref{eq:2nd_order}) to understand from what wavelength
a potential 2nd order spectrum would originate.
We note that $\lambda_1$ = 8000$\,${\AA} corresponds to
$\lambda_2$ = 4150$\,${\AA}, which is just where the order-blocking filter
starts to transmit.

Figure~\ref{fig:2nd_order} also shows four 2D spectra covering the
wavelength range $\lambda_1$ = 9200--9270$\,${\AA}
which is in the far red (only 8\% of the galaxy spectra go this red).
If a 2nd order spectrum is present in the figure, it will come
from $\lambda_2 \approx$ 4720--4760$\,${\AA} (cf.\ Eq.~\ref{eq:2nd_order}).
Panel~(b) shows a somewhat blue star [$(B-V) = 0.53, (V-I) = 0.81$]
observed in one of the MXU masks. A fainter 2nd order spectrum
located 4$\,$px above the 1st order spectrum is seen.
Panel~(c) shows an emission-line galaxy at $z = 0.5$.
No 2nd order spectrum is seen, presumably due to the
redder observed-frame colours.
This galaxy has $(V-I) = 2.09$, but a more relevant colour would be one
that compared 4700$\,${\AA} to 9200$\,${\AA}\@.
We note that for galaxies at $z>0.2$ the potential 2nd order contamination
comes from below the 4000$\,${\AA} break in the rest-frame of the galaxy,
even at the reddest observed wavelengths (9600$\,${\AA})\@.
Panel~(d) shows a very blue standard star [$(B-V) = 0.07$], and here
the 2nd order spectrum 
is dominant.
The spacing between the two spectral orders is 3 pixels for the MOS spectra,
so even in good seeing the two orders overlap.
Panel~(e) shows the less blue G0 standard star LTT7379 [$(B-V) = 0.62$],
which we used to establish the flux calibration. Here a modest 2nd order
contamination is present.
Figure~\ref{fig:2nd_order}
illustrates two points:
(1)~A modest 2nd order contamination is found in the used standard star spectra.
(2)~No significant 2nd order contamination is seen in the galaxy spectra.
Point~(2) is shown quantitatively below using colours synthesised from
the spectra.

\begin{figure*} % Two column figure
\makebox[\textwidth]{
  \makebox[0.500\textwidth]{
  \includegraphics[width=0.47\textwidth, bb = 15 231 468 745]
    {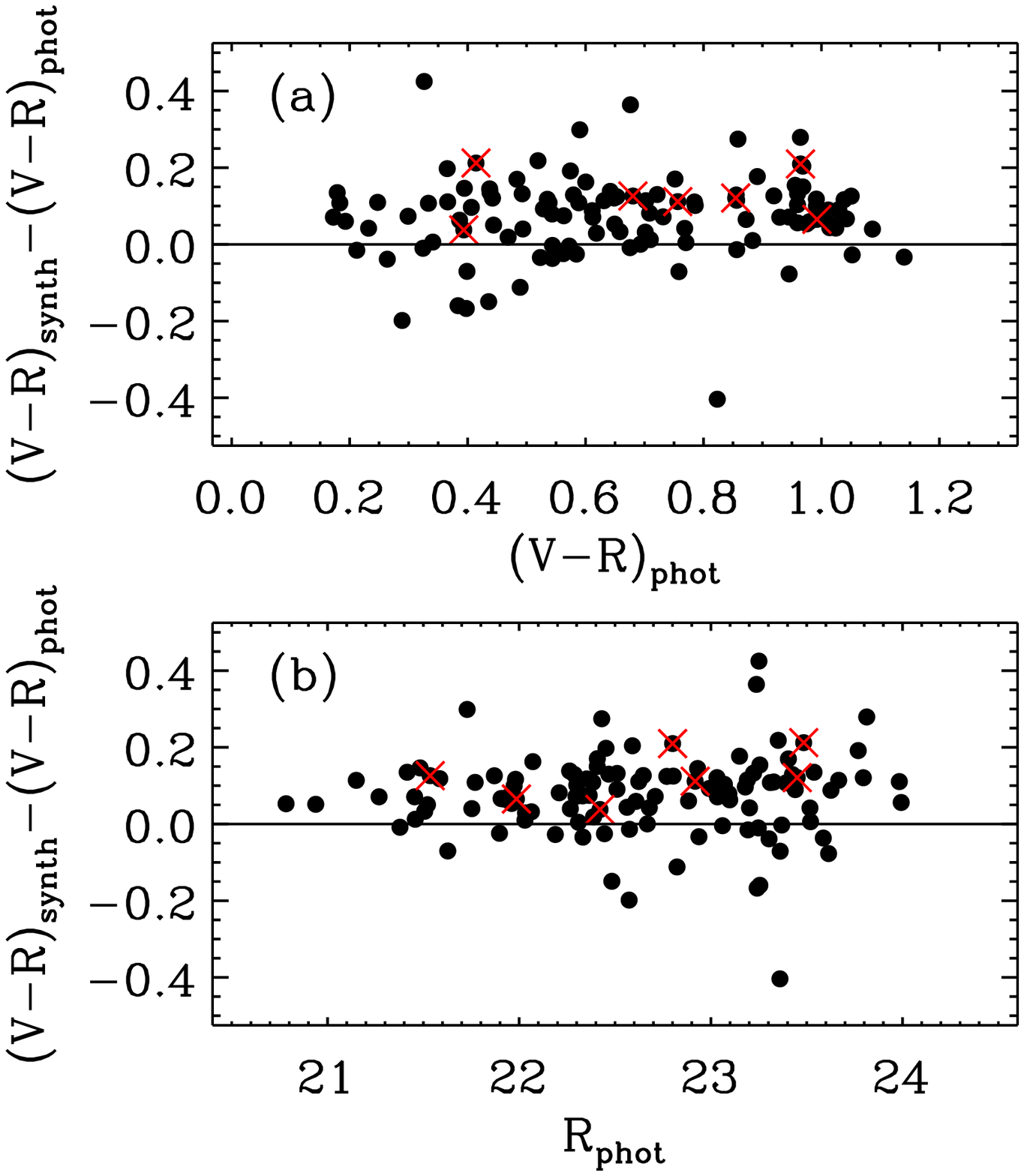}
  }
  \makebox[0.500\textwidth]{
  \includegraphics[width=0.47\textwidth, bb = 15 231 468 745]
    {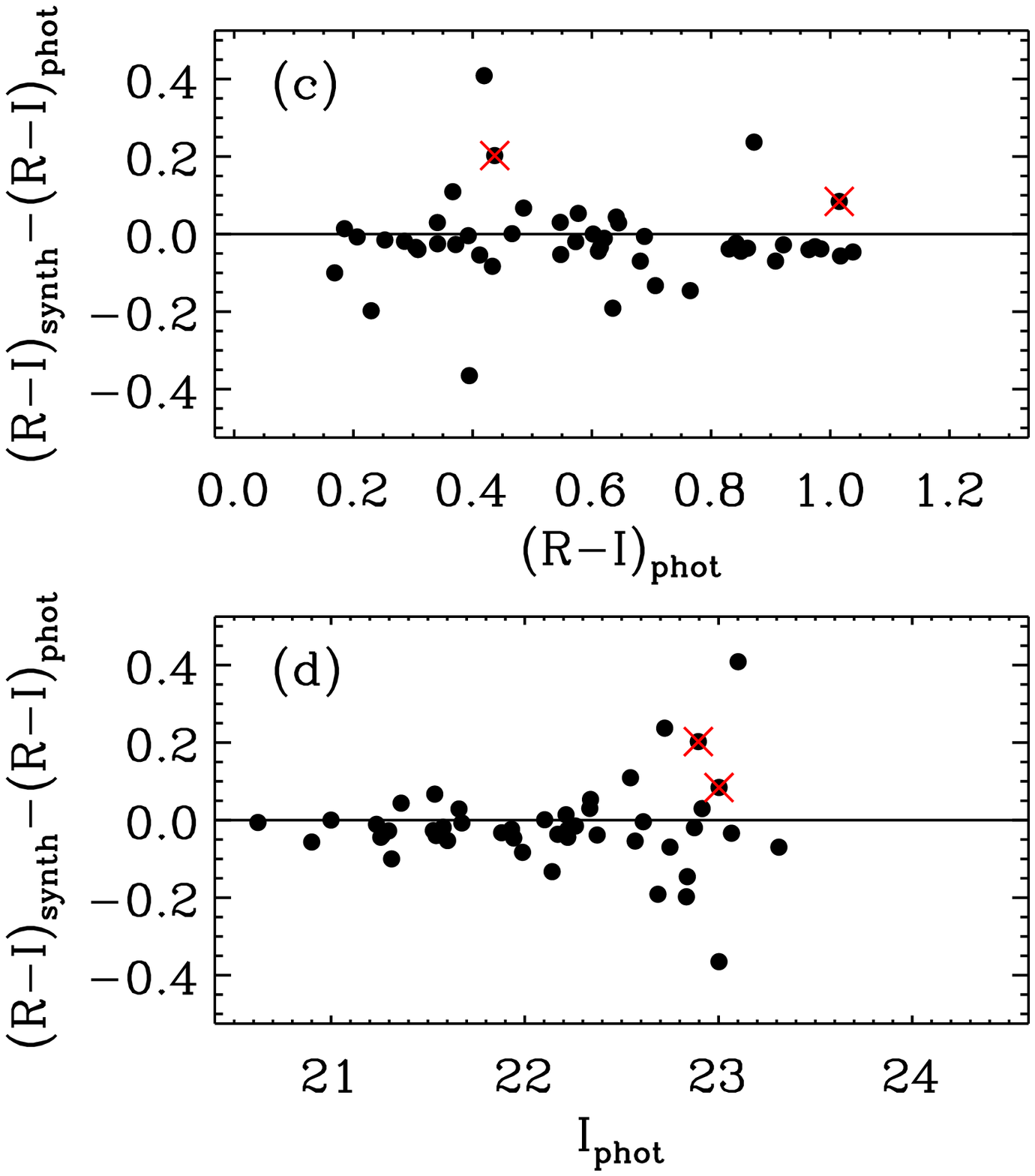}
  }
}
\caption[]{%
Comparison of colours synthesised from the spectra
with colours from the photometry.
Panels~(a) and (b) concern $(V-R)$, while
panels~(c) and (d) concern $(R-I)$\@.
To generate this figure, we only used spectra that spanned the wavelength range
4800--7500$\,${\AA} for panels (a) and (b), and
5700--8700$\,${\AA} for panels (c) and (d).
The different width of these minimum wavelength ranges
explains the different number of plotted points. % spectra available.
Only galaxies at \mbox{$z$ = 0.15--1.05} are shown.
Only spectra from fields with $VRI$ photometry are used.
Most of these are from run~3; the few spectra from run~4 (from 2~masks)
are shown with crosses.
All colours and magnitudes have been corrected for Galactic extinction
and are on the AB system \citep{Oke_Gunn:1983}.
The photometric colours and magnitudes have been measured within
a circular aperture of radius 1$''$ in images corrected to the same
fiducial seeing (cf.\ \citealt{White_etal:2005}).
The used transformations to AB are
$V_\mathrm{AB} = V_\mathrm{Vega} + 0.036$,
$R_\mathrm{AB} = R_\mathrm{Vega} + 0.216$, and
$I_\mathrm{AB} = I_\mathrm{Vega} + 0.438$.
}
\label{fig:synth_vs_phot}
\end{figure*}

In run~3 we observed 7 different standard stars, and
in run~4 we observed 2 different standard stars.
Typically, one star was observed at the start of the night
and another one at the end of the night.
There were no indications of night-to-night variations, so for each star
all the observations were combined and a sensitivity function was derived.
The sensitivity functions for the different stars all agreed until
8000$\,${\AA}, after which they diverged, with the blue stars
(e.g.\ white dwarfs)
indicating a higher sensitivity than the relatively red stars (e.g.\ G-stars),
consistent with the divergence being due to a varying degree of
2nd order contamination.
We note that a wide aperture was used to extract 1D spectra, so
all the flux from both spectral orders is included.
We decided to use the sensitivity function derived from the reddest star
observed in both runs, namely the G0 star LTT7379
(\citealt{Hamuy_etal:1992,Hamuy_etal:1994}; cf.\ Fig.~\ref{fig:flux_stars_LCR}).
In the 2D spectrum of this star the 2nd order spectrum \emph{is}
visible from about 8200$\,${\AA}
(cf.\ the upturn seen in Fig.~\ref{fig:flux_stars_LCR}a),
so we expect that the calibration is systematically off
redwards of 8000$\,${\AA}\@ % (maybe by 10--20\%). % at 9000$\,${\AA}
by an amount which is modest even at 9200$\,${\AA}
(cf.\ Fig.~\ref{fig:2nd_order}i).
Since we expect the spectra of the
high-redshift galaxies to have almost no 2nd order contamination
due to their much redder observed-frame SEDs (see also below),
a single correction function $f(\lambda)$ valid to a good approximation
for all high redshift galaxies should exist.
We note that all results published in this paper (e.g.\ redshifts)
are completely unaffected by this issue. % possible problem.

The spectra were corrected for atmospheric extinction.
The extinction curve for La Silla was used
\citep{Tug:1977,Schwarz_Melnick:1993}\footnote{Also available at
\texttt{http://www.ls.eso.org/lasilla/sciops/}
\texttt{observing/Extinction.html}}
since no extinction curve was available for Paranal
(ESO, priv.\ comm.). % private communication % 2005 
This is probably not a problem.
The La Silla extinction curve (measured over 41 nights in 1974--1976) can be
compared with the Paranal FORS1 broad-band extinction coefficients
(measured over 174 nights in 2000--2001) from \citet{Patat:2003}.
At 4300, 5500, 6600 and 7900$\,${\AA}, which we here take to represent $BVRI$,
the La Silla extinction curve gives
0.22, 0.11, 0.05 and 0.02$\,$mag$\,$airmass$^{-1}$, while
the Paranal extinction coefficients are
0.22, 0.11, 0.07 and 0.05 mag$\,$airmass$^{-1}$ on average,
with standard deviations of 0.01--0.02$\,$mag$\,$airmass$^{-1}$;
in other words, a formally perfect agreement for $B$ and $V$, and
a systematic difference for $R$ and $I$\@.
The latter is likely due to the extinction curve representing 
the part of the extinction that varies smoothly with wavelength
and which scales accurately with airmass
(specifically Rayleigh scattering, ozone absorption and aerosol scattering),
but not the telluric absorption bands from oxygen and water vapour
present in the $R$ and $I$ bands
(\citealt{Tug:1977}; F.~Patat, priv.\ comm.).
This is fortunate, since we will anyway in a separate step correct for the part
of the extinction that is due to the telluric absorption bands.
That correction is based on spectroscopic observations of hot stars.

Several hot stars were observed.
These stars are intrinsically practically featureless in the region where the
telluric absorption bands of interest are. % , 6800--8500$\,${\AA}\@.
Typically 1--2 stars were observed at the start of the night
and at the end of the night.
A 1$''$ wide longslit was used, giving spectra going to
8600$\,${\AA}\@. % orig. listed as 8450, but that's for centre FOV
The continuum was normalised to unity and the spectra from different
stars and nights were combined.
The spectral regions around the 4 telluric bands present in the hot star data
(the B--band near 6900$\,${\AA}, a weaker feature near 7200$\,${\AA},
 the A--band near 7600$\,${\AA}, and a weaker feature near 8200$\,${\AA})
were used to correct the MXU spectra for telluric absorption as follows.
For each mask, the spectra of a few bright stars in the mask
were located and used to derive the scaling and
the shift of the hot star spectrum around the telluric feature in question.
The typical rms in these fits was 0.05, as reported by the
\procedurename{telluric} IRAF task.
Each telluric feature was scaled and shifted individually,
apart from the weakest band (the one close to 8200$\,${\AA}), which was locked
to the A--band.
All the spectra from the given mask were then corrected using this
scaled and shifted continuum-normalised hot star spectrum.

As a test of the flux calibration and the two extinction correction steps,
we derived synthetic magnitudes from the spectra and
compared these with the photometry.
The wavelength range of the spectra varies
(cf.\ Sect.~\ref{sec:mask_creation}):
the bluest spectra start in the middle of the $B$--band and cover the
$V$ and $R$--bands, and the reddest spectra start a bit into the
$R$--band and end beyond the $I$--band.
This means that $(V-R)$ and $(R-I)$ colours can be synthesised from
two disjoint subsets of the spectra without extrapolation.
The high--$z$ fields with $VRI$ photometry are suited for such a
comparison, whereas the mid--$z$ fields with $BVI$ photometry are not.
In Fig.~\ref{fig:synth_vs_phot}, we show the results from the comparison.
Panel~(a) shows the colour difference $\VRsynth - \VRphot$ versus
the photometric colour, and panel~(b) shows the same colour difference
versus photometric magnitude.
Panels~(c) and (d) show $(R-I)$\@.
The scatter in colour difference is fairly small, namely
0.08$\,$mag for $(V-R)$ and 0.06$\,$mag for $(R-I)$,
after rejecting 3$\sigma$ outliers.
This scatter comes from the following sources:
(1)~photon noise in the spectra,
(2)~photon noise in the photometry,
(3)~possible differences due to the rectangular spectroscopic apertures
not being identical to circular photometric apertures,
and
(4)~possible spectrum-to-spectrum errors in the flux calibration.
The mean value of the colour difference is
$0.08\pm0.01\,$mag for $(V-R)$ and $-0.03\pm0.01\,$mag for $(R-I)$\@.
Ideally these values should be zero.
We have no explanation for the offset in $(V-R)$, but systematic
relative flux calibration uncertainties of the order of 10\% are
extremely difficult to avoid in multi-slit spectroscopic observations.
The negative offset in $(R-I)$ is qualitatively in agreement with the
above-mentioned systematic error in the flux calibration
redwards of 8000$\,${\AA}, since part of the $I$--band region of
the spectra is located there.
It is reassuring to see that there is no significant trend of
colour difference with colour (Fig.~\ref{fig:synth_vs_phot}a and c).
When a linear fit is performed, the slope is
$(0.031\pm0.026)$ for $(V-R)$ and
$(0.001\pm0.040)$ for $(R-I)$\@.
The fact that the $(R-I)$ slope is consistent with zero is compatible with
our conjecture that for high-redshift galaxies,
regardless of SED (colour), the 2nd order contamination is negligible.
Finally, we have compared the run~3 spectra, which constitute most of the
points in Fig.~\ref{fig:synth_vs_phot}, with the run~4 spectra
(shown as crosses). The run~4 points tend to lie higher in the plots
than the run~3 points, a difference that is marginally significant
(2--3~sigma). The difference may be due to the different screen flat lamp
setup used in run~4.

Our overall conclusion is that the accuracy of the flux calibration
is typically below 10\%, which is very good for this type of
multi-object spectroscopy.
The expected modest systematic error in the flux calibration
redwards of 8000$\,${\AA} can be corrected if it proves necessary.

\begin{figure*} % Two column figure
\centering
\includegraphics[width=0.8650\textwidth,bb = 74 201 512 745]
  {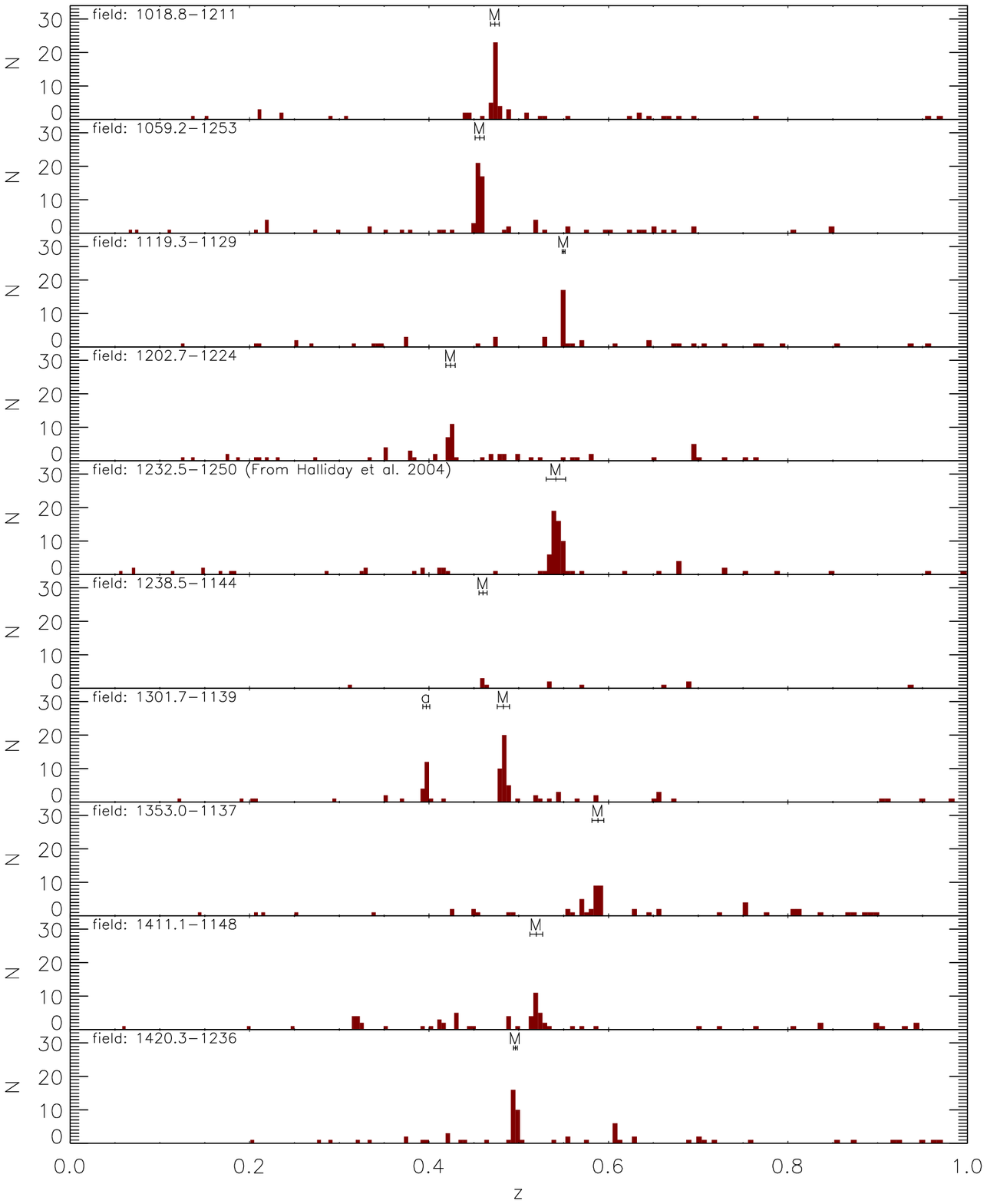}
\caption{%
Redshift histograms for the 10 mid--$z$ fields.
The clusters for which we have measured a redshift and a velocity dispersion
(\citealt{Halliday_etal:2004} or this paper) are indicated with the
$\pm3\sigmacl$ range shown. The labels are ``M'' for the main cluster
and ``a'' or ``b'' for secondary clusters.
Both 0-colon redshifts (``secure'') and
1-colon redshifts (``secure but with larger uncertainties'') have been used
in the plot.
The binsize in $z$, $\Delta z$, varies with $z$ in such a way that the
binsize in rest-frame velocity, $\Delta v_{\rm rest} = c \Delta z/(1+z)$,
is kept constant at 1000$\,$km$\,$s$^{-1}$.
This is achieved binning in $\log(1+z)$ space with a constant
binsize of $\log(\Delta v_{\rm rest}/c+1)$.
}
\label{fig:zhist_midz}
\end{figure*}

\begin{figure*} % Two column figure
\centering
\includegraphics[width=0.8650\textwidth,bb = 74 201 512 745]
  {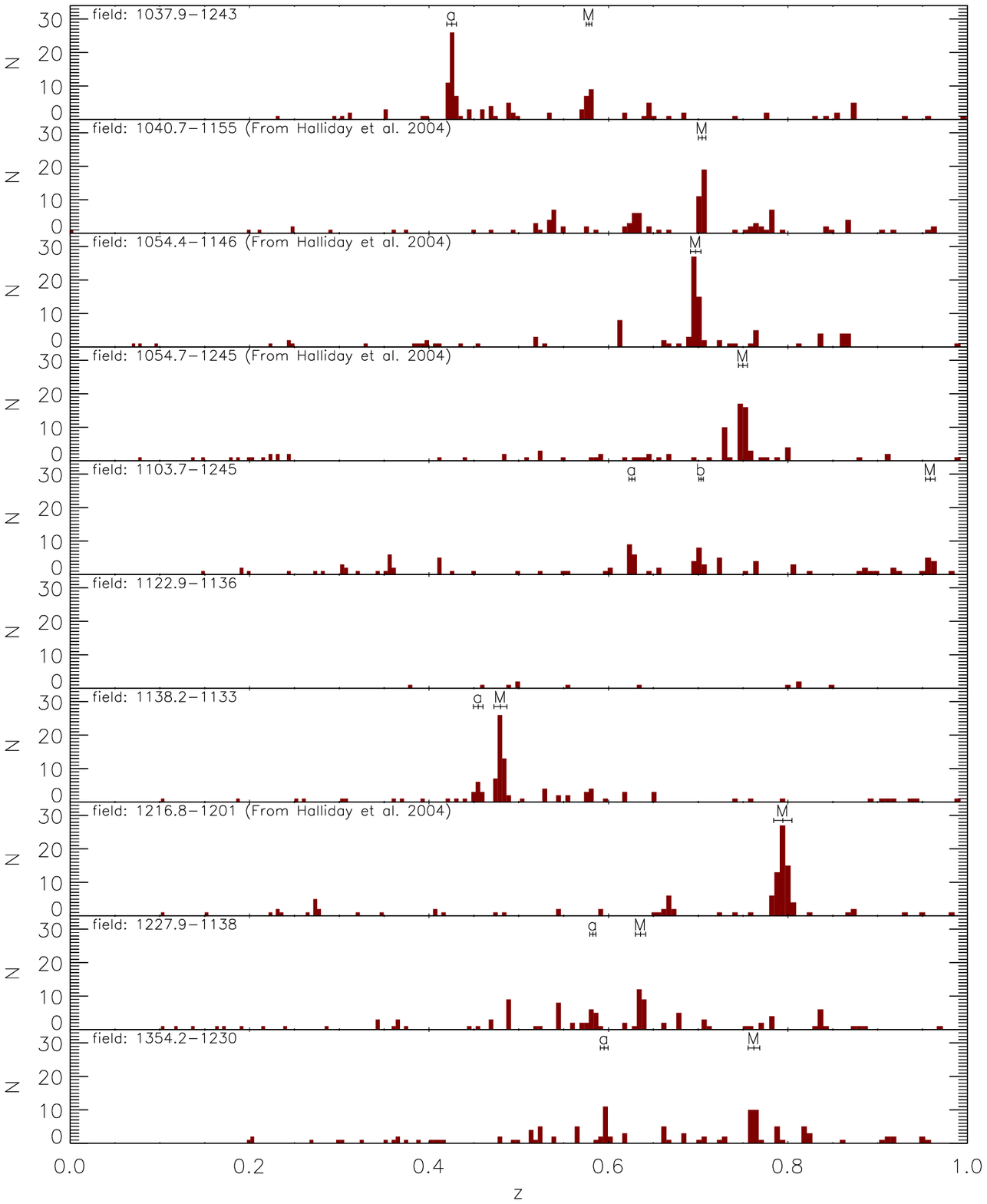}
\caption{%
Redshift histograms for the 10 high--$z$ fields.
Otherwise this figure is like Fig.~\ref{fig:zhist_midz}.
}
\label{fig:zhist_highz}
\end{figure*}

\subsection{Test of the wavelength calibration using skylines}
\label{sec:wlcal_verification}

To test the wavelength calibration, we measured the wavelength of
3 strong and almost unblended skylines
in all the 2D spectra.
The reference wavelengths of the 3 lines are
6300.30$\,${\AA}, 6363.78$\,${\AA} and 6863.96$\,${\AA}
\citep{Osterbrock_etal:1996},
where we have taken into account that the 
first and the last of these lines
are blends of a strong line and a 10--12 times weaker line
at our spectral resolution (FWHM $\approx$ 6$\,${\AA}).
For each of the 3 skylines and for each of the $\sim$2000 spectra
in the long masks,
the difference between the wavelength according to our wavelength calibration
and the reference wavelength,
$\Delta\lambda \equiv \lambda_\mathrm{wlcal} - \lambda_\mathrm{ref}$,
was calculated.
For the 3 lines, the mean values of $\Delta\lambda$ are 
$-0.11\,${\AA}, $-0.05\,${\AA} and $-0.33\,${\AA}, respectively.
The standard deviations are 0.67$\,${\AA}, 0.67$\,${\AA} and 0.65$\,${\AA},
respectively.
Since the redshift is given by
$z = \lambda_\mathrm{obs}/\lambda_\mathrm{rest} - 1$,
and since we typically measure the redshift from
spectral lines at wavelengths of approximately
$\lambda_\mathrm{rest} = 3800\,${\AA},
an error of 0.67$\,${\AA} in the observed wavelength $\lambda_\mathrm{obs}$
translates into an error in the redshift of 0.00018.
We note that this number is given with 5 decimals, whereas we normally provide
redshifts with only 4 decimals.
This error corresponds to a rest-frame velocity of 33$\,$km$\,$s$^{-1}$, 
at a typical redshift of $z=0.6$.
Since this error is rather small, we have chosen not to correct for it
(the zero points of the wavelength calibrations of the individual spectra
could in principle have been corrected using the measured values of
$\Delta\lambda$).
It is worth pointing out that the arc frames from which the wavelength
calibrations were established were taken during the day with the telescope
pointing at zenith. The fact that the typical error nevertheless is so small
testifies to the stability of the FORS2 instrument.

The measurement of the wavelength of the skylines was performed using
Gaussian fits. These fits also provide a measure of the width (FWHM)
of the lines.
For a subset of the masks, we tested how the line width depended on the
absolute value of the slit-tilt angle.
It was found that there was only a weak dependence:
from $0^\circ$ to $45^\circ$ the FWHM increased on average by just 4\%.

\section{Galaxy redshifts}
\label{sec:galaxy_redshifts}

\subsection{The redshift measurements}
\label{sec:redshift_measurements}

Spectroscopic galaxy redshifts were measured using emission lines
where possible, in particular the [O{\sc ii}]$\lambda$3727 line, or
the most prominent absorption lines, e.g.\ Calcium K and H lines at
3934$\,${\AA} and 3968$\,${\AA}.
The redshifts were manually assigned a quality flag.
The vast majority % ($\sim$95\%)
of the measured redshifts are of the highest
quality, and these redshifts are listed without colons in our data tables.
Secure redshifts but with larger uncertainties are listed with one colon,
and doubtful redshifts are listed with two colons.
For a small fraction of the objects
(3.3\%),
no redshift could be determined,
and these redshifts are listed as 9.9999 in our data tables.
For the objects targeted as possible cluster members in the 66 long masks,
the statistics are as follows:
2.8\%~stars, 93.9\%~galaxies and 3.3\%~without a determined redshift.
Of the galaxy redshifts, the quality distribution (i.e.\ 0, 1 or 2 colons) is
96.0\%, 2.6\% and 1.4\%, respectively.

We can estimate the typical redshift error using spectra for the galaxies
that have been observed more than once (i.e.\ in more than one mask).
In the long masks, we have 43 galaxies with 2 redshifts available,
and 2 galaxies with 3 redshifts available,
when only using redshifts without colons.
For each object and for each redshift, we first compute the difference
between the redshift and the mean of the redshifts available for the object.
For example, if 2 redshifts of 0.4704 and 0.4708 are available
we derive differences of $-0.0002$ and 0.0002; and
if 3 redshifts of 0.6960, 0.6962 and 0.6957 are available
we get differences of approximately 0.0000, 0.0002 and $-0.0003$.
We then divide the differences
by 0.7071 for the differences coming from 2 redshifts per object, and
by 0.8166 for the differences coming from 3 redshifts per object.
These scaling factors were calculated numerically
based on a Gaussian distribution. The factors correct for the fact that
we calculate the differences with respect to the mean of the observed values,
not with respect to the (unknown) mean of the parent distribution.
We finally calculate a biweight estimate of the dispersion % standard deviation
of the 92 scaled differences, giving 0.00030 (note: 5 decimals)
as the estimate of the typical redshift error.
This is the same value that was found in \citet{Halliday_etal:2004}.
This error corresponds to 56$\,$km$\,$s$^{-1}$ in rest-frame velocity
at a typical redshift of $z=0.6$.

\subsection{Redshift histograms and cluster names}
\label{sec:zhist}

\begin{table}
\caption{IDs for the preliminary BCGs for the additional secondary clusters}
\label{tab:BCGIDs}
\begin{center}
\renewcommand{\arraystretch}{1.05} % To make space for the errors; def. is 1.0
\setlength{\tabcolsep}{1.4pt} % Default is 6pt
\begin{tabular}{lcccc}
\hline
\hline
Cluster & Alt. & $\zcl$ & $z_{\rm BCG}$ & BCG ID \\
\hline
\multicolumn{5}{l}{Mid--$z$ fields:} \\
cl1301.7$-$1139a & G1\_1301 & 0.3969 & 0.3976 & EDCSNJ1301351$-$1138356 \\
\multicolumn{5}{l}{High--$z$ fields:} \\
cl1037.9$-$1243a & \ldots   & 0.4252 & 0.4278 & EDCSNJ1037523$-$1244490 \\
cl1138.2$-$1133a & C2\_1138 & 0.4548 & 0.4519 & EDCSNJ1138086$-$1136549 \\
cl1227.9$-$1138a & C2\_1227 & 0.5826 & 0.5812 & EDCSNJ1227521$-$1139587 \\
cl1354.2$-$1230a & \ldots   & 0.5952 & 0.5947 & EDCSNJ1354114$-$1230452 \\
\hline
\end{tabular}
\end{center}

\vspace*{2pt}

Notes --
For the remaining EDisCS clusters the BCG IDs are listed in
\citet{White_etal:2005}.
The column ``Alt.'' gives the name used in \citet{Poggianti_etal:2006}.
\end{table}

\begin{table*}
\caption{Illustration of the format of the spectroscopic catalogues}
\label{tab:sample_spectable}
\begin{center}
\begin{tabular}{lllllcc}
\hline
\multicolumn{1}{c}{Object ID} &
\multicolumn{1}{c}{RA (J2000)} &
\multicolumn{1}{c}{Dec (J2000)} &
\multicolumn{1}{c}{$\Ione$} &
\multicolumn{1}{c}{$z$} &
\multicolumn{1}{c}{Membership flag} &
\multicolumn{1}{c}{Targeting flag} \\
\multicolumn{1}{c}{(1)} &
\multicolumn{1}{c}{(2)} &
\multicolumn{1}{c}{(3)} &
\multicolumn{1}{c}{(4)} &
\multicolumn{1}{c}{(5)} &
\multicolumn{1}{c}{(6)} &
\multicolumn{1}{c}{(7)} \\
\hline
EDCSNJ1103355$-$1244515   & 11:03:35.53 & $-$12:44:51.5 & 20.585 & 0.6259   & 1a & 1 \\
EDCSNJ1103373$-$1246364   & 11:03:37.34 & $-$12:46:36.4 & 22.051 & 0.7030   & 1b & 1 \\
EDCSNJ1103420$-$1244409   & 11:03:41.99 & $-$12:44:40.9 & 22.488 & 0.9637   &  1 & 1 \\
EDCSNJ1103539$-$1248430   & 11:03:53.85 & $-$12:48:43.0 & 22.556 & 0.2727   &  0 & 2 \\
EDCSNJ1103538$-$1246324   & 11:03:53.76 & $-$12:46:32.4 & 22.828 & 0.7539   &  0 & 3 \\
EDCSNJ1018371$-$1214297   & 10:18:37.12 & $-$12:14:29.7 & 19.670 & 0.0000   & -- & 4 \\
EDCSNJ1103351$-$1249044   & 11:03:35.12 & $-$12:49:04.4 & 22.938 & 9.9999   & -- & 1 \\
EDCSNJ1103397$-$1246532   & 11:03:39.69 & $-$12:46:53.2 & 23.497 & 0.6246:  & 1a & 3 \\
EDCSNJ1103452$-$1245403   & 11:03:45.16 & $-$12:45:40.3 & 22.406 & 0.9383:: &  0 & 1 \\
EDCSNJ1227551$-$1136202:A & 12:27:55.07 & $-$11:36:20.2 & 21.475 & 0.6390   &  1 & 1 \\
EDCSNJ1227551$-$1136202:B & 12:27:55.07 & $-$11:36:20.2 & 21.475 & 0.5441   &  0 & 1 \\
EDCSXJ1103539$-$1244439   & 11:03:53.91 & $-$12:44:43.9 & 22.63  & 0.7025   & 1b & 1 \\
EDCSYJ1059032$-$1254311   & 10:59:03.21 & $-$12:54:31.1 & 99.99  & 0.4579   &  1 & 3 \\
\hline
\end{tabular}
\end{center}

\vspace*{0.1ex}

Notes --
This example table contains entries from several survey fields simply
to illustrate all relevant features of the tables published
electronically with this paper, in which the survey fields are not mixed.
\end{table*}

Redshift histograms for all 20 EDisCS fields are shown in
Fig.~\ref{fig:zhist_midz} (mid--$z$ fields) and
Fig.~\ref{fig:zhist_highz} (high--$z$ fields).
The binsize in terms of rest-frame velocity is kept constant at
1000$\,$km$\,$s$^{-1}$
for easier visual interpretation of the histograms.
We note that redshift histograms for 5 of the fields
(1232.5$-$1250,
1040.7$-$1155,
1054.4$-$1146,
1054.7$-$1245 and
1216.8$-$1201)
were already presented in \citet{Halliday_etal:2004},
but they are repeated here to give an overview of the full
set of EDisCS spectroscopy.

On the redshift histogram for the given field, we have indicated
the location of the cluster(s) for which we have determined a
velocity dispersion in \citet{Halliday_etal:2004} or this paper
(Sect.~\ref{sec:zcl_sigmacl}).
Most fields have a single, main cluster.
By main cluster is meant the cluster corresponding to the 
LCDCS detection that our survey originally targeted
(cf.\ \citealt{White_etal:2005}).
In some fields there is a secondary cluster labelled ``a'',
and in one field, there is also an additional secondary cluster labelled ``b''.
All the main clusters and two of the secondary clusters were already
discussed in \citet{White_etal:2005}.
In this paper, we identify 5 additional secondary clusters, chosen so that
we have measured a velocity dispersion $\sigmacl$ of all structures
with $\sigmacl \gtrsim 400$$\,$km$\,$s$^{-1}$.
The naming of the clusters is simple:
the main cluster is named ``cl'' plus the field name (e.g.\ cl1103.7$-$1245),
and the secondary ``a'' and ``b'' clusters have that letter added
(e.g.\ cl1103.7$-$1245a, cl1103.7$-$1245b).

In the $xy$ plots and velocity histograms presented in later
sections, we indicate the location of the BCG\@.
We therefore need to identify BCGs for the 5 additional secondary clusters.
We simply do this by provisionally identifying the brightest (in $\Itot$)
spectroscopic member without colons on the redshift
as the BCG, see Table~\ref{tab:BCGIDs}.
We note that the BCGs listed in \citet{White_etal:2005} for the other clusters
were identified in a more elaborate manner, considering also galaxies for
which we only have photometric redshifts. % have no spectroscopy.
In fact, 3 of the 5 BCGs listed in
Table~\ref{tab:BCGIDs} have quite blue colours compared to what is usual
for BCGs, making it likely that we simply have not obtained spectroscopy
for the true BCG\@. % for the given cluster.
Conversely, the listed BCGs for the clusters
cl1301.7$-$1139a and cl1037.9$-$1243a
have colours and total magnitudes in line with the BCGs
for the other EDisCS clusters listed in \citet{White_etal:2005}.
These 2 BCGs have been included in the study of the evolution
of the EDisCS BCGs \citep{Whiley_etal:EDisCS_BCG}.

A note should be made about the 1122.9$-$1136 field for which
we only have spectroscopy from an initial short mask 
(plotted in Fig.~\ref{fig:zhist_highz}).
The galaxy listed in \citet{White_etal:2005} as being the BCG and
as having $z = 0.6397$ does in fact have $z = 0.4995$.
The few redshifts at hand do not give substantial evidence of
a cluster in that field.
(The imaging \emph{does} show some evidence, see \citealt{White_etal:2005}.)

\subsection{The spectroscopic catalogues}
\label{sec:catalogues}

The spectroscopic catalogues for 5 fields were published in
\citet{Halliday_etal:2004}, and the spectroscopic catalogues for
14 fields are published in this paper.
The last of the 20 EDisCS fields has very little data and is not published.
The spectroscopic catalogues are published electronically at the CDS\@.
The format of the tables is illustrated in Table~\ref{tab:sample_spectable}.

Column~1 gives the object ID\@.
The spectroscopic target selection was based on photometric catalogues
created from the imaging available at the time.
Subsequently deeper imaging was obtained for some fields
(e.g.\ the total exposure time went from 1 to 2~hours),
and new photometric catalogues were created.
Both sets of catalogues used the $I$--band image for the object
detection and segmentation/deblending \citep{Bertin_Arnouts:1996}.
For the fields where the $I$--band image had changed due to getting
additional data, the object segmentation occasionally differed.
For example, a galaxy and a star seen next to each other in projection
might have been correctly segmented into two objects in the old catalogue
but merged into one object in the new catalogue, or vice versa.
The impact is as follows.
For 99.5\% of the objects targeted and observed spectroscopically
the object in the old catalogue is also found in the new catalogue,
and here we give the new ID (IDs starting with EDCS\textbf{N}J).
For the remaining 0.5\%, the object from the old catalogue is not found in
the new catalogue, and here we have constructed IDs starting with
EDCS\textbf{X}J\@.
Additionally, a handful of objects, all non-targeted (i.e.\
serendipitously observed), neither existed in the old nor in the new
catalogues, and we have given these IDs starting with 
EDCS\textbf{Y}J\@. % --- internally called NONAME!!!
We note that the EDCS\textbf{X}J and the EDCS\textbf{Y}J objects are not found
in the published photometric catalogues (e.g.\ the optical ones from
\citealt{White_etal:2005}), since these catalogues are the new ones,
but these objects can still be used for purposes that only use the
redshift (and position), such as determining cluster velocity dispersions
and substructure.

Another issue are the cases where a single object from the photometric
catalogue was found to correspond to two physically-distinct objects
in the obtained spectrum, i.e.\ at different redshifts.
In the cases where the two redshifts were close we have inspected the
available imaging, including HST imaging where available
\citep{Desai_etal:2007}, to check that it really was two distinct galaxies.
For these physically-distinct objects 
we have constructed unique IDs by appending ``:A'' and ``:B'' to the ID
from the photometric catalogue (with :A being the southernmost object).
We note that in \citet{Halliday_etal:2004}, where there was one such case,
we appended a colon to the IDs for both physical objects instead of :A and :B,
resulting in IDs that were not unique, which might be slightly misleading.

Column~2 gives the right ascension, and column~3 gives the declination
(J2000).

Column~4 gives $\Ione$, the I--band magnitude
(not corrected for Galactic extinction)
within a circular aperture of radius 1$''$.
This magnitude comes from the new catalogues
(published in \citealt{White_etal:2005}),
except for the EDCS\textbf{X}J objects where it comes from the
old catalogues.
No magnitude is available for the EDCS\textbf{Y}J objects, and a value of
99.99 is listed in the table. % \tmp{should that rather be ``--''?}

Column~5 gives the redshift, optionally with one or two colons appended
to signify lower quality, see Sect.~\ref{sec:redshift_measurements}.
In this paper, the redshifts are always given with 4 decimals.
A value of 0.0000 denotes a star, and 9.9999 denotes that no redshift
could be determined.

Column~6 gives the membership flag.
It is
  1    for members of the main cluster,
  1a   for members of the secondary ``a'' cluster,
  1b   for members of the secondary ``b'' cluster,
  0    for field galaxies, and
``--'' for stars and objects without a determined redshift.
The tables for the 14 fields contain flags indicating the 21 clusters
listed in Table~\ref{tab:zcl_sigmacl_Nmem} ahead.
Membership is defined as being within $\pm3\sigmacl$ from $\zcl$.

Column~7 gives the targeting flag, see Table~\ref{tab:targeting_flag_values}
and Sect.~\ref{sec:mask_creation}.

The published redshifts come from the 66 longs masks
(i.e.\ those listed in Table~\ref{tab:target_selection}).
In addition, 9 redshifts from the short initial masks
(cf.\ Sect.~\ref{sec:obs}) have been added:
8 galaxies which are members of cl1037.9$-$1243a ($\zcl = 0.4252$), and
1 galaxy which is a member of   cl1103.7$-$1245  ($\zcl = 0.9586$).

As mentioned in Sect.~\ref{sec:redshift_measurements}, some objects were
observed more than once.
In \citet{Halliday_etal:2004}, the published tables simply contained
all the observations. Here we publish two tables per field: one with
one entry per unique object ID, and one table with extra observations.
For example, if an object was observed 3 times, we place the best observation
in the unique object table, and the two other observations in the extra table.
What constitutes the best observation is determined from a set of rules,
for example that a redshift with 0~colons has priority over a redshift with
1~colon, and that a targeted observation has priority over a serendipitous
observation.

\section{Completeness, success rate and potential selection biases}
\label{sec:completeness}

\begin{figure*} % Two column figure
\centering
\includegraphics[width=1.00\textwidth,bb =  5 299 570 751]{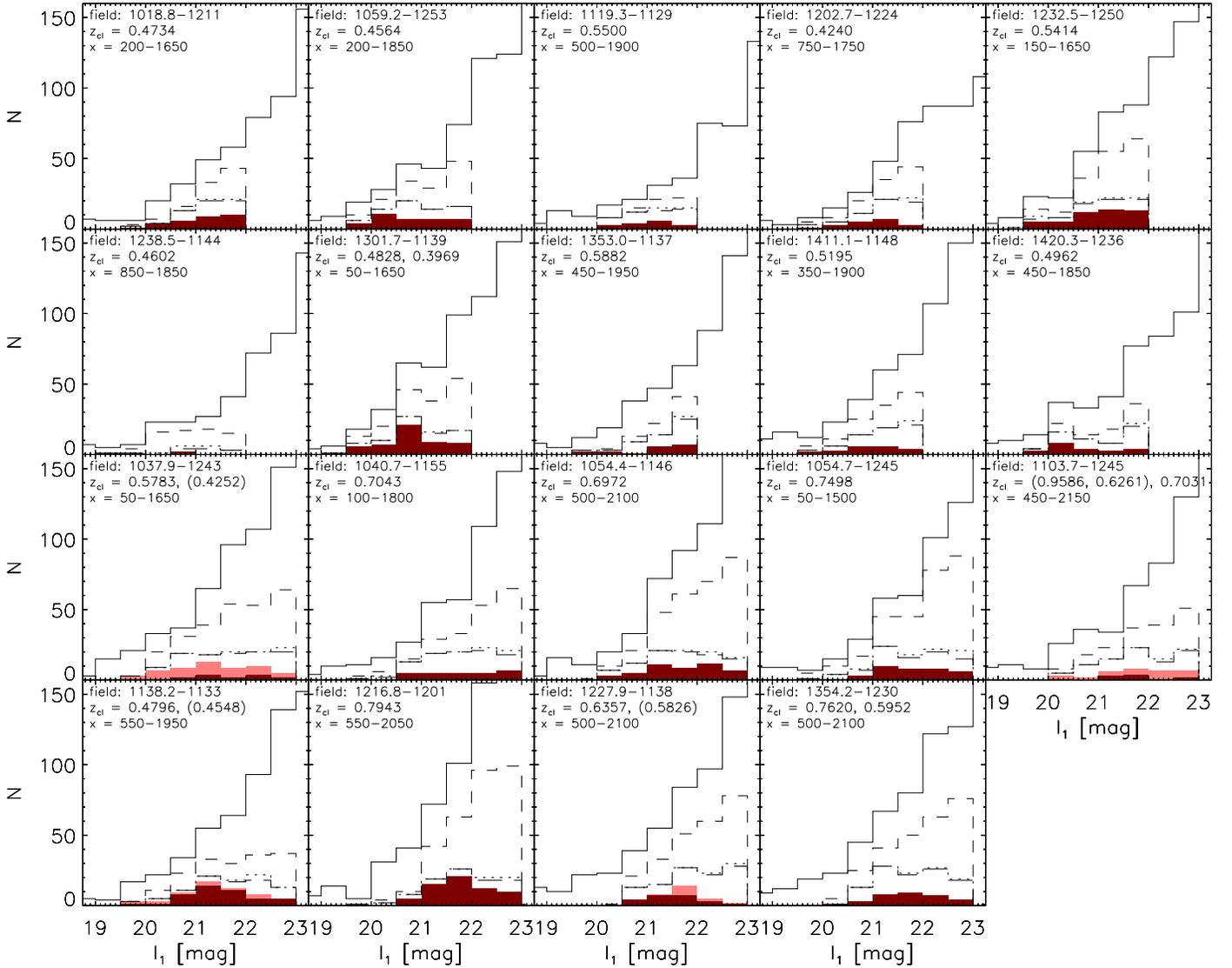}
\caption{%
The target selection process as a function of magnitude
for the EDisCS fields with long spectroscopic exposures
plus the 1238 field.
For each panel, 
the figure shows:
solid histogram: objects in the photometric catalogue,
dashed histogram: targets,
dotted histogram: observed targets,
long dashed histogram: targets with a secure redshift,
light red filled histogram: galaxies that are members of any cluster
in the field % for which we have measured a velocity dispersion
(cf.\ Table~\ref{tab:zcl_sigmacl_Nmem}),
dark red filled histogram: galaxies that are members of the
cluster(s) at the redshift(s) targeted by the $\zphot$--based
target selection.
The dotted and the long dashed histograms often coincide,
indicating that a secure redshift was obtained for all the observed targets.
All numbers have been computed within the region on the sky
spanned by % practically all
the spectroscopic observations,
with the range in $x$ being given on the panels,
and the range in $y$ being 100--1950$\,$px,
cf.\ the $xy$ plots (Fig.~\ref{fig:xy}).
The cluster redshifts are given on the figure.
If more than one value is given, the order is:
main cluster, secondary cluster ``a'', secondary cluster ``b''.
Values in parentheses are for clusters that are not at the
redshift(s) targeted by the $\zphot$--based target selection.
}
\label{fig:mag_hist}
\end{figure*}

\subsection{Completeness and success rate}
\label{sec:success_rate}

The target selection process, as a function of $\Ione$ magnitude,
is illustrated in Fig.~\ref{fig:mag_hist}.
Each panel corresponds to a given field.
To give a complete overview we also show the 5 fields from
\citet{Halliday_etal:2004}.
The starting point is a photometric catalogue (solid histogram).
Using the 4 target selection rules (Sect.~\ref{sec:target_selection}),
the target catalogue is created (dashed histogram).
Some of the targets are observed (dotted histogram).
For the vast majority of these (95\% on average),
a secure redshift is obtained (long dashed histogram).
Some of these objects are galaxies that are members of any of the
clusters in the given field, for which we have measured a velocity dispersion,
cf.\ Table~\ref{tab:zcl_sigmacl_Nmem} ahead (light red filled histogram).
Some galaxies are members of the cluster(s) at the redshift(s) 
targeted by the $\zphot$--based target selection
for the given field (dark red filled histogram).
The distinction between the two sets of clusters can be illustrated by
the 1037.9$-$1243 field: we have measured a velocity dispersion for both
the main cluster            (cl1037.9$-$1243,  $\zcl = 0.58$) and for % 0.5783
the secondary ``a'' cluster (cl1037.9$-$1243a, $\zcl = 0.43$).        % 0.4252
However, the $\zphot$--based target selection only targeted 0.58
(cf.\ Table~\ref{tab:target_selection}), i.e.\ only the main cluster.
The cluster redshifts are given on the figure, % Fig.~\ref{fig:mag_hist},
with the redshifts for the non-targeted clusters being given in parentheses.
All the histograms shown in the figure % Fig.~\ref{fig:mag_hist}
were computed within the region on the sky spanned by
the spectroscopic observations.
As can be seen from the $xy$ plots in
Fig.~\ref{fig:xy},
the objects observed spectroscopically occupy a region which does not
span the full width of the imaging. % in $x$ (here meaning east--west).
This is done in order to obtain a useful wavelength coverage in the spectra,
cf.\ Sect.~\ref{sec:mask_creation}.

\begin{figure*} % Two column figure
\centering
\includegraphics[width=1.00\textwidth,bb =  5 299 570 751]{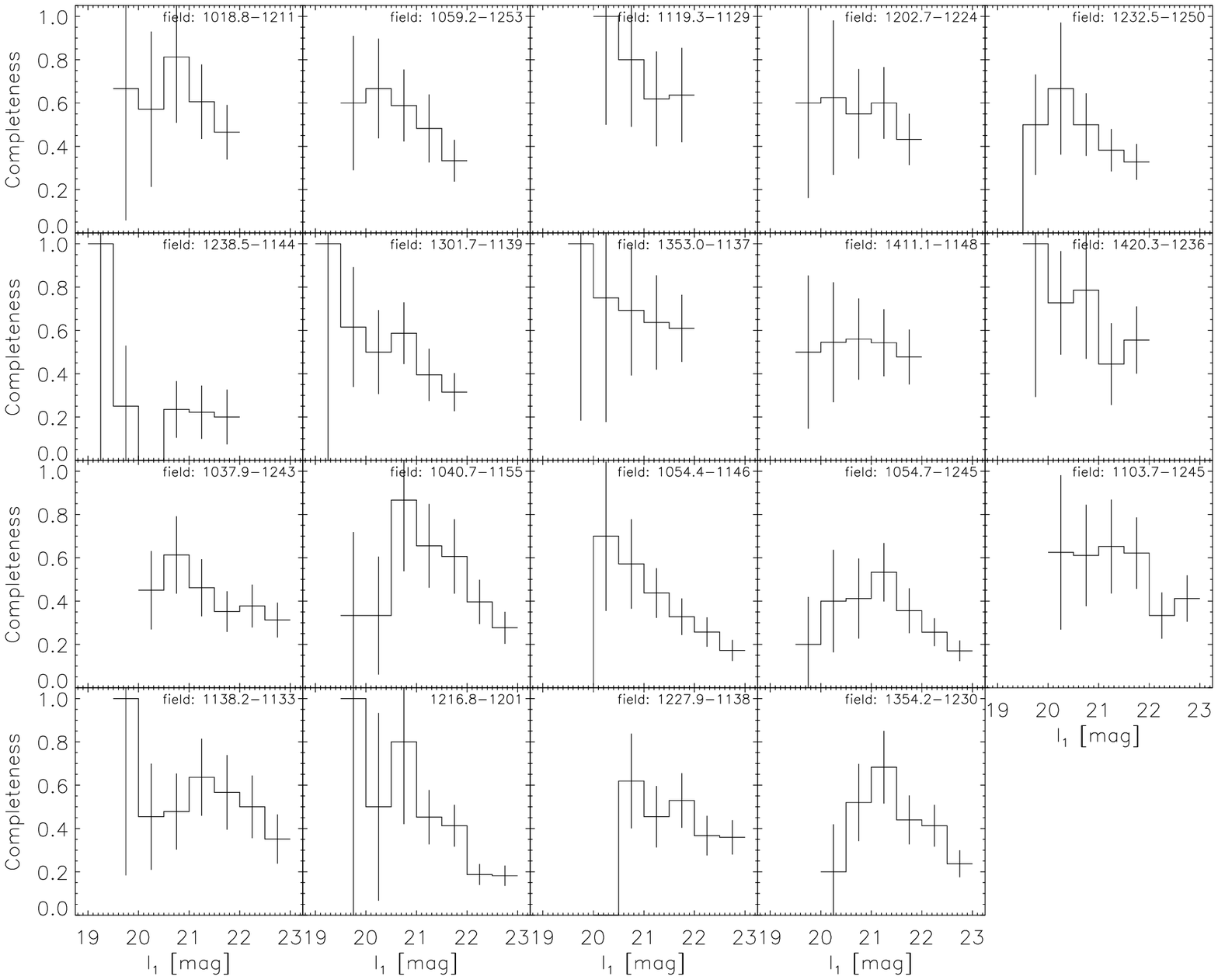}
\caption{%
The completeness as function of magnitude
for the EDisCS fields with long spectroscopic exposures plus the 1238 field.
The completeness is defined as $N_{\mathrm{secure}\,z}/N_\mathrm{targets}$,
with $N_{\mathrm{secure}\,z}$ being the number of targets
for which a secure redshift was obtained
(long dashed line in Fig.~\ref{fig:mag_hist}),
and $N_\mathrm{targets}$ being the number of targets
(dashed line in Fig.~\ref{fig:mag_hist}).
The shown error bars are too large, since they were calculated assuming
that the errors on $N_{\mathrm{secure}\,z}$ and $N_\mathrm{targets}$
are uncorrelated.
}
\label{fig:completeness}
\end{figure*}

The completeness, here defined as the fraction of the targets
for which a secure redshift was obtained,
is shown in Fig.~\ref{fig:completeness}.
The completeness typically decreases as function of magnitude.
This happens because the mask-design procedure (Sect.~\ref{sec:mask_creation})
gives priority to the brighter objects.
When using the spectroscopic sample to study properties that depend
on magnitude, such as the incidence of emission lines in the spectra,
a correction for the completeness as a function of magnitude can be made using
such
histograms, cf.\ \citet{Poggianti_etal:2006}.
It should be noted that Fig.~\ref{fig:mag_hist} and
\ref{fig:completeness} are based on the published redshift tables
(\citealt{Halliday_etal:2004} and this paper).
This means that a few secure redshifts for field galaxies and stars 
from the initial short masks are not included
(cf.\ Sect.~\ref{sec:obs} and \ref{sec:catalogues}), so the actual
completeness is slightly higher than shown in Fig.~\ref{fig:completeness},
particularly at bright magnitudes.

The histograms in Fig.~\ref{fig:mag_hist} also indicate the success rate
of the target selection,
i.e.\ the ratio between the number of galaxies that are members of 
the cluster(s) at the redshift(s) targeted
by the $\zphot$--based target selection (dark red filled histogram)
and the number of observed targets (dotted histogram).
In terms of the overall success rate (i.e.\ not as a function of magnitude),
for the 21 targeted clusters in all 19 fields with long masks,
the success rate is 37\% on average, ranging from
12\% for the 1103.7$-$1245 field ($\zcl = 0.70$),
to
63\% for the 1301.7$-$1139 field ($\zcl = 0.40$, 0.48).

Had we also considered the 5 clusters which happen to be located
in the fields but which were not targeted by our $\zphot$--based selection
(i.e.\ the 5 clusters with redshifts given in parentheses
on Fig.~\ref{fig:mag_hist}),
the success rate would have been 41\% on average for the 19 fields,
ranging from
21\% for the 1238.5$-$1144 field ($\zcl = 0.46$),
to
64\% for the 1138.2--1133 field ($\zcl = 0.48$, 0.45).
A note should be made about these 5 non-targeted clusters.
Three of them
(cl1103.7$-$1245a at $\zcl=0.63$,
 cl1138.2$-$1133a at $\zcl=0.45$, and
 cl1227.9$-$1138a at $\zcl=0.58$)
are at redshifts that are less than 0.1 from the redshifts targeted by
the $\zphot$--based selection (cf.\ Table~\ref{tab:target_selection}),
and any bias in the obtained spectroscopic samples of these clusters
is probably small.
The remaining two clusters are further than 0.1 from the targeted redshifts:
cl1037.9$-$1243a at $\zcl=0.43$, where the selection was
$\zphot \in [0.38,0.78]$ (i.e.\ centered on 0.58), and
cl1103.7$-$1245  at $\zcl=0.96$, where the selection was
$\zphot \in [0.50,0.90]$ (i.e.\ centered on 0.70).
For the latter cluster in particular, it is possible that the obtained
spectroscopic sample is biased with respect to the cluster members
as a whole, e.g.\ in terms of their spectral energy distributions,
because the observed sample is one of galaxies at $\zspec=0.96$
for which the photometry gives $\zphot$ in the range 0.50--0.90.
For this reason, the spectroscopic samples for those two clusters
have not been used in studies of
the [OII] emitting galaxies \citep{Poggianti_etal:2006}
and
the HST--based visual galaxy morphologies \citep{Desai_etal:2007},
even though both clusters have interesting properties
(a large number of confirmed members and high redshift, respectively).

\subsection{Failure rate}
\label{sec:failure_rate}

As described in Sect.~\ref{sec:target_selection}
the 4 target selection rules were:
1: the magnitude $\Ione$ had to be in a certain range;
2: the photometric redshift $\zphot$ had to be in a certain range or
   the probability % $P$
   of being at the cluster redshift had to be greater than 50\%;
3: the \procedurename{hyperz} galaxy and star flags had to have certain values
   designed to remove objects that were not galaxies and possibly stars;
and
4: the FWHM or the ellipticity had to be greater than certain values
   designed to remove small and round objects (likely stars; only in runs 3--4).
We want to know whether rules 2--4, on top of the simple
magnitude selection (rule~1), misses cluster members at the targeted redshifts.
We are able to test this because a smaller fraction of the observed objects
do not meet the target selection rules --- these are objects that were used to
fill the masks or which were serendipitously observed.
We proceed as follows. In the 66 long masks
(cf.\ Table~\ref{tab:target_selection}),
we select objects that meet rule~1 but not all of rules 2--4.
These are 154 (unique) objects.
We then remove the 6 objects that we know are blended, i.e.\ objects 
with colon-IDs, indicating that a single
photometric object turned out to be two spectroscopic objects,
typically two galaxies seen in projection (cf.\ Sect.~\ref{sec:catalogues}).
In this situation, the photometric redshift is not meaningful.
We also remove the 10 targeting flag 4 objects,
i.e.\ objects that were hand-picked to be stars, and placed into the masks
to help field acquisition and to measure seeing.
We are left with 138 objects, of which 122 objects
(comprised of 102 galaxies and 20 stars) have a measured redshift.
Of these 122 objects, 4 are
members of the cluster(s) at the targeted redshift(s) for the given field.
Therefore, a reasonable estimate of the failure rate of the target selection
is 4/122 $\approx$ 3\%.
Of the 4 failures, 3 failed rule~2.
The spectral types of these 3 galaxies differ:
one has an absorption-line spectrum,
one has an absorption-line spectrum possibly with some emission-filling
in H$\beta$, and
one has a spectrum with strong emission lines.
One galaxy failed rule~3; this galaxy has an absorption-line spectrum.
In conclusion, the failure fraction is low (about 3\%),
and the data for the small number of failures do not 
indicate that a bias towards a particular spectral type exists.
The target selection procedure worked effectively,
and for all intents and purposes we expect that our spectroscopic sample
of galaxies at the targeted redshifts behaves as an $I$--band selected sample.

\section{Cluster redshifts and velocity dispersions}
\label{sec:zcl_sigmacl}

The peculiar velocity of a galaxy with redshift $z$
in the rest-frame of a cluster with redshift $\zcl$ is given by
\begin{equation}
\vpecrest = c (z - \zcl) / (1 + \zcl)
\quad\quad ({\rm for}\,\, \vpecrest \ll c) \enspace  
\end{equation}
(e.g.\ \citealt{Carlberg_etal:1996}).
The dispersion of the $\vpecrest$ values for the cluster members
is the cluster rest-frame velocity dispersion $\sigmacl$.
Following standard practice, we will omit \emph{rest-frame}
and simply refer to $\sigmacl$ as the cluster velocity dispersion.

We have tested 2 methods for the determination of $\zcl$ and $\sigmacl$.
The data used in both cases are the set of 0-colon galaxy redshifts 
available for the given field.
Both methods employ an iterative $\pm3$sigma clipping scheme to determine
which galaxies are cluster members. This works as follows.
First, initial guesses of $\zcl$ and $\sigmacl$ are obtained.
Then, the following two steps are iterated until convergence on
$\zcl$ and $\sigmacl$ is reached:
(1)~Calculate $\vpecrest$ (which depends on $\zcl$) for all the galaxies.
(2)~For galaxies with $\vpecrest$ in the interval
$[-3\sigmacl,+3\sigmacl]$, calculate a new estimate of $\zcl$ and $\sigmacl$.
The details of the 2 methods are described below.

\begin{figure*} % Two column figure
\centering
\includegraphics[width=1.00\textwidth,bb = 0 549 564 745]
  {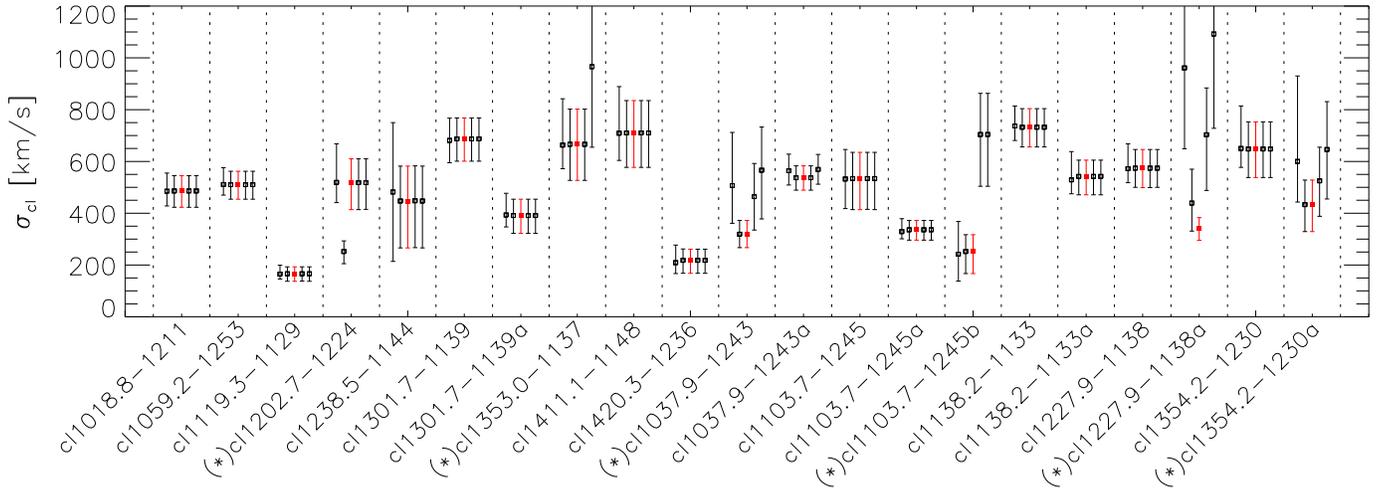}
\caption{%
Comparison of the cluster velocity dispersions determined using the 5 methods
that we have tested.
 From left to right for each cluster, the methods are:
method~1,
method~$2_{200}$,
method~$2_{300}$,
method~$2_{500}$ and
method~$2_{1000}$
(see Sect.~\ref{sec:zcl_sigmacl}).
For most clusters, the results from all methods agree, but for 6 clusters
marked with `(*)' this is not the case.
For these clusters, the 5 estimates of 
$ \sigmacl$
are illustrated in Fig.~\ref{fig:vpecrest_6clusters_5methods}.
}
\label{fig:comparison_5methods}
\end{figure*}

Method 1 is the method used in our previous paper \citep{Halliday_etal:2004}.
A first estimate of $\zcl$ is obtained from a visual inspection of the
redshift histogram. Galaxies with redshifts outside the region
$\zcl \pm 0.015$ are removed and cannot enter the analysis at a later stage.
The median redshift of the remaining galaxies is used as a new estimate
of $\zcl$, and this value is used to calculate $\vpecrest$.
The standard deviation of the $\vpecrest$ values is used as the initial
estimate of $\sigmacl$. %\tmp{correct???}
The iteration then starts, using the median to estimate $\zcl$, and
the biweight scale estimator \citep{Beers_etal:1990} to estimate $\sigmacl$.
In the event that the final number of cluster members is below 10, the
process is repeated using the gapper scale estimator \citep{Beers_etal:1990}
instead of the biweight scale estimator.
The 68\% asymmetric error bars on $\sigmacl$ are determined by
generating bootstrap samples from the final set of $\vpecrest$ values
for the cluster members.
For each bootstrap sample, a value of $\sigmacl$ is
measured without any iterative clipping.

\begin{table}
\caption{Cluster redshifts and velocity dispersions}
\label{tab:zcl_sigmacl_Nmem}
\renewcommand{\arraystretch}{1.25} % To make space for the errors; def. is 1.0
\setlength{\tabcolsep}{3pt} % Default is 6pt
\begin{tabular}{lcrrrr}
\hline
\hline
Cluster & $\zcl$ & \multicolumn{1}{c}{$\sigmacl$ [km$\,$s$^{-1}$]} &
$\Nmemzero$ & $\Nmemzeroone$ & $\Nmemzeroonetwo$ \\
\hline
\multicolumn{6}{l}{Mid--$z$ fields:} \\
cl1018.8$-$1211  & 0.4734 & $ 486^{~+59}_{~-63}$ & 32 & 32 & 33  \\
cl1059.2$-$1253  & 0.4564 & $ 510^{~+52}_{~-56}$ & 41 & 41 & 41  \\
cl1119.3$-$1129  & 0.5500 & $ 166^{~+27}_{~-29}$ & 17 & 17 & 17  \\
cl1202.7$-$1224  & 0.4240 & $ 518^{~+92}_{-104}$ & 19 & 19 & 19  \\
cl1238.5$-$1144  & 0.4602 & $ 447^{+135}_{-181}$ &  4 &  4 &  4  \\
cl1301.7$-$1139  & 0.4828 & $ 687^{~+81}_{~-86}$ & 34 & 35 & 35  \\
cl1301.7$-$1139a & 0.3969 & $ 391^{~+63}_{~-69}$ & 17 & 17 & 17  \\
cl1353.0$-$1137  & 0.5882 & $ 666^{+136}_{-139}$ & 18 & 20 & 20  \\
cl1411.1$-$1148  & 0.5195 & $ 710^{+125}_{-133}$ & 21 & 22 & 22  \\
cl1420.3$-$1236  & 0.4962 & $ 218^{~+43}_{~-50}$ & 22 & 24 & 24  \\
\multicolumn{6}{l}{High--$z$ fields:} \\
cl1037.9$-$1243  & 0.5783 & $ 319^{~+53}_{~-52}$ & 16 & 16 & 16  \\
cl1037.9$-$1243a & 0.4252 & $ 537^{~+46}_{~-48}$ & 43 & 44 & 45  \\
cl1103.7$-$1245  & 0.9586 & $ 534^{+101}_{-120}$ &  9 & 10 & 10  \\
cl1103.7$-$1245a & 0.6261 & $ 336^{~+36}_{~-40}$ & 14 & 15 & 15  \\
cl1103.7$-$1245b & 0.7031 & $ 252^{~+65}_{~-85}$ & 11 & 11 & 11  \\
cl1138.2$-$1133  & 0.4796 & $ 732^{~+72}_{~-76}$ & 45 & 48 & 49  \\
cl1138.2$-$1133a & 0.4548 & $ 542^{~+63}_{~-71}$ & 11 & 12 & 14  \\
cl1227.9$-$1138  & 0.6357 & $ 574^{~+72}_{~-75}$ & 22 & 22 & 22  \\
cl1227.9$-$1138a & 0.5826 & $ 341^{~+42}_{~-46}$ & 11 & 11 & 11  \\
cl1354.2$-$1230  & 0.7620 & $ 648^{+105}_{-110}$ & 20 & 21 & 21  \\
cl1354.2$-$1230a & 0.5952 & $ 433^{~+95}_{-104}$ & 14 & 14 & 15  \\
\hline
\end{tabular}

\vspace*{2pt}

Notes --
$\Nmemzero$ is the number of (unique) cluster members having
redshifts without colons (indicating ``secure'' redshifts).
$\Nmemzeroone$ also includes 1-colon redshifts (``secure but with
larger uncertainties'').
$\Nmemzeroonetwo$ also includes 2-colon redshifts (``not secure'').
We note that redshifts and velocity dispersions for 5 additional EDisCS clusters
are available in \citet{Halliday_etal:2004}.
\end{table}

\begin{figure*} % Two column figure
\centering
\includegraphics[width=1.00\textwidth,bb = 56 161 720 739]
  {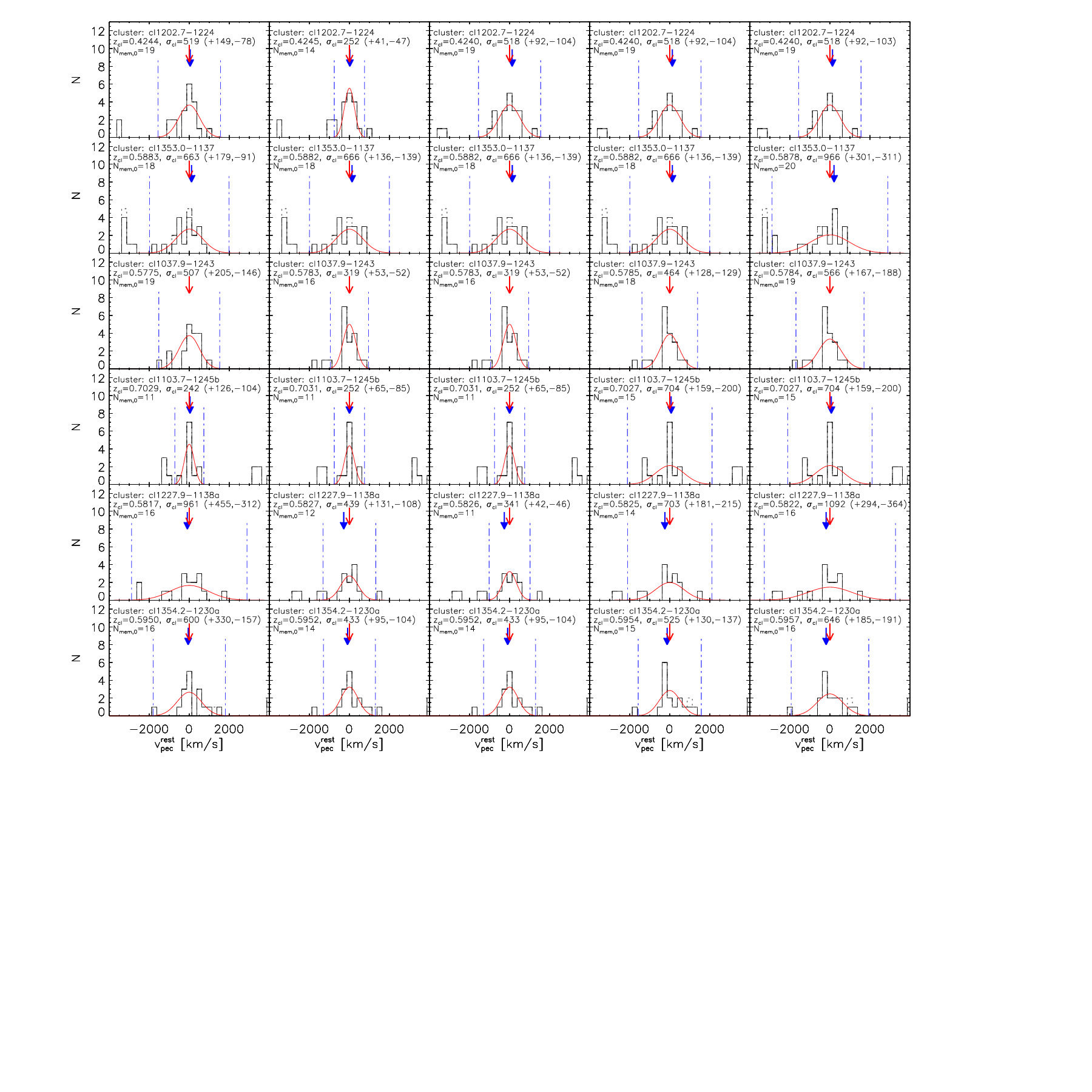}
\caption{%
Histograms of peculiar velocities in the cluster rest-frame,
$\vpecrest = c (z - \zcl) / (1 + \zcl)$,
for the 6 clusters for which the 5 methods of measuring the
cluster redshifts and velocity dispersions do not all agree
(cf.~Fig.~\ref{fig:comparison_5methods}).
Each row shows a given cluster. We note that all panels in a given row are
based on the same redshift data, but since the histograms show $\vpecrest$
which depends on the cluster redshift $\zcl$, the histograms do not
always look completely identical.
Each column shows a given method, from left to right:
method~1,
method~$2_{200}$, 
method~$2_{300}$, 
method~$2_{500}$ and
method~$2_{1000}$. 
For the analysis we adopt method $2_{300}$.
The remaining features of the plots are described in the
caption of Fig.~\ref{fig:vpecrest}.
}
\label{fig:vpecrest_6clusters_5methods}
\end{figure*}

Method 2 uses the final value of $\zcl$ from method~1 as the initial
guess of $\zcl$. It uses an initial guess of $\sigmacl$ of either
200, 300, 500 or 1000$\,$km$\,$s$^{-1}$, which gives rise to
four variants of method~2 referred to as methods
$2_{200}$, $2_{300}$, $2_{500}$ and $2_{1000}$.
In the iteration, the biweight location estimator \citep{Beers_etal:1990}
is used to estimate $\zcl$ and
the biweight scale estimator is used to estimate $\sigmacl$.
The 68\% asymmetric error bars on $\sigmacl$ are determined by
generating bootstrap samples from the final set of redshifts
for the cluster members. Each bootstrap sample is subjected to the
same iterative-clipping procedure as the original dataset itself.

The 5 methods were employed on the 21 clusters in the 14 fields.
The results in terms of the cluster velocity dispersions
are compared in Fig.~\ref{fig:comparison_5methods}.
It is seen that for most clusters the 4 variants of method~2 give
identical results (indicating that the initial guess of $\sigmacl$ has
no effect on the result), and the results from method~2 also agree with that
from method~1 to within a few per cent.
However, for the 6 clusters marked with `(*)' in
Fig.~\ref{fig:comparison_5methods} not all 5 methods agree.
For these clusters the results
(i.e.\ $\zcl$, $\sigmacl$ and number of cluster members $\Nmemzero$)
from the 5 methods are illustrated in
Fig.~\ref{fig:vpecrest_6clusters_5methods}.
This figure shows $\vpecrest$ histograms calculated for the given
value of $\zcl$. The overplotted Gaussians illustrate the given value
of $\sigmacl$, and the vertical dot-dashed lines indicate $\pm3\sigmacl$
and hence which galaxies were used in the measurement of $\sigmacl$
(i.e.\ the cluster members).

We have inspected the $\vpecrest$ histograms
(Fig.~\ref{fig:vpecrest_6clusters_5methods}) to determine which
value of $\sigmacl$ we consider to be the most ``correct'' one.
For the 6 clusters, our comments are as follows:
\newline
cl1202.7$-$1224: Two possible values: $\approx$ 250 and 520$\,$km$\,$s$^{-1}$.
The large value is driven by 4 galaxies on the blue side and 1 galaxy on
the red side. These 5 galaxies have a similar location on
the plane of the sky to the 14 galaxies in the central velocity peak.
Furthermore, the separation in velocity space between the
4 galaxies on the blue side and the 14 galaxies in the central peak is
small.
This makes us favour the larger value.
\newline
cl1353.0$-$1137: Two possible values: $\approx$ 660 and 970$\,$km$\,$s$^{-1}$.
The large value is driven by 2 galaxies which seem to belong to a different
peak, which makes us favour the smaller value.
\newline
cl1037.9$-$1243: Values range from $\approx$ 320 to 570$\,$km$\,$s$^{-1}$.
The 3 galaxies on the blue side which drive the difference
seem somewhat separated in velocity space from the remaining galaxies,
which makes us marginally favour the smaller value.
\newline
cl1103.7$-$1245b: Two possible values: $\approx$ 250 and 700$\,$km$\,$s$^{-1}$.
The 4 galaxies that drive the difference seem to constitute a separate peak,
which makes us favour the smaller value.
\newline
cl1227.9$-$1138a: Values range from $\approx$ 340 to 1090$\,$km$\,$s$^{-1}$.
The galaxies that drive the difference seem to constitute several
separate peaks, which makes us favour the small value.
It was the fact that method~1 gave us the larger value for this particular
cluster that made us test the other methods.
\newline
cl1354.2$-$1230a: Values range from $\approx$ 430 to 650$\,$km$\,$s$^{-1}$.
Some of the difference is driven by a single galaxy on the blue side which
seems quite far from the other galaxies in velocity space.
This makes us favour the smaller value.

The conclusion % from the above
is that method~$2_{300}$ --- as the only of the tested methods ---
in all cases provides the result that we consider to be the most
``correct'' one.
We therefore adopt this method throughout the rest of the paper.
However, while the method~$2_{300}$ results constitute our best guess,
the velocity dispersions for these 6 clusters should still be treated
as being more uncertain than for the rest of the clusters.

The adopted values of the cluster redshifts, velocity dispersions
and number of member galaxies for all 21 clusters in the 14 fields
are listed in Table~\ref{tab:zcl_sigmacl_Nmem}.
The values of the velocity dispersions are discussed in
Sect.~\ref{sec:discussion}.

\section{Cluster substructure}
\label{sec:substructure}

Possible cluster substructure is investigated using
velocity histograms (Sect.~\ref{sec:velhist}),
$xy$ plots (Sect.~\ref{sec:xyplots}) and a
Dressler--Shectman analysis (Sect.~\ref{sec:DStest}).

\begin{figure*} % Two column figure
\centering
\includegraphics[width=1.00\textwidth,bb = 56 -49 720 739]
  {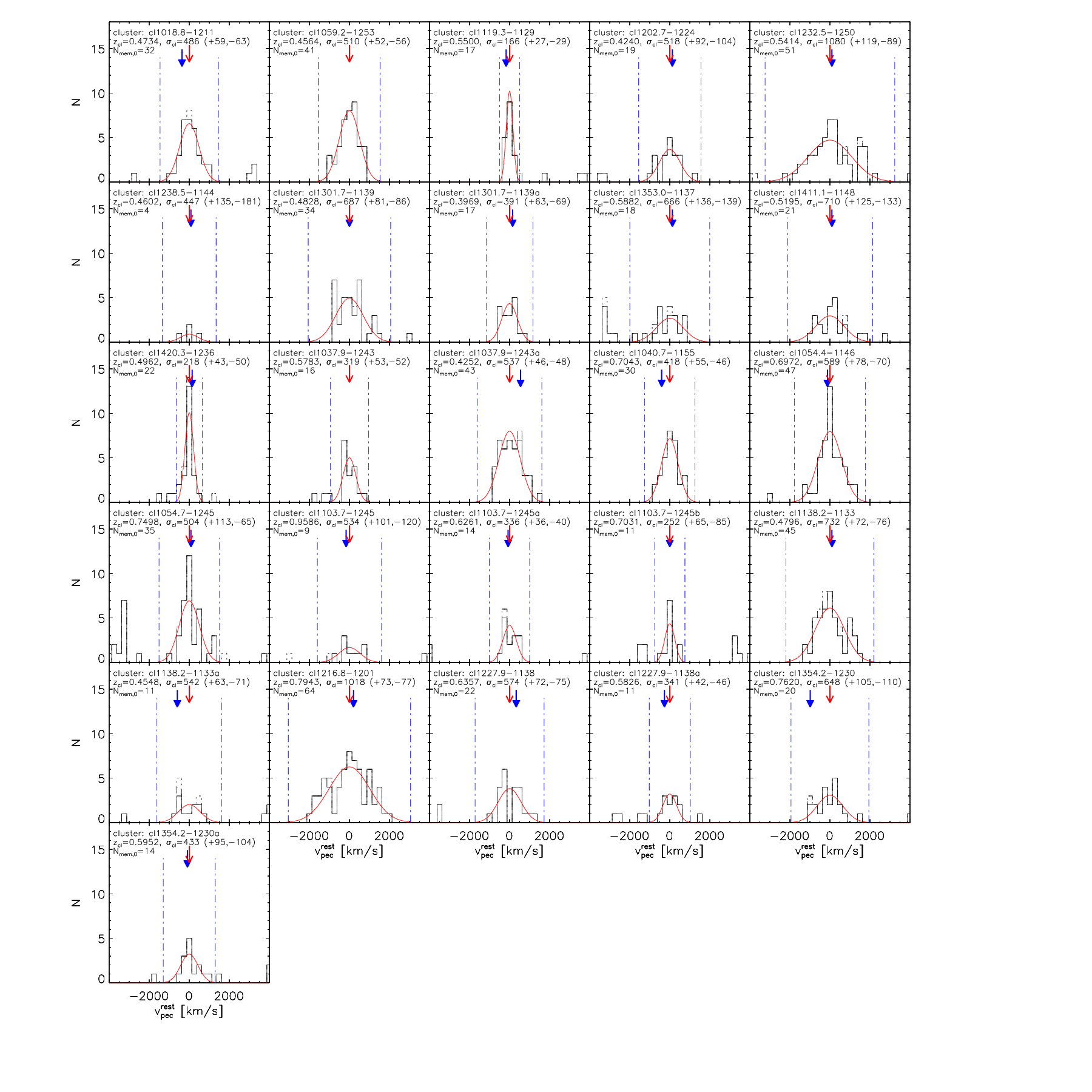}
\caption{%
Histograms of peculiar velocities in the cluster rest-frame,
$\vpecrest = c (z - \zcl) / (1 + \zcl)$,
for the 26 EDisCS clusters.
The solid histograms are for galaxies having redshifts without colons
(indicating ``secure'' redshifts).
The dashed histograms include the 1-colon redshifts (``secure but with
larger uncertainties'').
The dotted histograms include the 2-colon redshifts (``not secure'').
The binsize is 250$\,$km$\,$s$^{-1}$
and the plotted range is $\pm$4000$\,$km$\,$s$^{-1}$.
The overplotted Gaussians illustrate the measured velocity dispersion
$\sigmacl$ of the given cluster.
The vertical dot-dashed lines indicate $\pm3\sigmacl$, the limits
used to define cluster membership.
The number of cluster members having redshifts without colons is given
as $\Nmemzero$, and the area underneath the Gaussian
corresponds to this number.
The red skeletal arrows are located at $\vpecrest = 0$$\,$km$\,$s$^{-1}$ and
thus indicate the adopted cluster redshifts.
The blue filled arrows indicate % $\vpecrest$ for 
the adopted BCGs (except where no redshift is available).
}
\label{fig:vpecrest}
\end{figure*}

\subsection{Velocity histograms}
\label{sec:velhist}

Histograms of peculiar velocities in the cluster rest-frame, $\vpecrest$,
are shown for the 26 EDisCS clusters in Fig.~\ref{fig:vpecrest}.
These 26 clusters are the ones with a measured velocity dispersion
from \citet{Halliday_etal:2004} or this paper.
The velocity of the BCG is indicated with a blue/filled arrow,
except for the 2 BGCs without a spectroscopic redshift
(clusters cl1059.2$-$1253 and cl1037.9$-$1243, cf.\ \citealt{White_etal:2005}).
In a few cases, the adopted BCG has a substantial peculiar velocity,
e.g.\ in cl1354.2$-$1230, cf.\ e.g.\ \citet{Pimbblet_etal:2006}.
The overplotted Gaussians illustrate the measured
velocity dispersions. % \footnote{%
We note that velocity histograms for 5 of the clusters were already shown in
\citet{Halliday_etal:2004}, but those plots showed
observed-frame rather than rest-frame peculiar velocities,
whereas the overplotted Gaussians corresponded to the
rest-frame velocity dispersions.

 From these velocity histograms,
most
of our clusters appear to be
fairly well-described by Gaussian dispersions, particularly those with
many members or high velocity dispersions.  It is generally unclear
whether departures from Gaussianity are real or an effect of limited
statistics.  One feature that does stand out, however, is the
incidence of smaller galaxy associations close to our clusters, which
may be due to the tails of the true velocity distribution being longer
than Gaussian, but in many cases appear to be separate from the
cluster itself.  These may be interpreted as groups which surround,
and will eventually fall into, the main clusters.

\subsection{XY position diagrams}
\label{sec:xyplots}

\begin{figure*} % Two column figure

\vspace*{1.0em}

\centering
\includegraphics[width=1.00\textwidth,bb = 43 -48 720 739]
{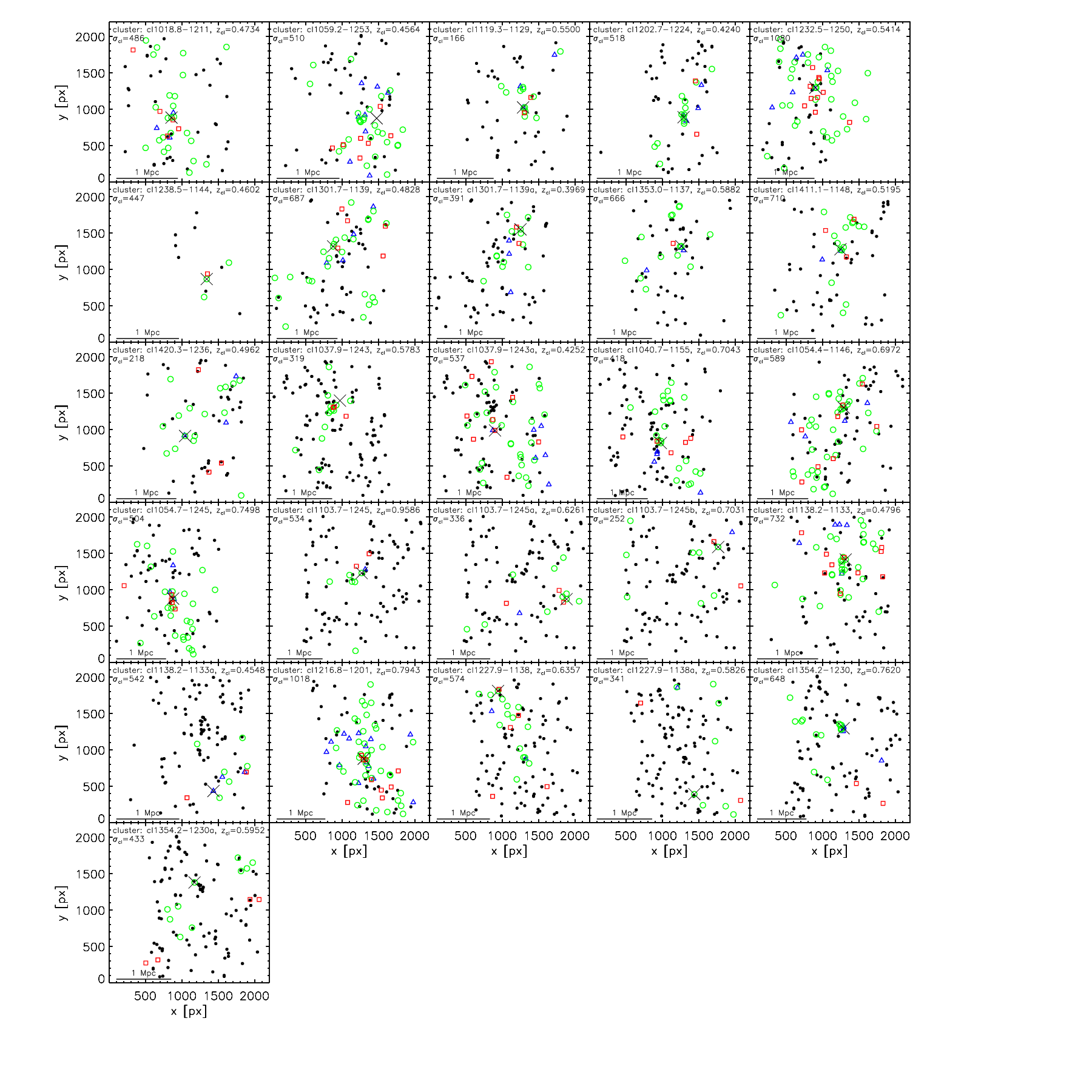}
\caption{%
$xy$ plots for the 26 EDisCS clusters.
North is up and east is to the left.
The units on the axes are pixels = 0.2$''$.
Only galaxies with no colons on their redshifts are shown.
The small dots are the non-members.
The large symbols are the cluster members. Depending on in which bin
$\vpecrest$ falls, the symbols are:
blue triangles: $[-3\sigmacl,-1\sigmacl[$,
green circles:  $[-1\sigmacl,+1\sigmacl]$,
red squares:  $]$$+1\sigmacl,+3\sigmacl]$.
The cross indicates the adopted BCG, which in the case of
cl1059.2$-$1253 and cl1037.9$-$1243
does not have any spectroscopy, cf.\ \citet{White_etal:2005}.
1~Mpc bars are shown for the assumed cosmology
($\Omega_\mathrm{m} = 0.3$, $\Omega_\Lambda = 0.7$
and $H_0 = 70\,\mathrm{km}\,\mathrm{s}^{-1}\,\mathrm{Mpc}^{-1}$).
}
\label{fig:xy}
\end{figure*}

Plots of the locations of the galaxies on the sky
($xy$ plots) for the 26 EDisCS clusters are shown in
Fig.~\ref{fig:xy}.
The cluster members are shown with large symbols.
The symbol type and colour indicate what bin the peculiar velocity
in the rest-frame of the cluster, $\vpecrest$, falls into.
Non-cluster members are shown with small dots.
The cross indicates the adopted BCG\@.
One of the 19 main clusters, namely cl1227.9$-$1138, has a BCG that
is close to the edge of the field, cf.\ \citet{White_etal:2005}.
We note that $xy$ plots for 5 of the clusters were already shown in
\citet{Halliday_etal:2004}, but they have been repeated here to provide an
overview of the full sample and because the plots here also show the
non-cluster members, thus illustrating over which region spectroscopy has
been obtained.

For the blended objects where one object in the photometric catalogue
turned out to be two physical objects in the spectrum
(cf.\ Sect.~\ref{sec:catalogues}),
the two objects (:A and :B) have identical $(x,y)$ coordinates,
namely the $(x,y)$ coordinates from the photometric catalogue.
To make both objects visible in the $xy$ plot, we have offset the
:A object by 1$''$ south and the :B object by 1$''$ north.

The clusters with velocity dispersions 
$\ga 400\,\mathrm{km}\,\mathrm{s}^{-1}$
generally display
a well-defined centre, usually coincident with the BCG\@.  However,
several of these clusters show signs of sub-clumps with coherent
motion, or possibly even, for cl1216.8$-$1201 and cl1037.9$-$1243a, an
overall rotation of the cluster.

\subsection{The Dressler--Shectman test}
\label{sec:DStest}

\begin{table}
\caption{Results from the Dressler--Shectman test}
\label{tab:DS}
\begin{center}
\renewcommand{\arraystretch}{1.05} % To make space for the errors; def. is 1.0
\setlength{\tabcolsep}{6pt} % Default is 6pt
\begin{tabular}{lcccc}
\hline
\hline
Cluster & $\zcl$ & $N_{\rm g}$ & $\Delta$ & $P$ \\
\hline
\multicolumn{5}{l}{Mid--$z$ fields:} \\
cl1018.8$-$1211  & 0.4734 & $32$ & $32.559$  & $0.264$ \\
cl1059.2$-$1253  & 0.4564 & $41$ & $50.193$  & $0.077$ \\
cl1301.7$-$1139  & 0.4828 & $34$ & $31.261$  & $0.586$ \\
cl1411.1$-$1148  & 0.5195 & $21$ & $13.914$  & $0.841$ \\
cl1420.3$-$1236  & 0.4962 & $22$ & $21.594$  & $0.482$ \\
\multicolumn{5}{l}{High--$z$ fields:} \\
cl1037.9$-$1243a & 0.4252 & $43$ & $59.027$  & $0.010$ \\
cl1138.2$-$1133  & 0.4796 & $45$ & $38.456$  & $0.631$ \\
cl1227.9$-$1138  & 0.6357 & $22$ & $14.428$  & $0.782$ \\
cl1354.2$-$1230  & 0.7620 & $20$ & $31.260$  & $0.001$ \\
\hline
\end{tabular}
\end{center}

\vspace*{2pt}

Notes --
$N_{\rm g}$ is the number of cluster members used in the test
($N_{\rm g}$ is identical to $\Nmemzero$ in Table~\ref{tab:zcl_sigmacl_Nmem}),
$\Delta$ is the Dressler--Shectman statistic, and
$P$ is the probability of there being
no substructure in the dataset; thus, a small value (e.g.\ less than 0.05)
indicates that substructure has been detected.
We note that Dressler--Shectman results for 5 additional EDisCS clusters
are available in \citet{Halliday_etal:2004}.
\end{table}

\begin{figure*} % Two column figure
\makebox[\textwidth]{
  \makebox[\fourthwidth]{
  \includegraphics[width=1.00\fourthwidth,bb = 13 103 471 587]{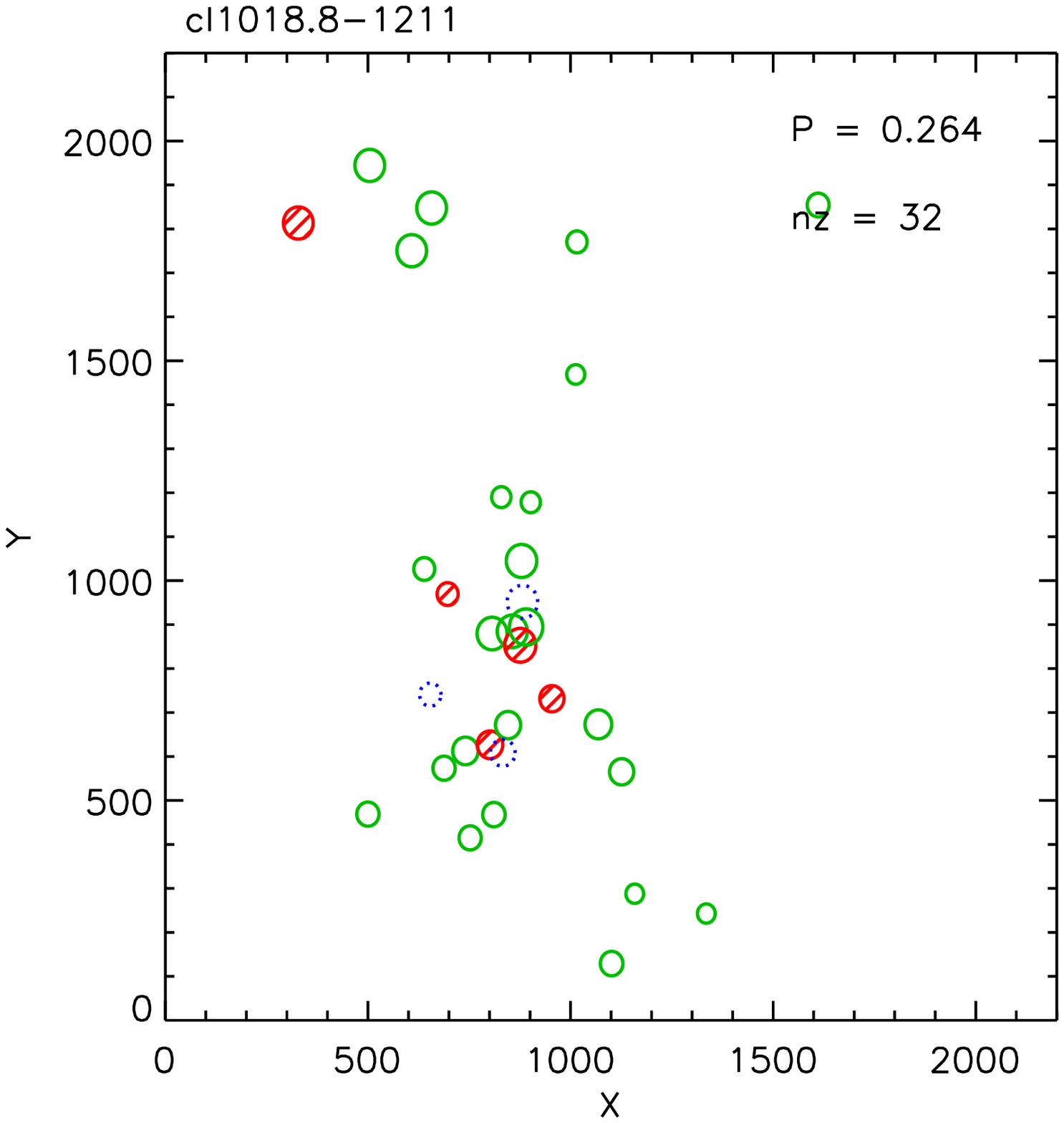}
  }
  \makebox[\fourthwidth]{
  \includegraphics[width=1.00\fourthwidth,bb = 13 103 471 587]{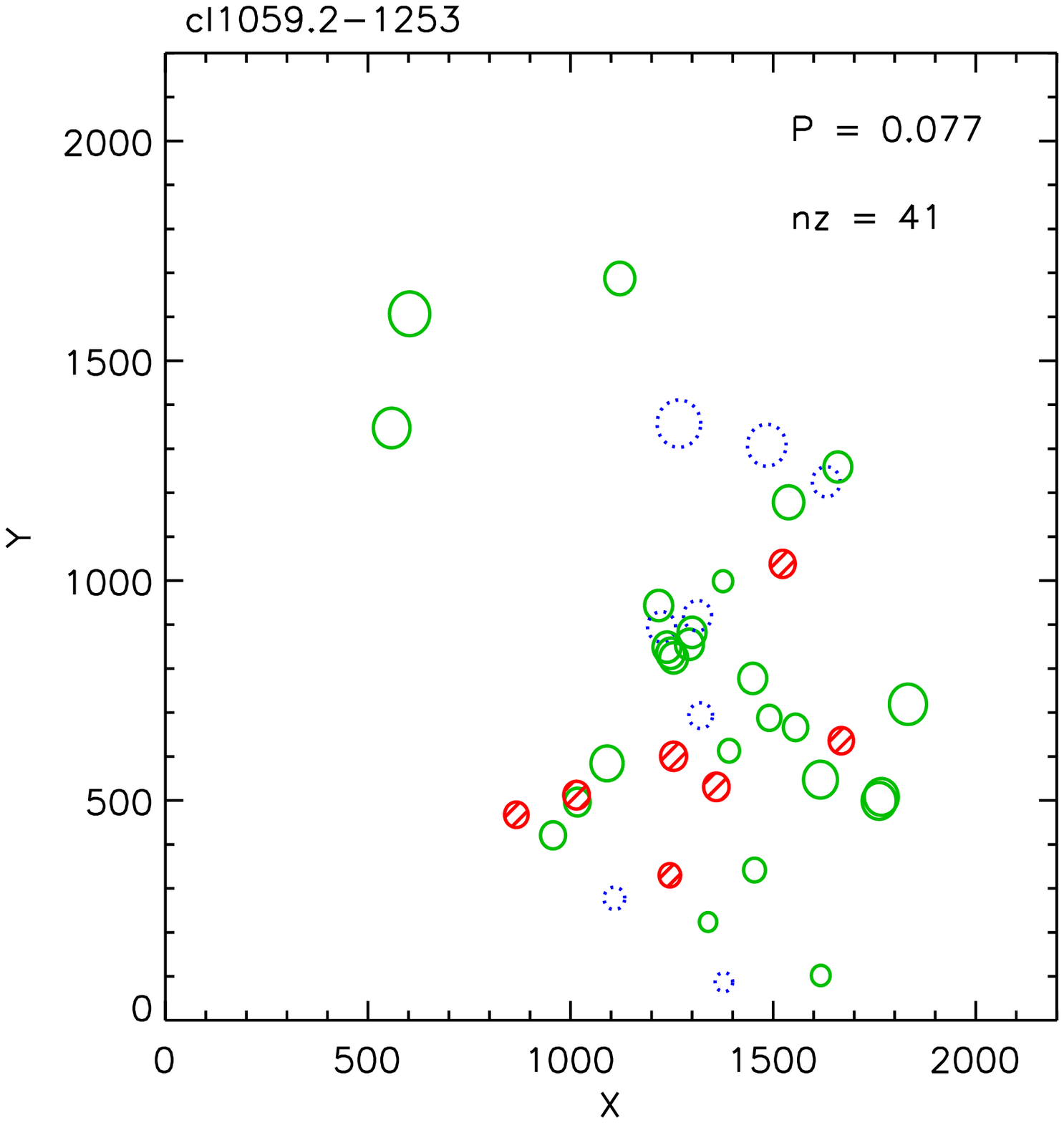}
  }
  \makebox[\fourthwidth]{
  \includegraphics[width=1.00\fourthwidth,bb = 13 103 471 587]{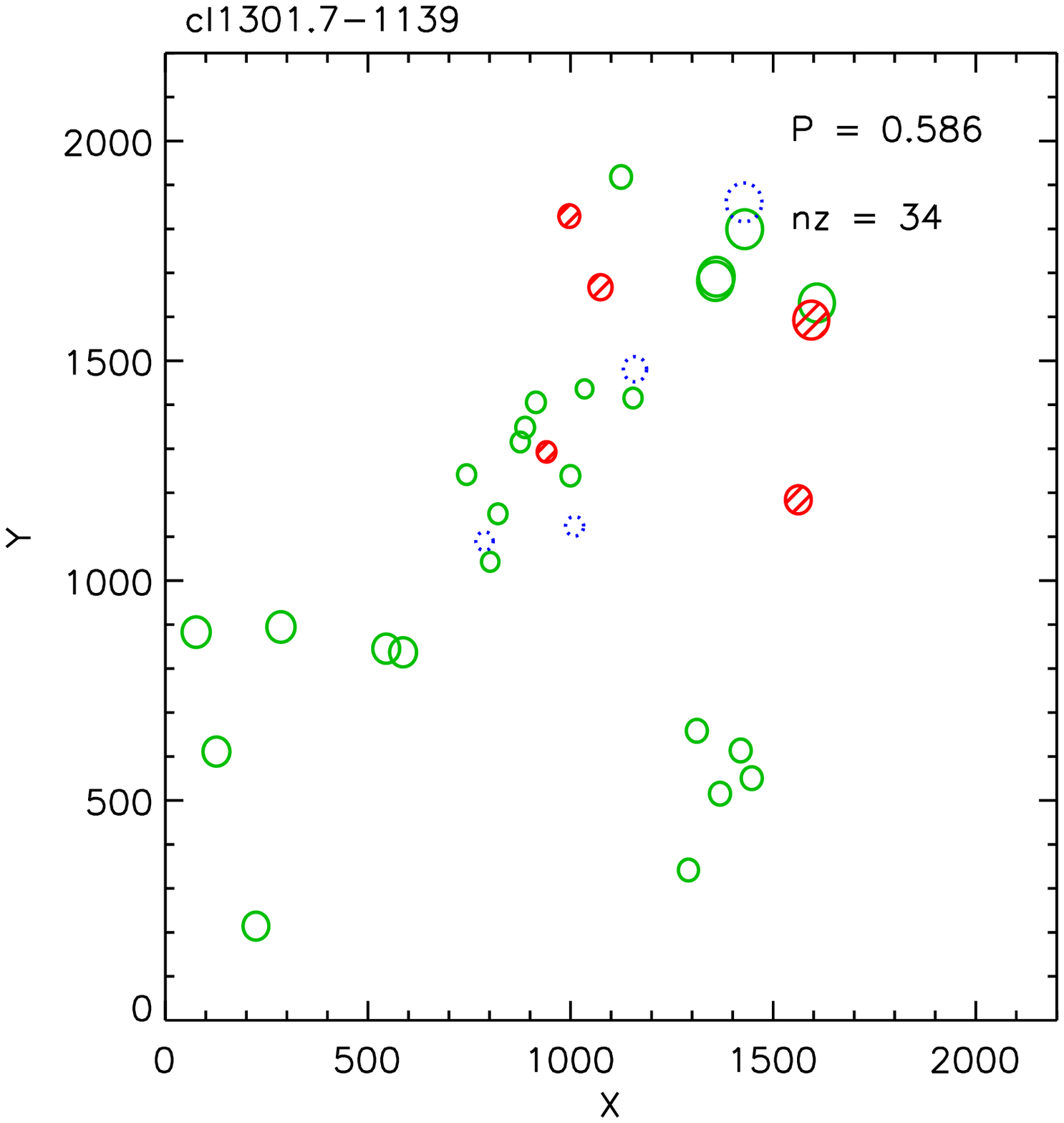}
  }
  \makebox[\fourthwidth]{
  \includegraphics[width=1.00\fourthwidth,bb = 13 103 471 587]{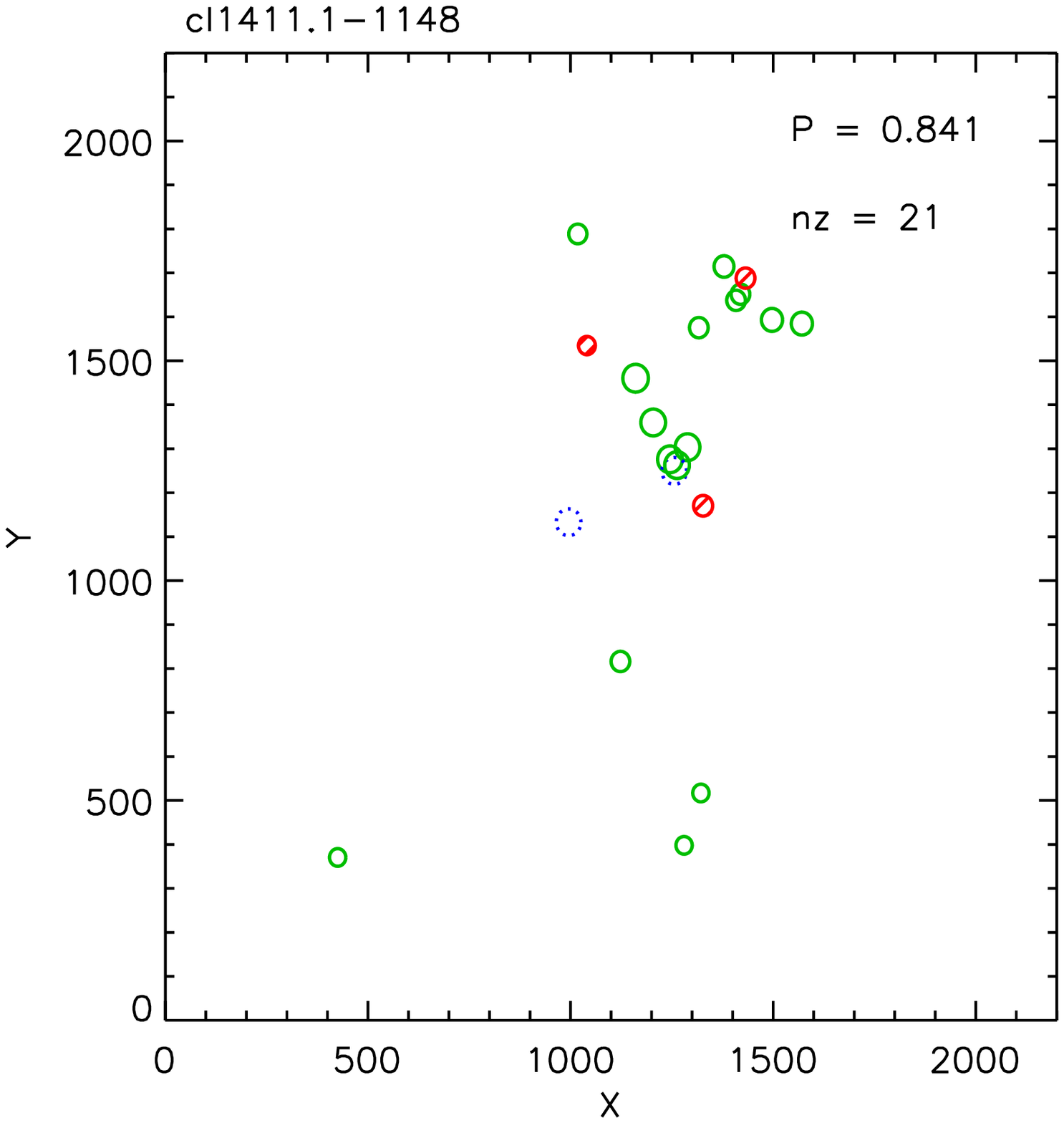}
  }
}
\makebox[\textwidth]{
  \makebox[\fourthwidth]{
  \includegraphics[width=1.00\fourthwidth,bb = 13 103 471 587]{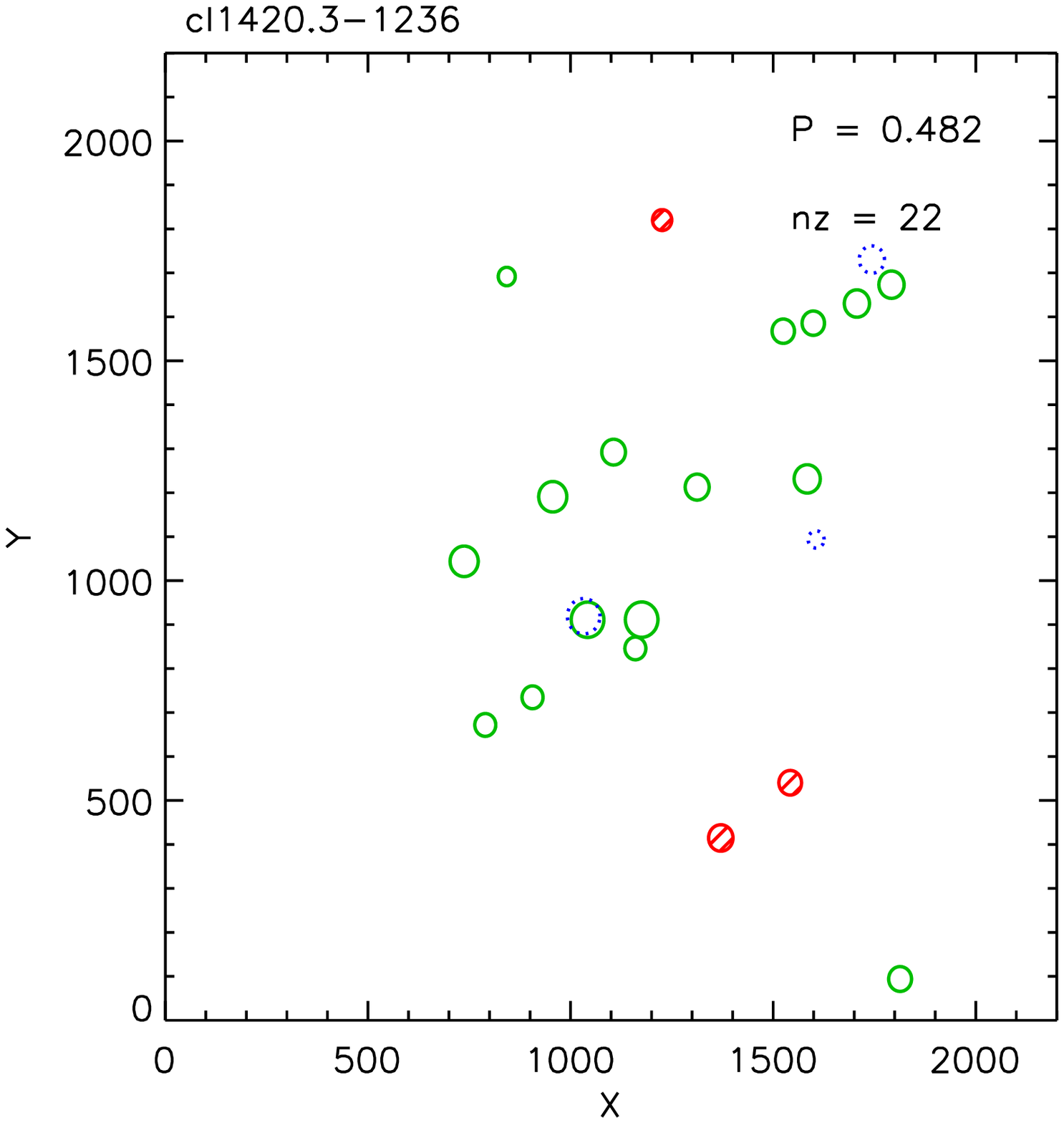}
  }
  \makebox[\fourthwidth]{
  \includegraphics[width=1.00\fourthwidth,bb = 13 103 471 587]{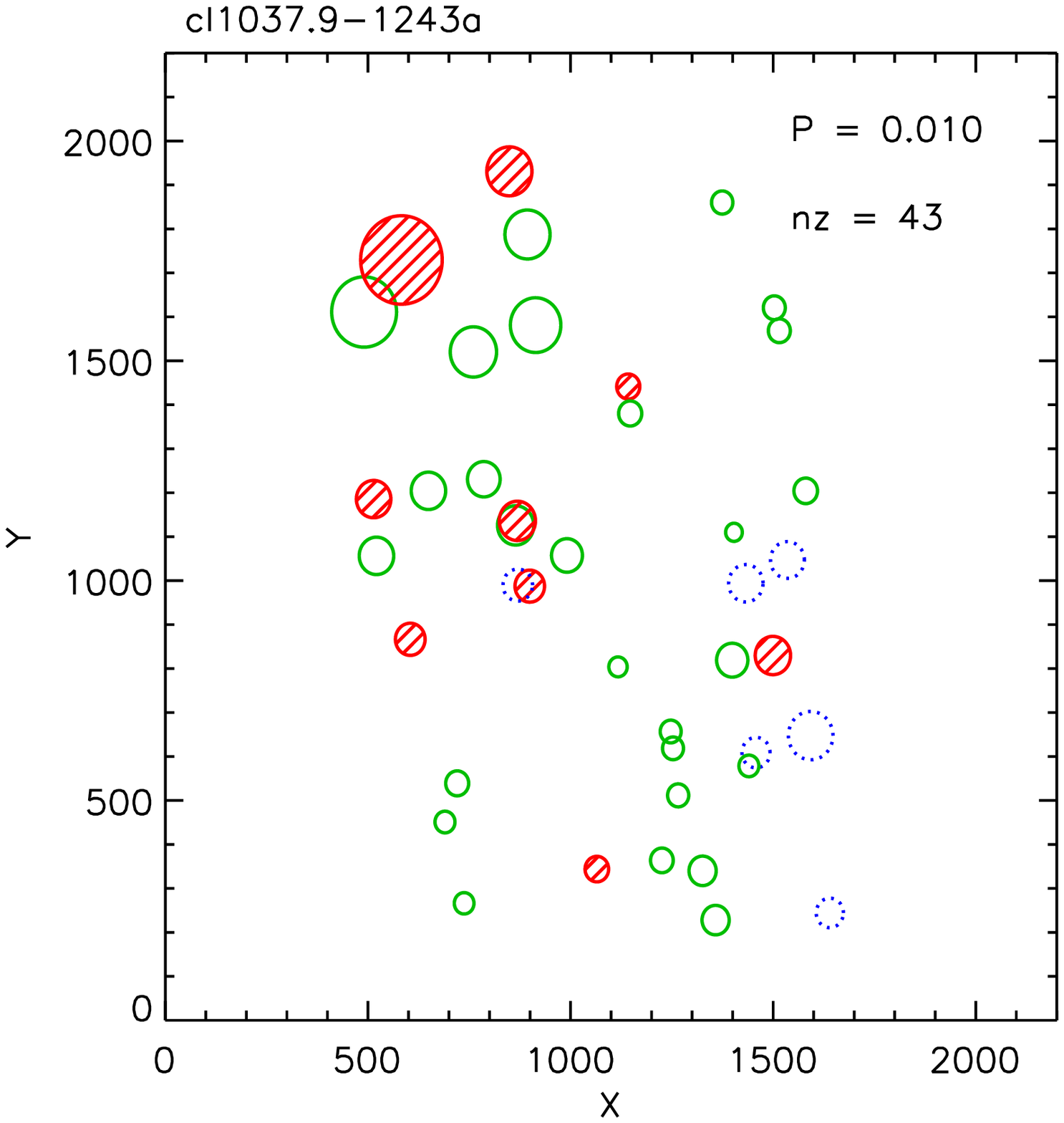}
  }
  \makebox[\fourthwidth]{
  \includegraphics[width=1.00\fourthwidth,bb = 13 103 471 587]{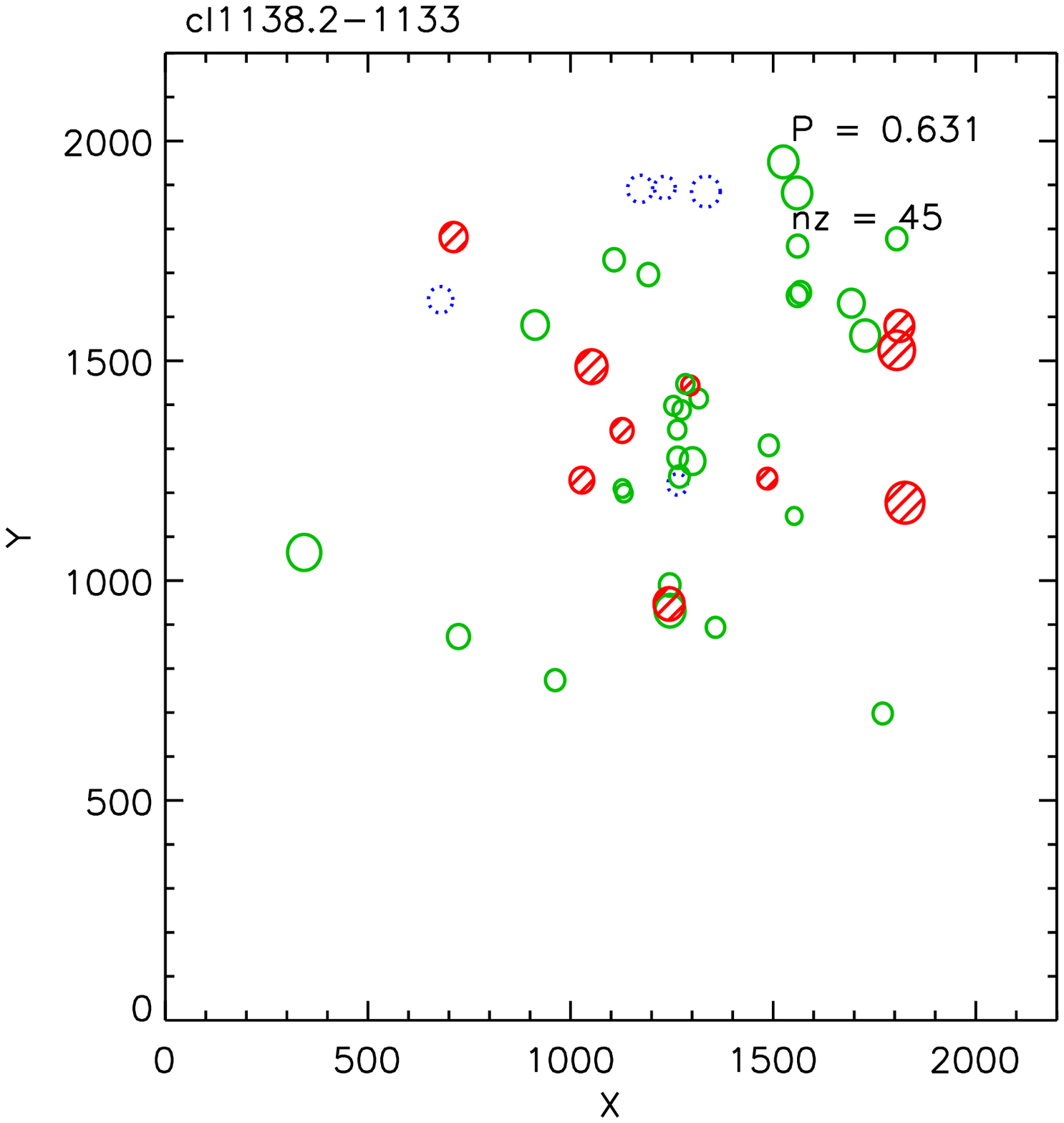}
  }
  \makebox[\fourthwidth]{
  \includegraphics[width=1.00\fourthwidth,bb = 13 103 471 587]{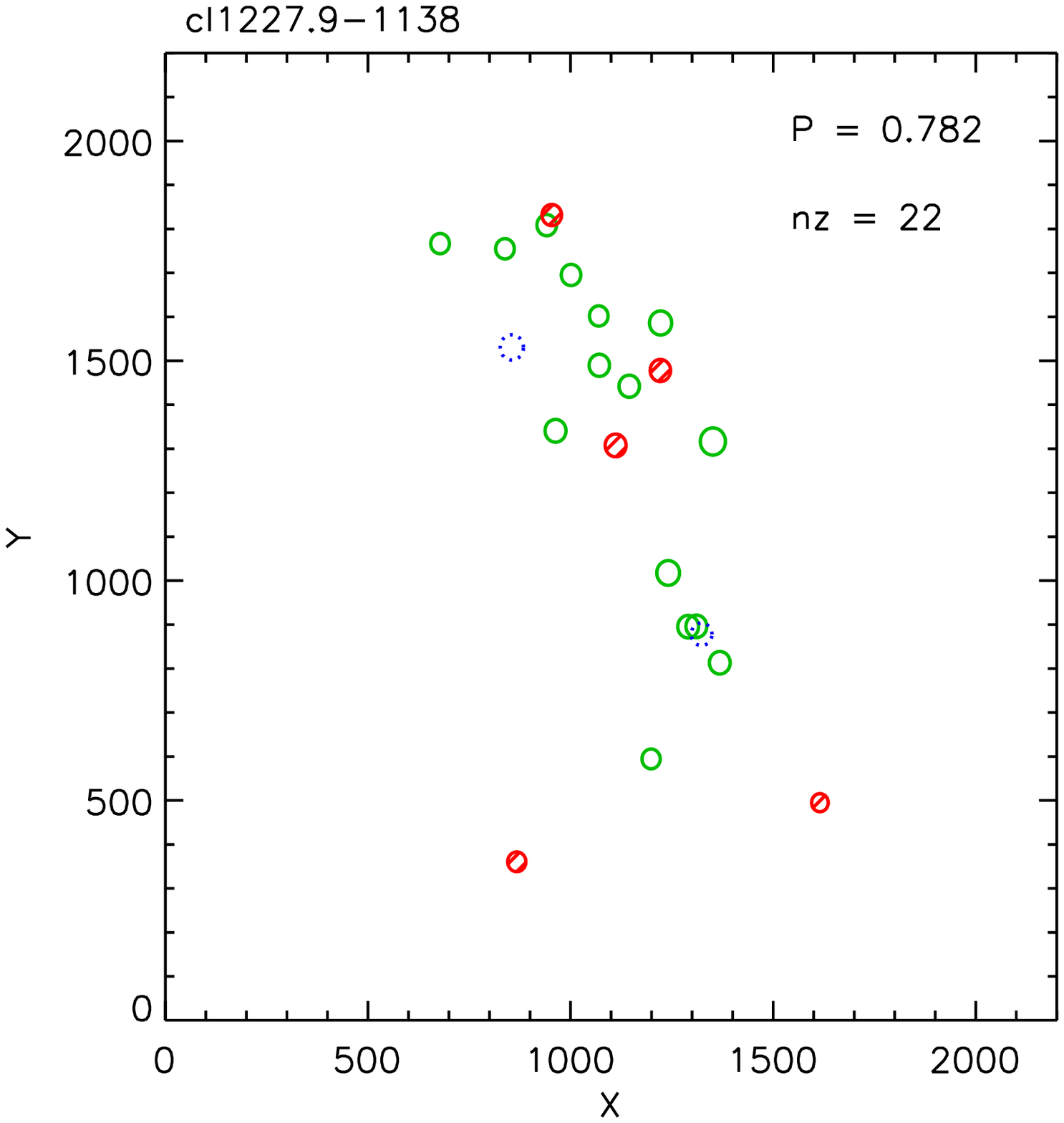}
  }
}
\makebox[\textwidth]{
  \makebox[\fourthwidth]{
  \includegraphics[width=1.00\fourthwidth,bb = 13 103 471 587]{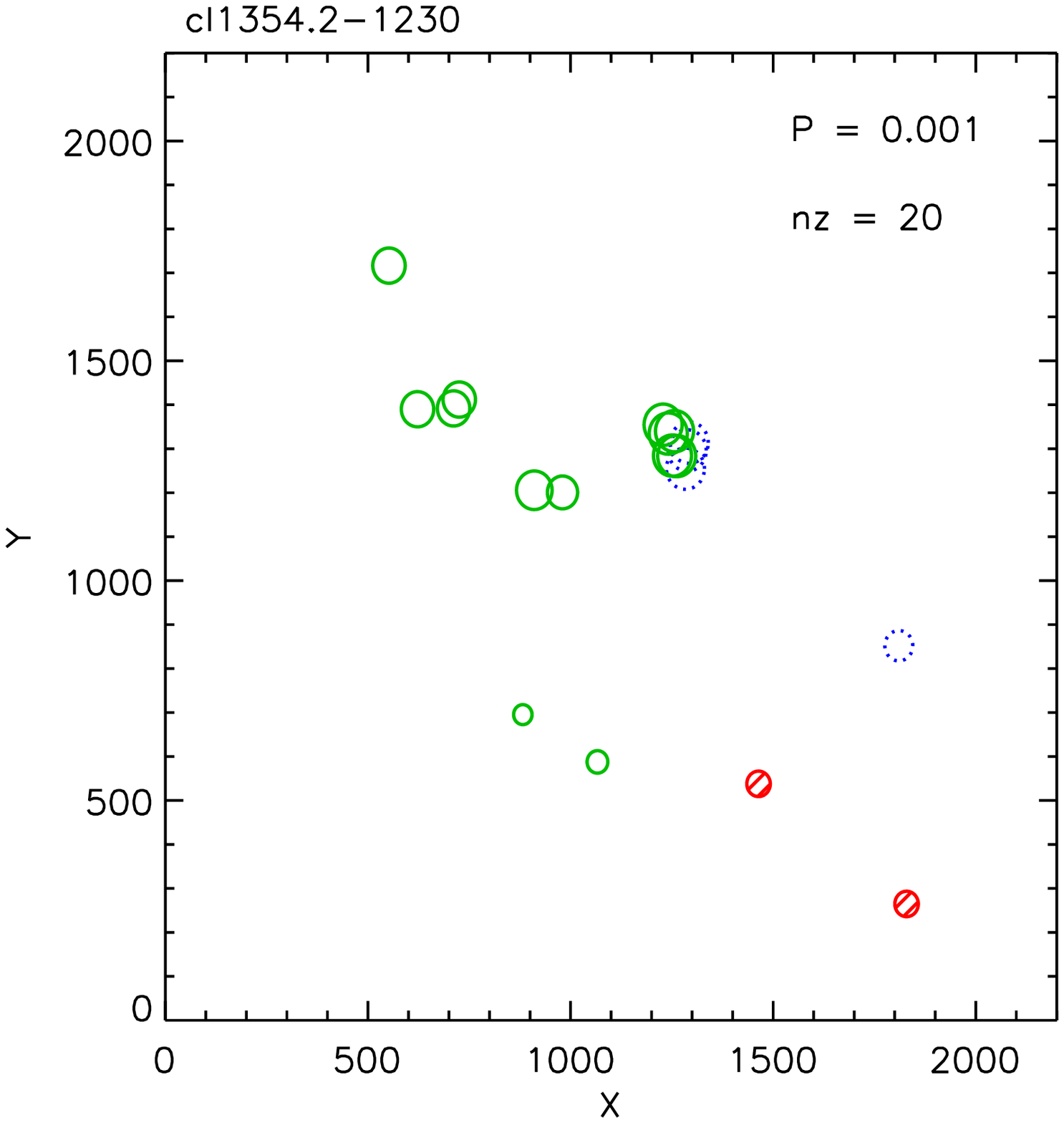}
  }
  \makebox[\fourthwidth]{
  ~
  }
  \makebox[\fourthwidth]{
  ~
  }
  \makebox[\fourthwidth]{
  ~
  }
}
\caption[]{%
Dressler--Shectman (DS) plots.
The DS analysis has only been performed on clusters
with at least 20 members.
The plots show the $x,y$ location of the cluster members.
The radii of the plotted
circles are equal to $e^{\frac{\delta}{2}}$ where $\delta$ is the DS
measurement of local deviation from the global velocity dispersion
and mean recession velocity (cf.\ Eq.~\ref{eq:DS}).
The blue/green/red circles (also shown as dotted/solid/hashed)
indicate velocity in the same way as in the $xy$ plots
(Fig.~\ref{fig:xy}).
The probability $P$ given on the figure is the probability of there being
no substructure in the dataset; thus, a small value (e.g.\ less than 0.05)
indicates that substructure has been detected.
The number of members is also given on the figure; only redshifts without
colons have been used.
The 9 clusters in this figure are shown in the same order as in
Table~\ref{tab:DS}.
We note that DS plots for 5 more clusters are found in
\citet{Halliday_etal:2004}.
}
\label{fig:DStest}
\end{figure*}

In order to check for the presence of substructure in the three-dimensional
space, we combine velocity and positional information by computing the
statistics devised by \citet{Dressler_Shectman:1988}.  The test works in the
following way:
for each galaxy that is a spectroscopic cluster member
(defined throughout this paper as being within $\pm3\sigmacl$ from $\zcl$),
the ten nearest
neighbours are found, and the local velocity mean and velocity dispersion are
computed from this sample of $11$ galaxies.  These quantities are compared to
the global dynamical parameters computed for the clusters by defining the
deviation $\delta$ as:
\begin{equation}
\delta^2 = (11/\sigma^2)\left[\left(\bar{v}_{\rm
      local}-\bar{v}\right)^2+\left(\sigma_{\rm local}-\sigma\right)^2\right]
\label{eq:DS}
\end{equation}
where $\bar{v}$ and $\sigma$ are the global dynamical parameters and
$\bar{v}_{\rm local}$ and $\sigma_{\rm local}$ are the local mean recessional
velocity and velocity dispersion, determined using the $10$ closest galaxies
(with spectroscopy available). % with measured radial velocities.   
Velocities and velocity dispersions were transformed to the rest-frame
of the cluster.

Dressler \& Shectman also define the cumulative deviation $\Delta$ as the
sum of the $\delta$ for all the cluster members $N_{\rm g}$. We note that the
$\Delta$ statistic is similar to a $\chi^2$: if the cluster velocity
distribution is close to Gaussian and the local variations are only random
fluctuations, then $\Delta$ will be of the order of $N_{\rm g}$.  

We have applied the above test to all structures with at least $20$ members,
as a conservative compromise between the formal minimum number
required to perform the test ($>11$), coupled to the desire to analyse
a sample of clusters as large as possible, and the need to find
statistically significant substructures. We note that
\citet{Dressler_Shectman:1988} apply the method to clusters with at
least 26 members.
The results of such an analysis are shown in Fig.~\ref{fig:DStest}.
In each panel, the 
size of the symbols is proportional to $e^{\delta/2}$  and the symbols are
coloured according to rest-frame peculiar velocity
in the same way as in the $xy$ plots (Fig.~\ref{fig:xy}).

In order to give a quantitative estimate of the significance of substructure,
we have performed $1000$ Monte
Carlo realizations for each structure by randomly shuffling the velocities of
the galaxies used for the analysis.  The significance of the occurrence of
dynamical substructure can be quantified using the ratio $P$ between the number
of simulations in which the value of $\Delta$ is larger than the observed
value, and the total number of simulations. 

In Table~\ref{tab:DS} we list, for each of the clusters used in this analysis,
the number of spectroscopic members, the measured $\Delta$ statistic, and the
probability $P$ of there being no substructure. % in the dataset.

Out of the 9 clusters tested in this paper, significant substructure
($P \le 5\%$) is detected in 2 clusters:
cl1037.9$-$1243a at $z = 0.43$ ($P = 1.0\%$) and
cl1354.2$-$1230  at $z = 0.76$ ($P = 0.1\%$).
In \citet{Halliday_etal:2004}, we tested 5 clusters and detected
significant substructure in 2 clusters:
cl1232.5$-$1250  at $z = 0.54$ ($P = 1\%$) and
cl1216.8$-$1201  at $z = 0.79$ ($P = 5\%$).
The fraction of EDisCS clusters with detected substructure is 4/14 = 29\%.
The same level, 21/67 = 31\%, was found by \citet{Solanes_etal:1999}
for a local ($z \la 0.1$) sample of clusters from
the ESO Nearby Abell Cluster Survey (ENACS)\@.
This sample is also optically-selected,
and the same substructure definition was used,
i.e.\ the \citet{Dressler_Shectman:1988} test with a $P = 5\%$ threshold.
More data are required to check this apparent lack of evolution in the
fraction of clusters with substructure from $z \sim 0.6$ to $z = 0.1$.

\section{Discussion}
\label{sec:discussion}

\begin{figure} % One column figure
\includegraphics[width=0.485\textwidth, bb = 10 495 263 728]
  {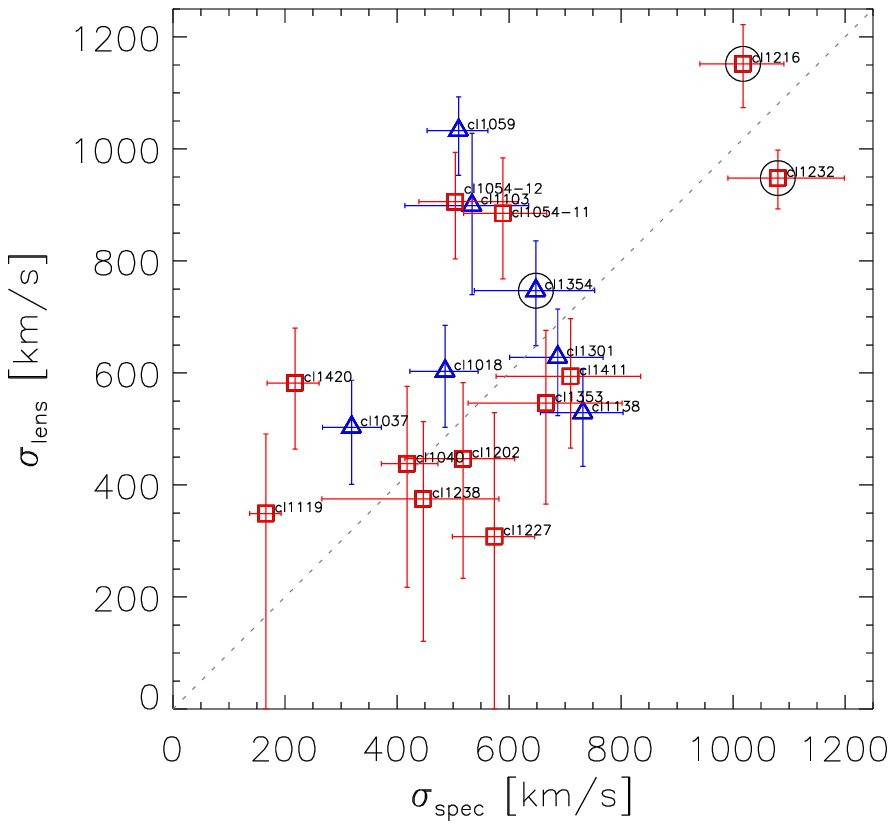}
\caption[]{%
Comparison of the velocity dispersions obtained from the
weak lensing analysis \citep{Clowe_etal:2006}, $\sigmalens$,
with the velocity dispersions obtained from the spectroscopy
(\citealt{Halliday_etal:2004} and this paper), $\sigmaspec$.
The figure shows the 19 main EDisCS clusters ($z$ = 0.42--0.96).
The blue triangles are the clusters with other structures near enough
to possibly affect the lensing measurements \citep{Clowe_etal:2006}, and
the red squares are the rest of the clusters.
The 3 circled clusters are those for which the Dressler--Shectman test
gives a significant detection of substructure
(\citealt{Halliday_etal:2004} and this paper).
The dotted line shows the one to one correspondence.
Abbreviated cluster names are given on the figure.
The 2 major outliers of the blue triangles are
cl1059.2$-$1253 and
cl1103.7$-$1245,
both of which were identified in \citet{Clowe_etal:2006} as having extremely
high mass-to-light ratios.
The 3 major outliers of the red squares are
cl1420.3$-$1236,
cl1054.7$-$1245 and
cl1054.4$-$1146.
}
\label{fig:lensing}
\end{figure}

\begin{figure} % One column figure
\includegraphics[width=0.48\textwidth]
  {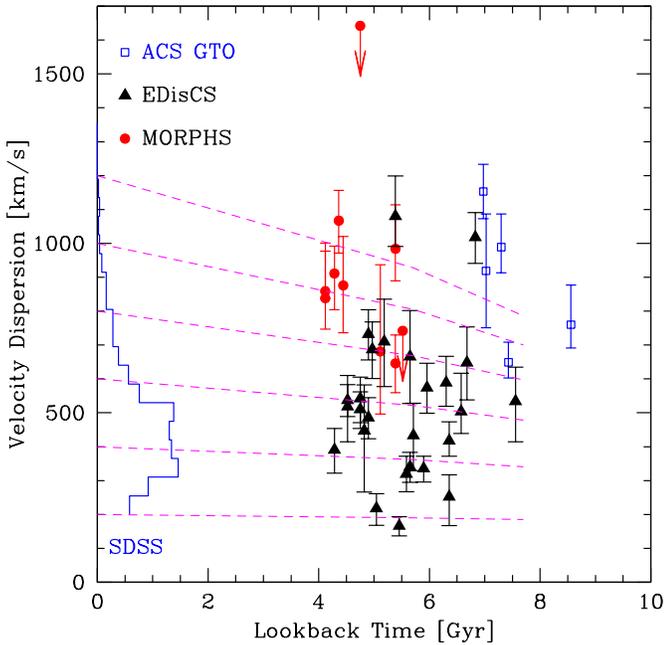}
\caption[]{%
The distribution of velocity dispersion $\sigma$
vs lookback time for EDisCS and for two other
well-studied cluster samples at similar redshifts,
as well as for a well-studied local sample.
The figure shows:
SDSS (blue histogram) at $z < 0.1$,
MORPHS (red circles) at $0.37 < z < 0.56$,
EDisCS (black triangles) at $0.40 < z < 0.96$, and
ACS GTO (blue squares) at $0.8 < z < 1.3$.
The EDisCS clusters fill the gap in
lookback time between the MORPHS and the ACS GTO
clusters and have a large range in $\sigma$.
The dashed lines show % the prediction of
how $\sigma$ is expected
to evolve with redshift from $z=1$ to $z=0$ \citep{Poggianti_etal:2006}.
 From these curves it is apparent that EDisCS is a high redshift cluster
sample for which a majority of the clusters can be progenitors
of ``typical'' low redshift clusters.
References for the plotted velocity dispersions:
SDSS: \citet{vonderLinden_etal:2007};
MORPHS:  \citet{Girardi_Mezzetti:2001};
EDisCS: \citet{Halliday_etal:2004} and this paper;
ACS GTO:
\citet{Gioia_etal:2004,Demarco_etal:2005,Gal_Lubin:2004,Demarco_etal:2004};
see also \citet{Postman_etal:2005}.
The shown SDSS sample contains 488 clusters selected to have
$\sigma > 200\,\mathrm{km}\,\mathrm{s}^{-1}$ and
$\sigma/\mathrm{uncertainty}(\sigma) > 4$
as measured by \citet{vonderLinden_etal:2007}
for a subset of the C4 cluster sample originally compiled by
\citet{Miller_etal:2005}.
}
\label{fig:sigma_vs_z}
\end{figure}

With spectroscopic velocity dispersions, $\sigmaspec$, available for all
the EDisCS clusters, we can compare these values with the
singular isothermal-sphere velocity dispersions from the weak-lensing analysis
from \citet{Clowe_etal:2006}, $\sigmalens$.
The weak-lensing analysis derived a velocity dispersion for the main cluster
in each field, and noted if additional mass peaks were present in the lensing
maps.
In Fig.~\ref{fig:lensing} we plot $\sigmalens$ vs $\sigmaspec$ for the
19 main clusters.
The blue triangles are the clusters with other structures sufficiently
close by in redshift-space, that may directly affect the lensing measurements
\citep{Clowe_etal:2006}, and the red squares are the rest of the clusters.
Visually there is a fairly convincing positive correlation between
the two velocity dispersion measurements. % (with substantial scatter).
Kendall and Spearman rank correlation tests (e.g.\ \citealt{Press_etal:1992})
lend some support to this:
the probability of \emph{no} correlation comes out to
9\% and 4\%, respectively.
There is no clear offset between the clusters with other peaks in the
lensing maps (blue triangles) and the rest of the clusters (red squares).
The 3 clusters in the plot with a significant spectroscopic detection
of substructure based on the Dressler--Shectman test
(probability of \emph{no} substructure $\le$ 5\%,
\citealt{Halliday_etal:2004} and Sect.~\ref{sec:DStest}) are indicated
with large circles.
(The fourth cluster with such a detection, cl1037.9$-$1243a, is not shown
in the plot since it is not a main cluster.)
One could have expected that $\sigmaspec$ for these clusters
would have been higher for their mass (and thus $\sigmalens$)
than for the other clusters, but the limited data in Fig.~\ref{fig:lensing}
do not indicate this.
This may indicate that the detected substructure
does not have a strong effect on the measured spectroscopic
velocity dispersions.
Of the remaining 16 clusters in the plot, the Dressler--Shectman test does
not find significant substructure for 10 of the clusters, and for the
last 6 clusters the test has not been performed due to the number of
spectroscopic members being less than 20.
Among the outliers in the plot the case of cl1103.7$-$1245 % ($z=0.96$)
can easily be explained: the extra lensing signal is likely due to
the secondary cl1103.7$-$1245a cluster % ($z=0.63$)
to the south.
A detailed analysis of the comparison between spectroscopic and lensing
velocity dispersions will be presented in Clowe et al., in prep.

The (spectroscopic) velocity dispersions for the % optically selected
EDisCS clusters are generally lower than the velocity dispersions for
other well-studied samples of clusters at similar redshifts.
This is illustrated in Fig.~\ref{fig:sigma_vs_z},
which plots velocity dispersion versus lookback time
for the EDisCS clusters as well as for
the MORPHS  clusters (e.g.\ \citealt{Smail_etal:1997:morph_types}) and 
the ACS GTO clusters \citep{Postman_etal:2005}.
The histogram on the left side shows the distribution of the
velocity dispersions of a sample of groups and clusters in the
SDSS, as described by \citet{vonderLinden_etal:2007}. This sample
is based on the C4 cluster sample \citep{Miller_etal:2005}, but
redefines the cluster centres and velocity dispersions. In
particular, the velocity dispersions are computed in a similar
fashion to those for the EDisCS sample.
The dashed lines show % the prediction of
how $\sigma$ is expected
to evolve with redshift \citep{Poggianti_etal:2006}. From these curves, it
is apparent that EDisCS is a high-redshift cluster sample for
which a majority of the clusters can be progenitors
of ``typical'' low-redshift clusters.

\section{Summary}
\label{sec:summary}

As part of the ESO Distant Cluster Survey (EDisCS), we have carried out
spectroscopic observations with VLT/FORS2 of galaxies in 20 survey fields.
In our first paper \citep{Halliday_etal:2004}, data for 5 fields were presented,
and in this paper we have presented the data for the remaining fields.
We have provided details of the target selection procedure, and we have
shown how a conservative use of photometric redshifts has given an
efficiency increase of almost a factor of 2, while only missing about 3\% of
the cluster members being targeted.
For all intents and purposes, we expect that our spectroscopic sample
of galaxies at the targeted redshifts behaves as an $I$--band selected sample.
In the data reduction, we have paid particular attention to the sky subtraction.
We have implemented the method from \citet{Kelson:2003}
of performing sky subtraction prior to any rebinning/interpolation of the data.
This method delivers photon-noise-limited results, whereas the traditional
method of subtracting the sky after the data have been rebinned/interpolated
results in substantially larger noise for spectra from tilted slits
(about half of our slits are tilted to be along the major axes of the
galaxies).
The difference between the two methods is found
where the gradient in the sky background is large,
i.e.\ at the edges of the skylines (cf.\ \citealt{Kelson:2003}).
For our data, the difference in noise can reach a factor of 10.
The difference increases with the total number of
collected sky counts, indicating that the longer the total exposure time is,
the more of a problem the excess noise in the traditional sky subtraction
becomes.
We provide data tables containing position, redshifts and $I$--band magnitude
for galaxies in 14 fields.
Cluster redshifts and velocity dispersions are presented for 21 clusters
located in these fields.
Together with the clusters from \citet{Halliday_etal:2004},
velocity dispersions in the range
$166\,\mathrm{km}\,\mathrm{s}^{-1}$--$1080\,\mathrm{km}\,\mathrm{s}^{-1}$
are available for 26 EDisCS clusters with redshifts in the range 0.40--0.96.
For clusters with at least 20 spectroscopically-confirmed members
(9 clusters out of the 21 clusters from this paper),
we have performed the Dressler--Shectman test for cluster substructure.
Significant detections were obtained for 2 of the clusters.
Combined with the results from \citet{Halliday_etal:2004}
substructure is detected at the 95\% confidence level for 4 clusters
out of 14 clusters tested.
We have taken a first look at the comparison between
the velocity dispersions from the weak-lensing analysis
\citep{Clowe_etal:2006}, and those derived using spectroscopic redshifts.
The two quantities show a reasonable agreement.
The few clusters with detected substructure do not show an offset from
the rest of the clusters, possibly indicating that the detected substructure
does not have a strong effect on the measured spectroscopic
velocity dispersions.
A detailed analysis of the comparison between lensing and spectroscopic 
velocity dispersions will be presented in a future paper
(Clowe et al., in prep.).
We have finally noted that the EDisCS clusters, of which many have fairly
modest velocity dispersions ($\sim$$500\,\mathrm{km}\,\mathrm{s}^{-1}$),
is a high-redshift cluster sample for
which a majority of the clusters can be progenitors
of ``typical'' low-redshift clusters. 

Therefore, both this property and the large range of masses
spanned qualify the EDisCS cluster sample as an unprecedented and unique
dataset to  study the processes affecting cluster galaxy evolution as a
function of cluster mass.
Future papers include studies of
the optical and NIR luminosity functions,
the stellar masses,
the stellar populations,
the spectral types,
the gas phase metallicities,
the star formation histories,
the dependence of galaxy properties on density,
the bar fractions,
the Fundamental Plane and
the Tully--Fisher relation.

\begin{acknowledgements}
The ESO Paranal staff is thanked for their assistance during the observations.
Frank Valdes from the IRAF Project % IRAF Programming Group at NOAO
is thanked for writing the task \procedurename{fceval} for us.
Scott Burles and David Schlegel are thanked for making their
B-spline fitting procedures in the \procedurename{idlutils} library
available to to the community.
We thank the referee Dr.\ Florence Durret for constructive comments
that helped improve the presentation.
BMJ acknowledges financial support from the Villum Kann Rasmussen foundation.
BMJ, SN \& RPS acknowledge financial support from the Deutsche
Forschunggemeinschaft (DFG), SFB 375 (Astroteilchenphysik). % \tmp{correct??}
The Dark Cosmology Centre is funded by the Danish National Research
Foundation.
This research has made use of NASA's Astrophysics Data System Bibliographic
Services.
This research used the facilities of the Canadian Astronomy Data Centre
operated by the National Research Council of Canada with the support of
the Canadian Space Agency.
\end{acknowledgements}

\bibliographystyle{aa} % style aa.bst
\bibliography{papers_cited_by_milvang,intro_extra_claire,intro_extra_steven}

\appendix

\section{Quantitative comparison of the performance
of the two sky subtraction methods}
\label{appendix:skysub_compare}

\begin{figure*} % Two column figure
\includegraphics[width=1.00\textwidth,bb = 5 339 569 670]
  {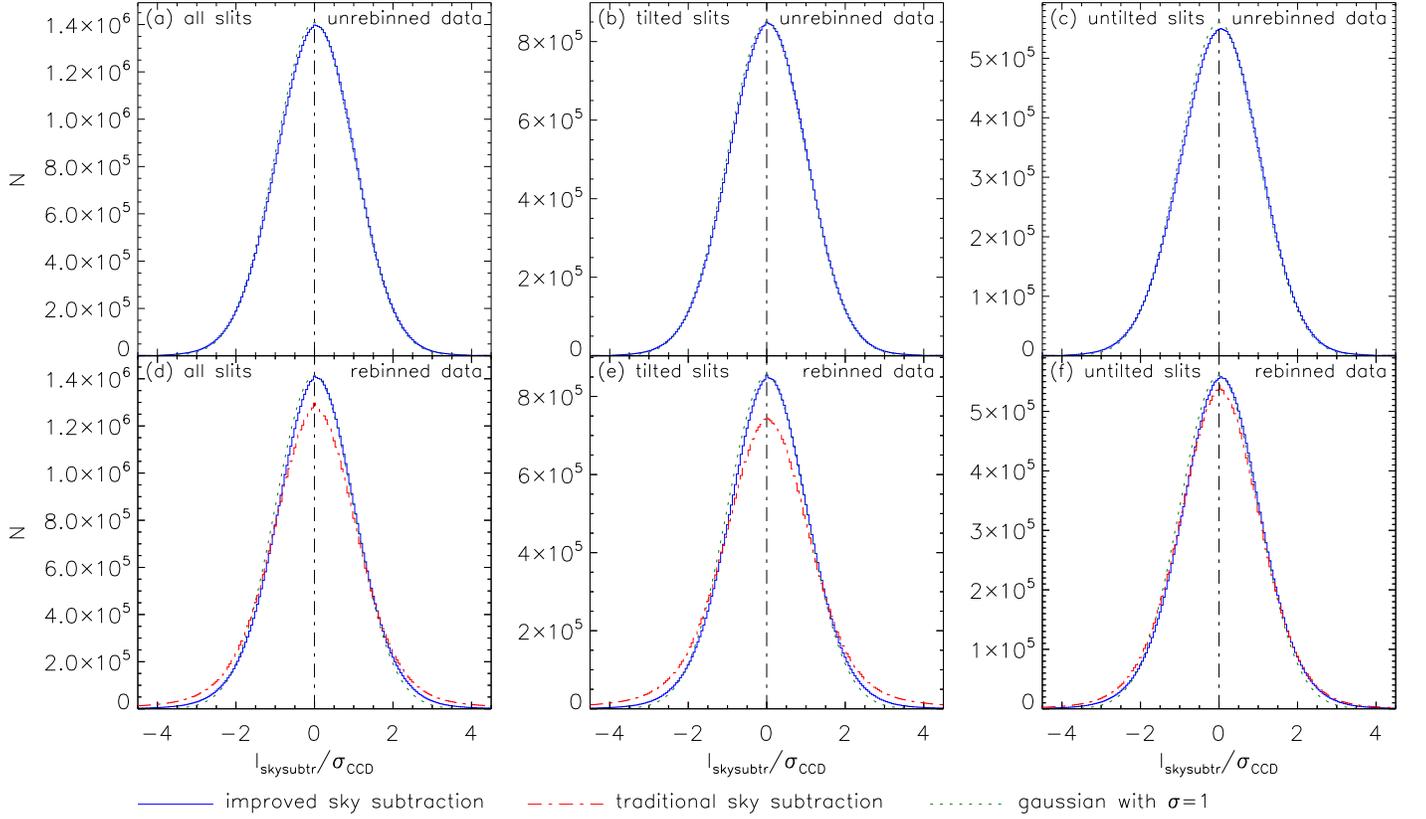}
\caption[]{%
Illustration of the performance of the two sky subtraction methods
for the spectra from all the masks and for all wavelengths.
The figure shows
histograms of $\Iskysubtr/\sigmaCCD$, where $\Iskysubtr$ are the pixel values
in the background-subtracted 2D galaxy spectra, and $\sigmaCCD$ are the
corresponding 2D spectra giving the standard deviation expected from the CCD
noise model (photon noise and read-out noise;
Eq.\ \ref{eq:sigmaCCDunrebinned} and \ref{eq:sigmaCCDrebinned}).
Only pixels in the manually determined background regions were used
to make the histograms, thus excluding practically all signal from known
objects in the spectra (galaxies, stars).
The histograms therefore illustrate the scatter caused
by both natural noise sources (photon noise and read-out noise) and by
possible imperfections in the sky subtraction.
Solid/blue histograms: improved sky subtraction (i.e.\ sky subtraction
performed on the unrebinned data);
dash-dotted/red histograms: traditional sky subtraction (i.e.\ sky subtraction
performed on the rebinned data).
Dotted/green curves: Gaussians with $\sigma = 1$, for reference.
Panels (a)--(c) show unrebinned data 
(where sky-subtracted frames are only available for
the improved sky subtraction) while
panels (d)--(f) show rebinned data (where sky-subtracted frames 
are available for both types of sky subtraction).
Panels (a)+(d) show data from all slits, while the data have been split in
tilted and untilted slits in panels (b)+(e) and (c)+(f), respectively.
The main conclusions from this figure are:
(i)~The result from the improved sky subtraction is close to the
CCD noise limit, since the data in panel (a) [solid/blue curve]
agree so well with a $\sigma = 1$ Gaussian [dotted/green curve];
(ii)~The improved sky subtraction is better than the traditional one,
with a large improvement for tilted slits (panel e) and a smaller improvement
for untilted slits (panel f).
}
\label{fig:backsub_hist}
\end{figure*}

\begin{figure*} % Two column figure
\includegraphics[width=1.00\textwidth,bb = 3 184 572 753]
  {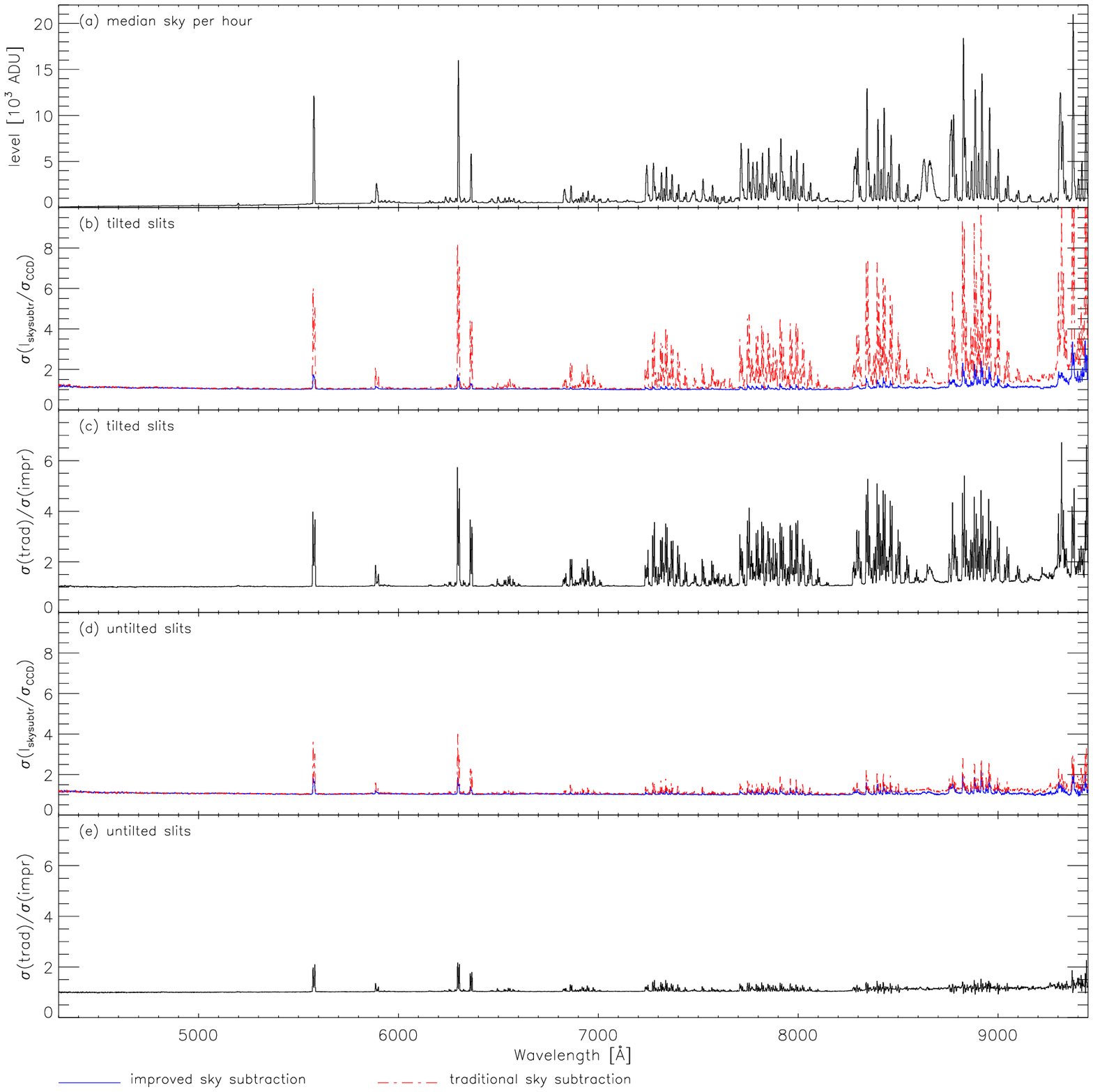}
\caption[]{%
Illustration of the performance of the two sky subtraction methods
for the spectra from all the masks as function of wavelength.
The data are plotted against wavelength in bins of 1.6$\,${\AA}\@.
Panel (a) shows a sky spectrum for reference.
Panel (b) shows $\sigma(\Iskysubtr/\sigmaCCD)$, which is
the standard deviation of the $\Iskysubtr/\sigmaCCD$ values in the given bin.
Results from both sky subtraction methods are plotted.
A value of 1 represents the noise floor set by photon noise and read-out noise.
Panel (c) shows the ratio of $\sigma(\Iskysubtr/\sigmaCCD)$
for the two methods, illustrating that the traditional sky subtraction
has several times more noise than the improved sky subtraction
at the locations of skylines. This plot is based on all masks; had we only
plotted the masks with the longest exposure times (and hence the largest
number of collected sky counts) the difference between the two methods
would have been larger.
Panels (b) and (c) are for spectra from tilted slits, whereas
panels (d) and (e) are for spectra from untilted slits.
}
\label{fig:backsub_lambda}
\end{figure*}

To quantify the performance of the two sky subtraction methods
(traditional and improved),
we will use the values in the sky-subtracted 2D spectra ($\Iskysubtr$).
We will only use the pixels located in the manually-determined background
regions, which are located away from the known objects on the slits.
In these regions, there is practically no signal from astronomical objects
(galaxies, stars), so in the sky-subtracted spectra the pixel values will 
scatter around zero. The scatter will come from two sources:
``natural'' sources found in any CCD frame (photon noise and read-out noise)
and possible extra noise from an imperfect subtraction of the skylines.
The scatter will vary greatly with wavelength due to the emission-line
nature of the sky background spectrum. To normalise things, we will divide
the sky-subtracted 2D spectra by the corresponding 2D spectra giving
the noise expected from the CCD noise model (photon noise and read-out noise).
For \emph{uncorrelated pixel values}, such as those in the 
combined but unrebinned (uninterpolated) spectra
pixelised in the original coordinates $(x,y)$,
the expected noise (in ADU) from the CCD noise model is
\begin{equation}
\sigmaCCDunrebinned = \sqrt{
  \frac{\Inonskysubtrunrebinned}{\nave K} +
  \left(\frac{R_\mathrm{ADU}}{\sqrt{\nave}}\right)^2
} \quad ,
\label{eq:sigmaCCDunrebinned}
\end{equation}
where $\Inonskysubtrunrebinned$ (in ADU) represents the combined,
non-sky-subtracted 2D spectrum,
a spectrum that was created as an average of $\nave$
individual exposures each with conversion factor $K$ (in e$^-$/ADU)
and read-out noise $R_\mathrm{ADU}$ (in ADU)\@.
For our dataset, we have $K = 0.70\,$e$^-$/ADU and
$R_\mathrm{ADU}$ = 4.14$\,$ADU for chip~1, and 4.50$\,$ADU for chip~2.

For the rebinned (interpolated) spectra 
pixelised in $(\xr,\yt)$, things are more complicated
due to the correlated errors introduced by the interpolations
(first in $y$ to remove the spatial curvature, and then in $x$ to apply
the 2D wavelength calibration).
In principle, one could calculate the expected noise
in the rebinned 2D spectrum, $\sigmaCCDrebinned$,
by following how the errors propagate and become correlated
through the two interpolations (rebinnings).
In practice, this is complicated, so we will take a simpler approach
and calculate a quantity $\sigmaCCDrebinnedtilde$ that is equal to
$\sigmaCCDrebinned$ on average and thus equally suitable for
statistical comparisons.
Imagine two pixels in the unrebinned spectrum,
with values $f_1$ and $f_2$ drawn from
identical Gaussian parent distributions with standard deviation $\sigma_f$.
We do a linear interpolation defined by
\begin{equation}
g = \alpha f_1 + (1-\alpha) f_2, \quad 0 \le \alpha \le 1
\label{eq:linear_interpolation}
\end{equation}
to derive the value $g$ in the rebinned pixel.
The expected standard deviation of $g$, $\sigma_g$, can be calculated
from Eq.~(\ref{eq:linear_interpolation}) using the
propagation of errors formula for uncorrelated errors,
which gives $\sigma_g^2 = \alpha^2 \sigma_f^2 + (1-\alpha)^2 \sigma_f^2$.
This reduces to
\begin{equation}
\frac{\sigma_g}{\sigma_f} = \sqrt{2\alpha^2 + 1 - 2\alpha} \quad .
\label{eq:linear_interpolation_sigma}
\end{equation}
This has the following well known consequences:
For $\alpha = 0$ (i.e.\ no interpolation), we obtain $\sigma_g/\sigma_f = 1$,
meaning that the noise does not change (trivial).
And for $\alpha = 0.5$ (i.e.\ taking the average of two values), we obtain
$\sigma_g/\sigma_f = 1/\sqrt{2}$, meaning that the noise goes down by
a factor of $\sqrt{2}$, at the expense of inheriting correlated errors
with the neighbouring pixel.
The two values 1 and $1/\sqrt{2}$ are the extremes of
Eq.~(\ref{eq:linear_interpolation}). 
The mean value is found by integrating over $\alpha$ from 0 to 1 and
comes out to $\approx$0.81.
When we perform another linear interpolation orthogonal to the first one
the same arguments apply, and the noise goes down by another factor of
$\approx$0.81 on average, i.e.\ by a factor of $\approx$0.66 in total.
The following 2D spectrum
\begin{equation}
\sigmaCCDrebinnedtilde \equiv 0.66 \sqrt{
  \frac{\Inonskysubtrrebinned}{\nave K} +
  \left(\frac{R_\mathrm{ADU}}{\sqrt{\nave}}\right)^2
} \quad 
\label{eq:sigmaCCDrebinned}
\end{equation}
will therefore on average provide the correct expected noise in the
rebinned spectrum (i.e.\ when averaging over all the pixels in the spectrum),
but for individual pixels the correct factor may not be 0.66 but somewhere
between 0.5 and 1.

We note that we are concerned with the expected noise in a single pixel.
If we had wanted to calculate the expected noise in, e.g., the sum of
the values in a box of $10\times10$ pixels, the answer would have been
different.

In the following, we will simplify the notation and use $\sigmaCCD$ to denote
$\sigmaCCDunrebinned$ (Eq.~\ref{eq:sigmaCCDunrebinned})
when dealing with the unrebinned data, and
$\sigmaCCDrebinnedtilde$ (Eq.~\ref{eq:sigmaCCDrebinned})
when dealing with the rebinned data.

Our basic quantity for the analysis of the performance
of the two sky subtraction methods is $\Iskysubtr/\sigmaCCD$
(for pixels in the background regions, which will be implicit from now on).
Figure~\ref{fig:backsub_hist} shows histograms $\Iskysubtr/\sigmaCCD$\@.
In the first row of panels, the unrebinned data have been used,
and here only the improved sky subtraction is available.
Also shown are Gaussians with $\sigma = 1$, for reference.
Panel (a) is for all the slits, whereas panels (b) and (c) shows data from
tilted and untilted slits, respectively.
The histograms of $\Iskysubtr/\sigmaCCD$ in panels (a)--(c) agree very well
with the $\sigma = 1$ Gaussians, which indicates that the improved
sky subtraction is very close to the noise floor set by photon noise and
read-out noise.
The second row of panels are for the rebinned data. Here we have used the
approximate formula for $\sigmaCCD$ (Eq.~\ref{eq:sigmaCCDrebinned}). The
histograms for the improved sky subtraction (blue solid histograms)
still resemble the
$\sigma = 1$ Gaussians quite well, indicating that the used approximation is
valid on average. For this reason, we will only use the rebinned data in
the following figures
(Fig.~\ref{fig:backsub_lambda}--\ref{fig:backsub_collected_sky_6300}),
since here we can compare the two sky subtraction methods
(the traditional sky subtraction is by its nature only available
for the rebinned data).
The second row of panels of Fig.~\ref{fig:backsub_hist} also show
histograms of $\Iskysubtr/\sigmaCCD$ for the traditional sky subtraction
(red dash-dotted histograms).
It is seen that these histograms are wider than those for the
improved sky subtraction, showing that the traditional sky subtraction
has larger noise than the improved sky subtraction. This is particularly
the case for spectra coming from tilted slits (panel e), as expected.

Figure~\ref{fig:backsub_lambda} plots $\Iskysubtr/\sigmaCCD$ in a different
way. The data are split in bins of 1.6$\,${\AA} in wavelength. Instead of
plotting a histogram of the $\Iskysubtr/\sigmaCCD$ values, a robust
(biweight) estimate of their standard deviation,
$\sigma(\Iskysubtr/\sigmaCCD)$, is calculated and plotted versus
wavelength, see panel (b) (tilted slits) and (d) (untilted slits).
The results from both sky subtraction methods are plotted, and their
ratio is plotted in panels (c) and (e).
Panel (a) shows a sky spectrum for reference, and it is seen that the
traditional sky subtraction has several times larger noise than the
improved sky subtraction at the location of the skylines.
It is also seen that this difference in noise increases with the strength of
the skyline. This indicates that the extra noise found in the traditional
sky subtraction is a stronger function of the sky level than the
square root which enters $\sigmaCCD$ (Eq.~\ref{eq:sigmaCCDrebinned}).

Figure~\ref{fig:backsub_lambda_zoom_selmask}
is akin to a zoom of Fig.~\ref{fig:backsub_lambda}
centered at the strong 6300$\,${\AA} skyline.
The figure shows that the traditional sky subtraction method
has the highest noise at the edges of the skylines, i.e.\ where the
gradient in the sky background is the largest (cf.\ \citealt{Kelson:2003}).
Panel (c) shows that for tilted slits in this particular mask
the noise in the traditional sky subtraction is 7--8 times larger 
than the noise in the improved sky subtraction
at the edges of this skyline.

Figure~\ref{fig:backsub_collected_sky_6300} shows
$\sigma(\Iskysubtr/\sigmaCCD)$ versus the number of collected sky counts
for both sky subtraction methods and for tilted slits (panel a) and
untilted slits (panel b).
As before, the quantity $\sigma(\Iskysubtr/\sigmaCCD)$ would be unity if
the noise in the sky-subtracted spectra was at the noise floor
set by photon noise and read-out noise.
What is seen most clearly in panel~(a) is that the improved sky subtraction
(blue crosses) is almost at the noise floor, with just a small excess noise
that increases weakly with the number of collected sky counts.
The traditional sky subtraction (red triangles) is much above the noise floor,
and $\sigma(\Iskysubtr/\sigmaCCD)$ increases with the number of collected
sky counts in a way resembling a square root function.
Since $\sigma(\Iskysubtr/\sigmaCCD)$ has already been divided by
$\sigmaCCD$ (which essentially is proportional to the square root of the
number of collected sky counts), the plot indicates that the extra noise in
the traditional sky subtraction goes linearly with the number of collected
sky counts.
This has the implication that for increasingly long total exposure times,
the systematic extra noise in the traditional sky subtraction becomes larger
and larger compared to the photon noise.

\begin{figure} % One column figure
\makebox[0.48\textwidth]{
  \includegraphics[width=0.48\textwidth,bb = 3 396 302 753]
    {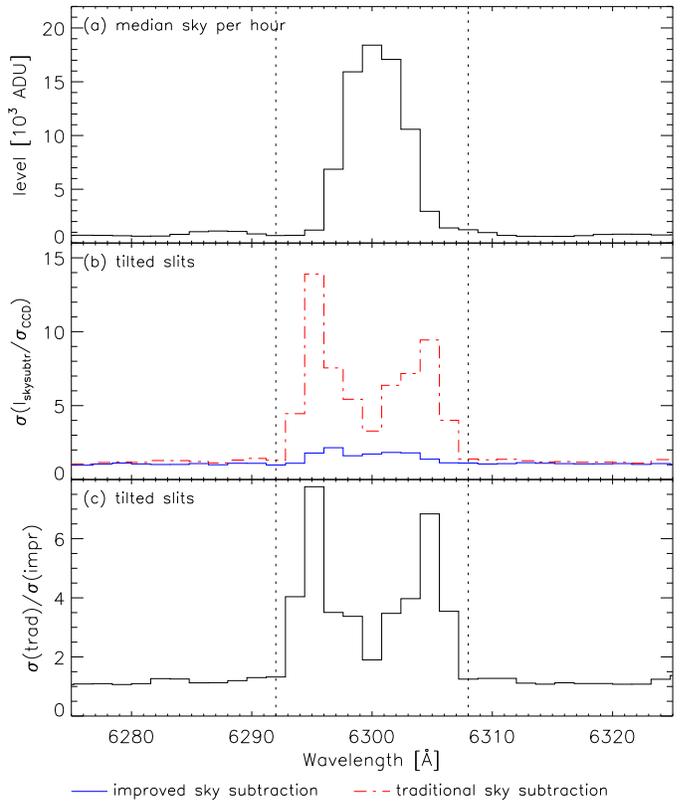}
}
\caption[]{%
Illustration of the fact
that the traditional sky subtraction method
has the highest noise at the edges of the skylines, i.e.\ where the
gradient in the sky background is the largest
(cf.\ \citealt{Kelson:2003}).
This figure is akin to a zoom of Fig.~\ref{fig:backsub_lambda}
centered at the strong 6300$\,${\AA} skyline, but only data
from a single mask have been used
(and we note that the $y$--axis range for panel b has been increased).
If this figure had been made using all 51 masks it would have looked
rather similar, but the peaks in panel (b) and (c) would not have been so
sharp due to the small wavelength shifts that exist between the masks
due to instrument flexure.
The dotted lines indicate the wavelength region used
for the statistics shown in Fig.~\ref{fig:backsub_collected_sky_6300}.
}
\label{fig:backsub_lambda_zoom_selmask}
\end{figure}

\begin{figure} % One column figure
\makebox[0.48\textwidth]{
  \includegraphics[width=0.40\textwidth,bb = 25 369 234 751]
    {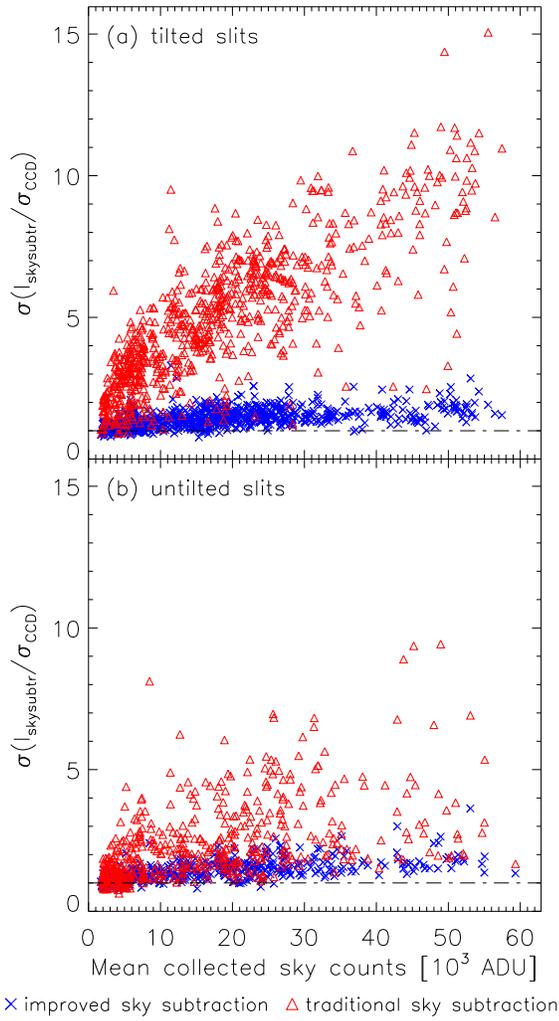}
}
\caption[]{%
Illustration of how the noise in the sky-subtracted spectra
depends on the collected sky counts.
Only data in the narrow wavelength range $6300\pm8\,${\AA} have been used
(cf.\ Fig.~\ref{fig:backsub_lambda_zoom_selmask}).
Data from all 51 masks have been used.
The $x$--axis shows the mean collected sky counts over the total
exposure time for the given spectrum,
with the mean being taken over the used wavelength range.
The quantity on the $x$--axis thus depends linearly
on the total exposure time and on the sky brightness at 6300$\,${\AA}\@.
The $y$--axis shows $\sigma(\Iskysubtr/\sigmaCCD)$
which was also used in Fig.~\ref{fig:backsub_lambda} and
\ref{fig:backsub_lambda_zoom_selmask}, just here computed 
in the single wavelength bin of $6300\pm8\,${\AA} instead of in 
multiple bins of 1.6$\,${\AA}\@.
The quantity on the $y$--axis is the noise relative to 
$\sigmaCCD$ (the noise expected from photon noise and read-out noise).
The horizontal dot--dashed line at 1 represents
the noise floor set by photon noise and read-out noise.
The points for the traditional sky subtraction show a square root like
behaviour. Since the quantity on the $y$--axis has already been divided by
$\sigmaCCD$ (which essentially is proportional to the square root of the
quantity on the $x$--axis) the plot indicates that the extra noise in
the traditional sky subtraction goes linearly with the number of collected
sky counts.
}
\label{fig:backsub_collected_sky_6300}
\end{figure}

\end{document}